\documentclass[fleqn,usenatbib]{mnras}

\usepackage{newtxtext}

\usepackage[T1]{fontenc}
\usepackage{hyperref}

\DeclareRobustCommand{\VAN}[3]{#2}
\let\VANthebibliography\thebibliography
\def\thebibliography{\DeclareRobustCommand{\VAN}[3]{##3}\VANthebibliography}

\usepackage{graphicx}	
\usepackage{subcaption}
\usepackage{amsmath}	
\usepackage{amssymb}	
\usepackage{soul}
\usepackage[normalem]{ulem}
\usepackage{float}
\usepackage{multirow}
\usepackage[dvipsnames]{xcolor,colortbl}
\usepackage{orcidlink}
\graphicspath{{./}{figures/}}

\def\mseventwelve{\mbox{$M_\ast\sim 7\times 10^{12}M_\odot$}} 
\def\mstentotheten{\mbox{$M_\ast\sim10^{10}M_\odot$}} 
\def\mssimtombar{\mbox{$M_\ast\sim 0.16\times M_\text{vir}$}} 
\def\mslessmbar{\mbox{$M_\ast< 0.16\times M_\text{vir}$}} 
\def\zrangetwotosix{\mbox{$2\lesssim z\lesssim6$}}

\title[How massive can galaxies get in $\Lambda$CDM?]{Empirical estimates of how massive galaxies can be in $\Lambda$CDM}

\author[M. Enríquez-Vargas et al.]{
\parbox{17cm}{
Miguel Enríquez-Vargas\orcidlink{0000-0001-8537-5079},$^{1}$\thanks{E-mail: mvargas@astro.unam.mx (MEV)}
Aldo Rodríguez-Puebla{\orcidlink{0000-0002-0170-5358}},$^{1}$\thanks{E-mail: apuebla@astro.unam.mx (ARP)}
Aditya Manuwal{\orcidlink{0000-0003-2893-2793}},$^{1}$
L. Y. Aaron Yung{\orcidlink{0000-0003-3466-035X}},$^{2}$
Vladimir Avila-Reese{\orcidlink{0000-0002-3461-2342}}$^{1}$
and Carlo Cannarozzo{\orcidlink{0000-0003-3843-7366}}$^{3,4}$ 
}
\\ 
\\
$^{1}$Universidad Nacional Autónoma de México, Instituto de Astronomía, A. P. 70-264, 04510, Ciudad de México, México\\
$^{2}$Space Telescope Science Institute, 3700 San Martin Drive, Baltimore, MD 21218, USA\\
$^{3}$New York University Abu Dhabi, PO Box 129188, Abu Dhabi, United Arab Emirates \\
$^{4}$Center for Astrophysics and Space Science (CASS), New York University Abu Dhabi\\ 
}

\date{Accepted XXX. Received YYY; in original form ZZZ}

\pubyear{2026}

\begin{document}
\label{firstpage}
\pagerange{\pageref{firstpage}--\pageref{lastpage}}
\maketitle

\begin{abstract} 
Using Extreme Value Statistics applied to the observed galaxy stellar mass and the UV luminosity functions, we empirically estimate masses and luminosities of the most extreme galaxies in cosmological surveys, including the full sky. We incorporate uncertainties in stellar mass measurements (Eddington bias) and the scatter in the stellar–halo mass relation to derive empirical limits for galaxies residing in the most massive halos. The maximum observed $M_\ast$ strongly depends on survey area and redshift, ranging from \mseventwelve{} for full-sky surveys at $z\sim0$ to \mstentotheten{} at $z\sim16$. Massive galaxies, particularly at high redshift, approach the theoretical maximum baryonic mass available in halos, \mssimtombar{}, consistent with previous claims. Accounting for measurement uncertainties significantly reduces the inferred maximum $M_\ast$ by up to $\sim1$ dex at $z\gtrsim10$, yielding stellar masses consistent with \mslessmbar{} at all redshifts. Assuming a perfect rank-order correspondence between the most massive halos and galaxies would guarantee this inequality at all redshifts. At \zrangetwotosix{}, the most massive galaxies have stellar masses comparable to the total cold gas reservoir from cold and cooling flows, suggesting near-maximal star formation efficiencies, SFEs. At higher redshifts, halos are predicted to host galaxies undergoing starburst phases. When accounting for dust attenuation and adopting empirically inferred SFEs, we find good agreement between the model and the brightest observed UV galaxies at high redshifts. At lower redshifts, however, observed UV galaxies are too bright. Overall, our results indicate that current observations remain broadly consistent with $\Lambda$CDM once statistical and observational effects are properly accounted for.
\end{abstract}


\section{Introduction} \label{sec:intro}

The Lambda Cold Dark Matter ($\Lambda$CDM) model provides a robust cosmological framework for describing the hierarchical assembly of cosmic structures, from the Cosmic Microwave Background (CMB) anisotropies to the formation of dwarf, normal, and massive galaxies in clusters \citep[see][for a recent discussion on the status of $\Lambda$CDM; see also \citealp{DESI2025}, which presents indications of a time-evolving equation of state for dark energy]{Peebles2025}. Within this framework, galaxy formation is fundamentally linked to the growth and evolution of dark matter halos, whose abundances are well described by N-body cosmological simulations constrained through precise observational measurements \citep{Planck+2015,DESI2025}. In particular, the dark and baryonic matter content of the universe has been constrained to within a few percent, with baryons comprising approximately $\sim16\%$ of the total matter. Consequently, this framework sets clear theoretical limits on the baryonic mass that can accumulate within halos, providing a baseline for the maximum stellar masses expected throughout cosmic history.

As cosmology enters an era of increasingly precise observational constraints driven by cutting-edge experiments, galaxy studies are undergoing a similar transformation. In its first years of operation, the James Webb Space Telescope (JWST) has already significantly contributed to our understanding of galaxy formation and evolution. Recent JWST observations have revealed a population of unexpectedly massive galaxies at high redshifts \citep[$6\lesssim z \lesssim 10$,][]{Labbe+2023}, some with inferred stellar masses exceeding $M_\ast\sim10^{10} M_\odot$. In parallel, results from the JWST survey indicate a surprisingly shallow evolution in the galaxy number density, suggesting a higher abundance of massive systems at early cosmic times than anticipated \citep{Castellano+2025}. Early JWST studies have also reported hints of an overabundance of galaxies in the high-redshift universe \citep{Finkelstein+2023,Labbe+2023,Finkelstein+2025}. These discoveries collectively point toward a notably accelerated stellar mass assembly shortly after the Big Bang. Interpreting these systems within the standard $\Lambda$CDM framework has led to suggestions that they may require star formation efficiencies approaching or even exceeding the available baryonic budget \citep{Boylan-Kolchin_2023,Dekel+2023,Lovell+2023}. However, recent work has shown that, when accounting for systematic uncertainties and revised modeling assumptions, the inferred stellar masses and efficiencies may remain consistent with the cosmic baryon budget and do not necessarily imply extreme star formation efficiencies \citep{Yung+2024,Yung+2025,Somerville+2025}. Alternative interpretations have also explored extensions to the standard cosmological model, such as early dark energy scenarios \citep{Shen+2024}.  

Concerns about the existence of massive galaxies at high redshift and their compatibility with theoretical expectations from $\Lambda$CDM have been raised before. An early example is \citet{Steinhardt+2016}, who highlighted galaxies at $z \sim 4{-}6$ with masses seemingly too large for standard galaxy formation models \citep[but see][]{Behroozi-Silk2018}. The latest JWST observations have reignited debates on whether these massive, early-forming galaxies pose a fundamental challenge to standard cosmological models or whether the apparent tension arises from observational uncertainties in stellar mass estimates \citep{Behroozi-Silk2018,Chen+2023,Rodriguez-Puebla+2025}, selection biases \citep{Furlanetto+2023,Kocevski+2023,Zavala+2023,Labbe+2025}, or alternative interpretations of the observed sources, such as active galactic nuclei \citep[AGN,][]{Kocevski+2023,Desprez+2024,Trussler+2024} or even brown dwarfs \citep{Desprez+2024}.

One of the main challenges in assessing whether high-redshift massive galaxy observations genuinely conflict with the predictions of the $\Lambda$CDM model lies in the uncertainties surrounding stellar mass estimates. At high redshifts, these estimates are highly sensitive to assumptions about star formation histories, dust attenuation laws, the choice of initial mass function (IMF), and stellar population synthesis models, all of which can introduce substantial systematic biases in the observed galaxy stellar mass function \citep[GSMF;][see Section 4 of \citealp{Rodriguez-Puebla+2017} for a discussion]{Conroy+2009,Behroozi+2010,Pforr+2012,Grazian+2015,Tacchella+2022,Harvey+2025,Weibel+2024}. Furthermore, the intrinsic scatter in the stellar-to-halo mass relation (SHMR), combined with observational uncertainties, can significantly inflate the inferred abundance of massive galaxies due to Eddington bias \citep{Behroozi+2010,Rodriguez-Puebla+2017,Behroozi-Silk2018,Chen+2023,Rodriguez-Puebla+2025}. As a result, the apparent tension with $\Lambda$CDM may not necessarily indicate a fundamental failure of the model, but it could instead arise from systematic and random errors in stellar mass measurements.

This issue has been previously explored by \citet{Behroozi-Silk2018} and more recently by \citet{Chen+2023}, both of whom emphasize the importance of properly accounting for these uncertainties in the context of observations before and during the JWST era, respectively.  In this paper, we revisit this question by setting empirical limits on how massive galaxies can be in the $\Lambda$CDM framework, taking into account observational uncertainties and survey areas. Unlike previous studies, our analysis is fully empirical and extends to cover a broad range of redshifts ($0 \lesssim z \lesssim 17$), including local galaxies, as well as future large-area surveys and full-sky constraints.

To do so, we use the extreme value statistics (EVS) framework, which allows us to quantify the expected distribution of the most massive galaxies as a function of redshift and survey area, independently of detailed galaxy formation models. Furthermore, we compare these distributions with the theoretical upper limits on stellar mass assembly set by the halo mass function, the cosmic baryon fraction and the expected baryonic content available for star formation. By incorporating effects such as random errors in stellar mass estimates and the intrinsic scatter in the SHMR, our goal is to reassess whether current JWST observations are in tension with $\Lambda$CDM or remain consistent within the model’s statistical expectations.

The paper is organized as follows. In Section \ref{sec:framework} we introduce the statistical framework, focusing on the EVS and its connection to galaxies and dark matter halos. Section \ref{sec:EVS_GSMF} presents the empirical EVS distribution of massive galaxies and defines the most massive object, while Section \ref{sec:EVS_HMF} analyzes the EVS of dark matter halos and estimates their baryonic content. Section \ref{sec:connecting_massive_gals_to_halos} connects the distributions of galaxies and halos, and Section \ref{sec:expected_baryons_and_stellar_mass} compares the most massive galaxies with the baryonic reservoirs of their host halos. Together, these sections address the questions: how massive can galaxies become within $\Lambda$CDM? and How efficient can they be in this framework? Finally, Sections \ref{sec:discussion} and \ref{sec:summ_and_concl} present the discussion and conclusions.

\section{The Statistical Framework}
\label{sec:framework}

In this section, we describe the statistical framework used in this paper to address our main questions: How massive can galaxies become within the $\Lambda$CDM model? And how efficient can massive galaxies get in this framework? As part of the goals of this paper, we explore these questions empirically within the context of the galaxy–halo connection.

\subsection{EVS: Extreme Value Statistics}
\label{sec:EVS}

We study how massive galaxies can become in the $\Lambda$CDM model from different perspectives. Unlike previous studies \citep[e.g.,][]{Lovell+2023,Shen+2024}, an important part of our attention in this paper is devoted on understanding the most extreme galaxies from an \emph{empirical} perspective, based on predictions from the \emph{observed} GSMF. Specifically, we first use the EVS framework to empirically determine the most massive galaxies expected from the observed GSMF as a function of redshift and for a given survey area. A key advantage of EVS is that it does not require knowledge or estimation of the data’s selection function \citep{Lovell+2023}, as it focuses on the statistical properties of extreme events drawn from a probability distribution function (PDF) in this case, the GSMF. In the remainder of this section, we follow the framework described by \citet[][see also \citealp{Harrison+2011,Harrison+2012,Holz_Perlmutter2012,Waizmann+2012}]{Lovell+2023}.

In the EVS framework, the cumulative distribution function, CDF, of the maximum stellar mass, $M_{\ast,\text{max}}$, drawn from the PDF associated with the GSMF, is given by
\begin{equation}
    \Phi_\ast(M_{\ast,\text{max}} \leq M_\ast; N) = [F_\ast(M_\ast)]^N.
    \label{eq:CDF_EVS}
\end{equation}
Here, $F_\ast(M_\ast)$ is the CDF associated with the GSMF, and $N$ is the number of independent galaxies in the volume. It follows immediately that the PDF of the most massive galaxy, $M_{\ast,\text{max}}$, is obtained by differentiating the CDF:
\begin{equation}
        \begin{aligned}
        \theta(M_{\ast,\text{max}}) \equiv & \frac{d}{d \log  M_\ast}\Phi_\ast(M_{\ast,\text{max}} = M_\ast; N) \\
        = & N f_\ast(M_\ast) [F(M_\ast)]^{N-1},
        \end{aligned}
    \label{eq:most_massive_pdf}
\end{equation}
where $f_\ast$ is the PDF associated with the GSMF.

To estimate the empirical PDFs and CDFs associated with the GSMF ($\phi_\ast$, in units of $\text{Mpc}^{-3} \ \text{dex}^{-1}$), we begin by defining the total cumulative GSMF:

\begin{equation}
    n_{\ast,\text{tot}} = \int_{-\infty}^{\infty} \phi_\ast (M_\ast) d\log M_\ast,
    \label{eq:n_gal}
\end{equation}
where the integration is performed over $\log M_\ast$, such that the lower limit $-\infty$ corresponds to $M_\ast \to 0$. From which the PDF $f_\ast(M_\ast)$ is associated with the GSMF as:
\begin{equation}
    f_\ast(M_\ast) = \frac{\phi_\ast(M_\ast)}{n_{\ast,\text{tot}}}, 
    \label{eq:pdf_from_gsmf}
\end{equation}
while the corresponding CDF is
\begin{equation}
    F_\ast(M_\ast) = \frac{1}{n_{\ast,{\text{tot}}}} \int_{-\infty}^{M_\ast} \phi_\ast (M) d\log M.
    \label{eq:cdf_from_gsmf}
\end{equation}
Equations (\ref{eq:CDF_EVS})-(\ref{eq:cdf_from_gsmf}) define the empirical EVS framework used to predict the most massive galaxies expected at different redshifts.

Similar to \citet{Lovell+2023}, we account for the fact that galaxy surveys are conducted in light-cones rather than in fixed-redshift volumes, as is the case for snapshots in simulations. To do this, we modify Eqs. (\ref{eq:n_gal})–(\ref{eq:cdf_from_gsmf}) as follows:
\begin{equation}
    n_{\ast,\text{tot}} = \int_\Omega  \int_{z_\text{min}}^{z_\text{max}}\int_{-\infty}^{\infty} \phi_\ast (M_\ast,z) \frac{d^2V_c}{dzd\Omega}  dz \ d\Omega \ d \log M_\ast,
\end{equation}
where $V_c$ is the comoving volume enclosed within the redshift interval $z_\text{min}$–$z_\text{max}$ and solid angle $\Omega$. Similar modifications are applied to $f_\ast$ and $F_\ast$ to account for the observational solid angle. 

For clarity, the total number of galaxies $N$ is computed as
\begin{equation}
N = \left( \int_{M_{\ast,\min}}^{\infty} \phi_\ast(M_\ast)\, d\log M_\ast \right) V_c \ \ .
\end{equation}

For the second part of this paper, we will use the halo mass function to estimate the most massive halos allowed by $\Lambda$CDM and assess whether this is consistent with the logical upper limit set by the universal baryon fraction $f_{\text{bar}} = \Omega_{\text{bar}} / \Omega_{\text{m}} \sim 0.16$, by comparing it with the most massive empirical galaxies observed. In addition, we use a simple model to estimate the total \emph{cold gas} mass expected in the central region of a halo at a given redshift (from cooling and cold accretion), which will be available for star formation, $M_\text{cold}$. The total cold gas mass is expected to be comparable to, or less than, the total baryonic content of dark matter halos, $M_\text{cold} \lesssim f_{\text{bar}} M_\text{vir}$. We emphasis that $M_\text{cold}$ is the most natural quantity to compare with the stellar mass of the galaxy.

The most massive halos, $M_\text{vir,max}$, allowed by the $\Lambda$CDM will be inferred in a similar fashion using Eqs. (\ref{eq:most_massive_pdf})-(\ref{eq:n_gal}), but this time making the following replacement $\phi_\ast \rightarrow\phi_{\text{vir}}$, for which the EVS will be denoted by
\begin{equation}
    \Phi_\text{vir}(M_\text{vir,max}\leq M_\text{vir};N)
     = \left[  F_\text{vir}(M_\text{vir}) \right]^{N},
\end{equation}
and 
\begin{equation}
    \theta_\text{vir} (M_\text{vir,max}) \equiv \frac{d}{d \log M_\text{vir}} \Phi_\text{vir}(M_\text{vir,max}\leq M_\text{vir};N).
\end{equation}

\subsection{The galaxy-halo connection}
\label{sec:gal_halo_conn}

In this paper, we use a generalized (sub)halo abundance matching model, see \citet{Wechsler+2018} for a discussion, that accounts for the scatter around the SHMR. The scatter in the SHMR introduces an effect analogous to the \citet{Eddington1913,Eddington1940} bias. Accounting for this scatter implies that the GSMF (see Eq. \ref{eq:gsmf}) is the result of a convolution between an \emph{intrinsic} GSMF, $ \phi_{\ast,\text{intr}}$, which is directly linked to dark matter halos, and the distribution of scatter and the SHMR. Note that in this subsection we ignore random errors associated with stellar mass; we will return to this issue in Section \ref{sec:GSMF}. As it will become clear below, we define the resulting GSMF from the above convolution as the true GSMF, $\phi_{\ast\text{,true}}$.

Here, we follow the approach described in \citet[][and references therein]{Rodriguez-Puebla+2025}, assuming that the scatter in the SHMR is 0.15 dex and independent of redshift and draw from a Gaussian distribution $\mathcal{H}_\text{gal-halo}$ that represents the width of the scatter around the mean of the SHMR:

\begin{equation}
\begin{split}
\phi_{\ast,\mathrm{true}}(M_\ast) = \int_{-\infty}^{\infty}
\mathcal{H}_{\mathrm{gal-halo}}(\log M_\ast' - \log M_\ast)\, \\
\times \ \phi_{\ast,\mathrm{intr}}(M_\ast')\, d\log M_\ast' \ ,
\end{split}
\label{eq:intrinsci_GSMF}
\end{equation}
where $\log M_\ast'$ is the integration variable in logarithmic stellar mass.
We then deconvolve this equation to derive the intrinsic GSMF.
The intrinsic GSMF is connected to the HMF through the SHMR as
\begin{equation}
\phi_{\ast,\mathrm{intr}}(M_\ast) =
\phi_\text{vir}(M_\text{vir})\left|\frac{d\log M_\text{vir}}{d\log M_\ast}\right|,
\end{equation}
where $\phi_\text{vir}(M_\text{vir})$ is the HMF and the mapping between $M_\ast$ and $M_h$ is given by the mean SHMR.

Using Eq. (\ref{eq:intrinsci_GSMF}) to establish the connection between the observed GSMF and the underlying intrinsic GSMF, we use the deconvolved $\phi_{\ast,\mathrm{intr}}$ to compute the extreme value statistics of the most massive galaxies, $\theta_\ast(M_{\ast,\max})$. We then ask, what is the maximum cumulative efficiency, $\epsilon_\text{max} \equiv M_{\ast,\text{max}}/ (f_\text{bar} \; M_\text{vir,max})$, in the most massive dark matter halos? To estimate this, within the EVS and galaxy-halo connection context, we combine the EVS distributions $\theta_\ast(M_{\ast,\text{max}})$ and $\theta_\text{bar}(f_\text{bar}M_\text{vir,max})$ to derive the EVS distribution of the maximum efficiency in dark matter halos: $\theta_\text{eff}(\epsilon_\text{max})$. This allows us to empirically determine the maximum cumulative efficiency that dark matter halos can achieve for the $\Lambda$CDM model. 

\begin{table*}
    \centering
    \begin{tabular}{c c l}
    \hline
    Variable & Definition & Description \\
    \hline
        $\Phi(X)$ & Eq. (\ref{eq:CDF_EVS}) & The EVS cumulative distribution function of a random variable $X$. \\
        $\theta(X)$ & Eq. (\ref{eq:most_massive_pdf}) & The EVS differential PDF of a random variable $X$ at the interval $\log X \pm d\log X/2$. \\
        $R$ & $M_*/M_{\mathrm{vir}}$  & Stellar-to-halo mass ratio. \\
        $\mathcal{M}$ & $\log \left( M_*/M_\odot \right)$ & Logarithm of the stellar mass. \\
        $\mathcal{H}^1$ & $\log \left( M_{\mathrm{vir}}/M_\odot \right)$ & Logarithm of the halo mass. \\
        $\mathcal{R}$ & $\log R = M - H$ & Logarithm of the stellar-to-halo mass ratio. \\
        $\mathcal{E}$ & $\log \epsilon = \mathcal{R} - \log f_{\mathrm{bar}}$ &  Logarithm of the cumulative star formation efficiency.  
    \end{tabular}
    \caption{Several definitions that are employed in this paper. $^1$Do not confuse with the variable defined as $\mathcal{H}_\text{gal-halo}$ defined above as the Gaussian distribution representing the scatter around the SHMR.}
    \label{tab:definitions}
\end{table*}

We note that Eq. (\ref{eq:intrinsci_GSMF}) accounts for the impact of the intrinsic scatter in the SHMR on the GSMF at the population level. However, it does not uniquely determine how the most massive galaxies and the most massive dark matter halos are paired in the extreme realization. Therefore, within the EVS framework, additional assumptions are required to describe the joint statistics of $M_{*,\max}$ and $M_{\mathrm{vir},\max}$.

We estimate $\theta_\text{eff}(\epsilon_\text{max})$ in two ways. First, we assume that both $M_{\ast,\text{max}}$ and $M_\text{vir,max}$ are independent random variables. This assumption allows for the possibility that the most massive halo does not necessarily host the most massive galaxy, reflecting a fully stochastic pairing between galaxy and halo extremes. By convolving the corresponding EVS distributions $\theta_\ast(M_{\ast,\text{max}})$ and $\theta_\text{bar}(f_\text{bar}M_\text{vir,max})$, we can derive the EVS distribution of the maximum ratio $M_{\ast,\text{max}}/ M_\text{vir,max}$. To do so, we define the variables describe in Table \ref{tab:definitions}.

Since $\theta_\ast(M_{\ast,\text{max}})$ and $\theta_\text{vir}(M_{\text{vir},\text{max}})$ were derived per logarithmic bin, the EVS distributions remain invariant under the change of variables. In particular,
\begin{align*}
\theta_\ast(\mathcal{M}_\text{max})\, d\mathcal{M}_\text{max} &= \theta_\ast(M_{\ast,\text{max}})\, d\log M_{\ast,\text{max}}, \\
\theta_\text{vir}(\mathcal{H}_\text{max})\, d\mathcal{H}_\text{max} &= \theta_\text{vir}(M_{\text{vir},\text{max}})\, d\log M_{\text{vir},\text{max}}, \\
\theta_\text{eff}(\mathcal{R}_\text{max})\, d\mathcal{R}_\text{max} &= \theta_\text{eff}(R_\text{max})\, d\log R_\text{max}, \\
\theta_\text{eff}(\epsilon)\, d\log \epsilon &= \theta_\text{eff}(\mathcal{E})\, d\mathcal{E}.
\end{align*}
Their corresponding cumulative distributions are $\Phi_\ast(\mathcal{M}_\text{max})$, $\Phi(\mathcal{H}_\text{max})$ $\Phi_\text{eff}(\mathcal{R}_\text{max})$ and $\Phi_\text{eff}(\mathcal{E}_\text{max})$.

Thus, our problem reduces to finding the distribution of a random variable composed as the difference of two other random variables. This distribution is given by the following convolution:
\begin{equation}
    \theta_\text{eff}(\mathcal{R}_{\text{max}}) = \int_{-\infty}^{\infty} \theta_\ast(\mathcal{M}_{\text{max}})\theta_\text{vir}(\mathcal{M}_{\text{max}}-\mathcal{R}_{\text{max}}) d \mathcal{M}_{\text{max}}.
    \label{eq:SHMR_ratio_max}
\end{equation}
Note that these functions depend on redshift, but, for simplicity, this dependence is not explicitly shown. Additionally, notice that $\theta_\text{eff}(\mathcal{E}_{\text{max}})d\mathcal{E}_{\text{max}} = \theta_\text{eff}(\mathcal{R}_{\text{max}})d\mathcal{R}_{\text{max}}$. 

Our second method for estimating either $\theta_\text{eff}(\mathcal{R}_{\text{max}})$ or $\theta_\text{eff}(\mathcal{E}_{\text{max}})$ is to rank-order the probabilities such that, at a given redshift, the \emph{most} massive halo drawn from $\theta_\text{vir}(M_\text{vir,max})$ corresponds to the \emph{most} massive galaxy drawn from $\theta_\ast(M_{\ast,\text{max}})$:
\begin{equation}
    \int_{M_\text{vir,max}}^{\infty} \theta_\text{vir}(\mathcal{H}_\text{max}) d \mathcal{H}_\text{max} = \int_{M_{\ast,\text{max}}}^{\infty} \theta_\ast(\mathcal{M}_\text{max}) d \mathcal{M}_\text{max}.
\end{equation}
This rule helps to avoid issues that arise in the stochastic case described by Eq. (\ref{eq:SHMR_ratio_max}), such as small baryonic halos hosting unphysically massive galaxies, i.e. $M_{\ast,\text{max}} \gtrsim f_\text{bar} M_\text{vir}$. We denote the resulting stellar-to-halo mass ratio and cumulative efficiencies denoted respectively by $\mathcal{R}_\text{max,rank}$ and $\mathcal{E}_\text{max,rank}$.

Finally, we note that a third option can be considered by solving the relation
\begin{equation}
    \int_{M_\text{vir,max}}^{\infty} \theta_\text{vir}(\mathcal{H}_\text{max}) d \mathcal{H}_\text{max} =  \int^{M_{\ast,\text{max}}}_{-\infty} \theta_\ast(\mathcal{M}_\text{max}) d \mathcal{M}_\text{max}.
\end{equation}
In this case, the \emph{most} massive halo drawn from $\theta_\text{vir}(M_\text{vir,max})$ corresponds to the \emph{least} massive galaxy drawn from $\theta_\ast(M_{\ast,\text{max}})$. While we briefly mention this case in Section \ref{sec:connecting_massive_gals_to_halos}, we do not explore it in detail, as it would increase the likelihood of finding small baryonic halos hosting unphysically massive galaxies. This case will be also denoted by $\mathcal{R}_\text{max,inv-rank}$ and $\mathcal{E}_\text{max,inv-rank}$ from the resulting ratio and cumulative efficiencies.

Before presenting our predictions, we discuss errors affecting the galaxy-halo connection.

\subsection{Errors affecting the galaxy-halo connection}
\label{sec:random_errors}

In this section, we discuss uncertainties affecting the GSMF and HMFs, which are often ignored in the context of the massive galaxies problem \citep[but see][]{Chen+2023}. This oversight is significant because, as we will see below, errors in galaxy mass estimates can introduce large variations due to \citet{Eddington1913,Eddington1940} bias, similar when accounting the scatter around the SHMR. Indeed, several studies have shown that this bias is one of the main sources of uncertainty, particularly for massive galaxies, in efforts to understand the galaxy-halo connection \citep[see for example][for a comprehensive analysis]{Behroozi+2010}. For halos, the HMF is affected by uncertainties related to halo finders, finite-volume simulations, cosmological parameters, and other factors. Below, we explore these uncertainties in more detail.

\subsubsection{The Galaxy Stellar Mass Function}
\label{sec:GSMF}

As extensively discussed in \citet[][see also \citealp{Conroy2013}]{Conroy+2009a}, individual stellar mass estimates are subject to random errors. Previous studies \citep{Behroozi+2010, Cattaneo+2008, Wetzel+2010, Behroozi+2013, Moster+2013, Rodriguez-Puebla+2017,Rodriguez-Puebla+2025} have shown that these random errors cause the observed galaxy count at a given stellar mass, $M_\ast$, to deviate from the true count due to the Eddinton bias. Specifically, random errors tend to scatter more low-mass galaxies into higher mass bins because of the steep decline in the GSMF. Therefore, the \emph{observed} GSMF is the result of the convolution between random errors over the \emph{true} GSMF \citep{Cattaneo+2008,Behroozi+2010,Ilbert+2013,Grazian+2015,Rodriguez-Puebla+2017,Rodriguez-Puebla+2020,Rodriguez-Puebla+2025,Bernardi+2010,Bernardi+2013,Bernardi+2017,Leja+2020,McLeod+2021}

\begin{equation}
    \phi_{\ast,\text{obs}}(M_\ast) = \int_{-\infty}^{\infty} \mathcal{P}_\text{error}(x-\log M_{\ast}) \, \phi_{\ast,\text{true}}(x) d x,
    \label{eq:gsmf}
\end{equation}
where $\mathcal{P}_\text{error}(x)$ is the probability distribution of errors.
We assume that the error distribution, $P_{\mathrm{error}} \equiv G$, follows a Gaussian form. The observed GSMF is therefore given by the convolution of the intrinsic GSMF with this Gaussian kernel, which accounts for uncertainties in stellar mass estimates.

We notice that by combining Eqs. (\ref{eq:gsmf}) and (\ref{eq:intrinsci_GSMF}), the observed GSMF, $\phi_\text{obs}$, takes the form of a double convolution \citep[see e.g.,][]{Behroozi+2013,Rodriguez-Puebla+2017}:

\begin{equation}
\begin{aligned}
    \phi_{\ast,\text{obs}}(M_\ast) = \int \int & \mathcal{P}_\text{error}(x-\log M_\ast) \times \, \\
    & \mathcal{H}_\text{gal-halo}(y-x)  \phi_{\ast,\text{intr}}(y) d y  dx.
\end{aligned}
\label{eq:intrinsci_gsmf}
\end{equation}

Under the assumption that both $\mathcal{P}_\text{error}$ and $\mathcal{H}_\text{gal-halo}$ follow Gaussian distributions, this simplifies to
\begin{equation}
    \phi_{\ast,\text{obs}}(M_\ast) = \int \mathcal{G}(x-\log M_\ast) 
     \phi_{\ast,\text{intr}}(x) 
    dx,
    \label{eq:convo_observed_GSMF}
\end{equation}
where $\mathcal{G}$ represents the convolution of $\mathcal{P}_\text{error}$ and $\mathcal{H}_\text{gal-halo}$, i.e. $\mathcal{P}_\text{error} \circ \mathcal{H}_\text{gal-halo}$.

\begin{figure}
		\includegraphics[width=1.05\columnwidth]{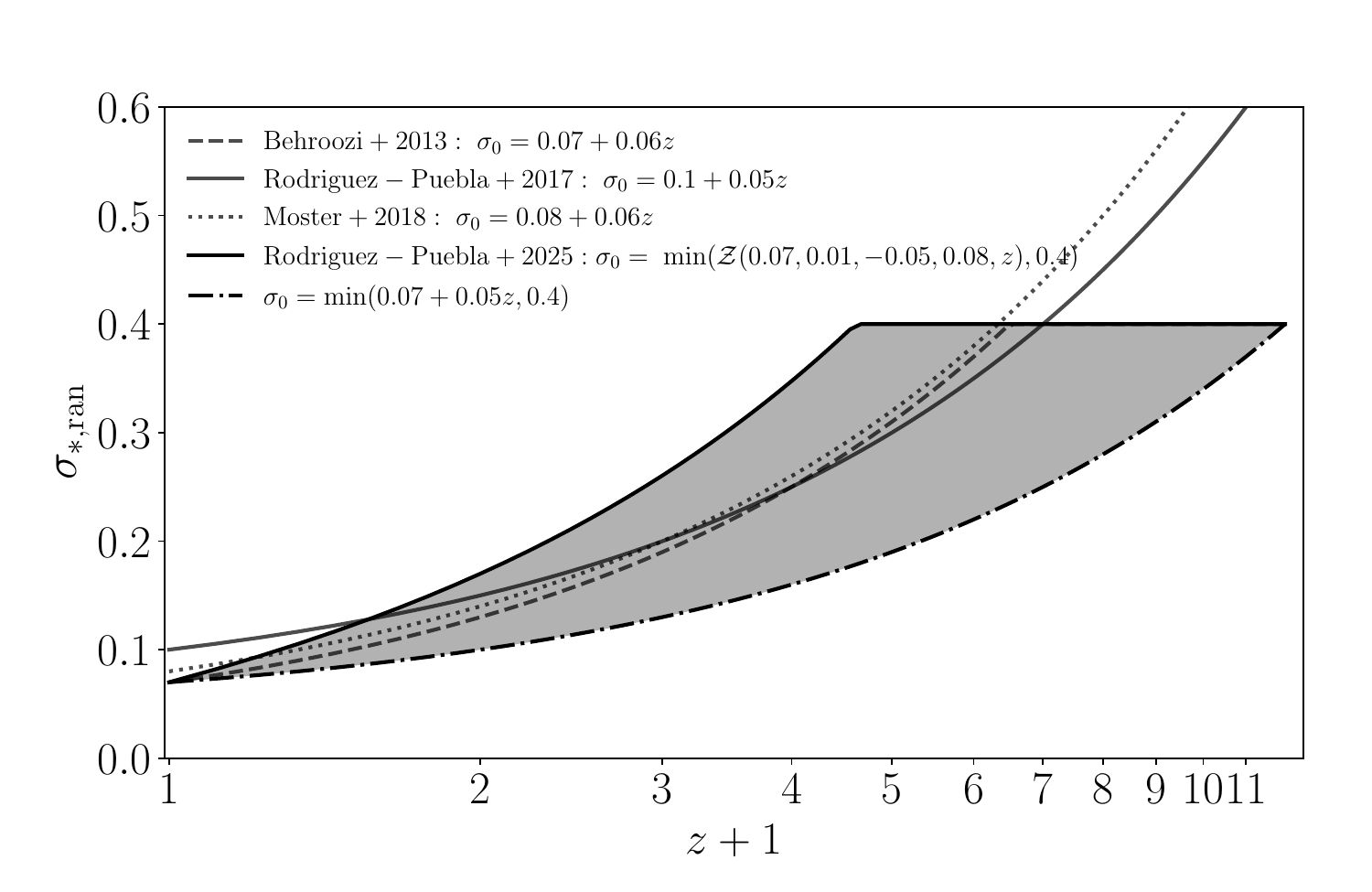}
			\caption{Redshift evolution of the 1$\sigma$ random stellar mass uncertainties, $\sigma_{\ast,\mathrm{ran}}$. Lines show different literature prescriptions, and the shaded region denotes the range adopted in this work.
 		}
	\label{fig:ran_ms}
\end{figure}

As recently discussed in \citet{Rodriguez-Puebla+2025}, the 1-$\sigma$ distribution of stellar mass uncertainties reported in various studies tends to increase with redshift. Figure \ref{fig:ran_ms} presents different models for random errors as a function of redshift, as reported by \citet{Behroozi+2013,Rodriguez-Puebla+2017,Moster+2018}, and \citet{Rodriguez-Puebla+2025}, which have been employed in modeling the galaxy–halo connection. In general, the models from \citet{Behroozi+2013,Rodriguez-Puebla+2017}, and \citet{Moster+2018} show random errors increasing from $\sim$0.1 dex at low redshifts to $\sim$0.4 dex at $z \sim 5$, with the latter two increasing further to values as high as $\sim$0.6 dex at $z \sim 7-8$. The new model proposed by \citet[][refer to their figure 1]{Rodriguez-Puebla+2025}, the solid line in Figure \ref{fig:ran_ms}, suggests that previous estimates may underestimate the true uncertainties. We note, however, that other studies \citep[e.g.,][]{Leja+2020,Weibel+2024} report significantly smaller 1-$\sigma$ random errors, suggesting disagreement with all of the aforementioned models. The dot-dashed line in Figure \ref{fig:ran_ms}, described by the equation shown in the label, is intended to represent this alternative trend, consistent with the findings of \citet{Leja+2020} and \citet{Weibel+2024}. This model will be further discussed in Rodriguez-Puebla et al. (in prep). Rather than adopting a single model, in this paper we consider the shaded region in Figure \ref{fig:ran_ms} to represent the plausible range of stellar mass uncertainties, with its upper bound defined by the \citet{Rodriguez-Puebla+2025} model. It is important to note that the model used here assumes random errors that depend only on redshift and are independent of galaxy color or star formation activity. In the figures presented throughout this paper, whenever results are associated with Figure \ref{fig:ran_ms}, we display two curves with the same line styles as those enclosing the shaded region in Figure \ref{fig:ran_ms}. These curves represent the propagated uncertainties, as discussed in the context of the corresponding figures. Nonetheless, we adopt the upper curve as our fiducial model.

Recent JWST-based analyses have further highlighted the importance of accounting for systematic uncertainties in stellar mass estimates. For example, it has been shown that variations in SED-fitting assumptions, including the use of parametric versus non-parametric star formation histories, different dust attenuation curves, \citep[see e.g.,][]{Carnall+2024,Harvey+2025}, as well as the IMF \citep{Cheng2026}, can significantly affect inferred stellar masses. In particular, the stellar mass density at $z\sim 10.5$ may vary by up to 0.75 dex depending on the assumed SFH model, and adopting a top-heavy IMF can reduce mass estimates by $\sim$0.5 dex without noticeably degrading the fit quality \citep[see e.g.,][]{Harvey+2025}. Likewise, the analysis of a large NIRCam-selected sample of over 30,000 galaxies from  \citet{Finkelstein+2025}, PRIMER \citep{Donnan+2024}, and JADES \citep{Hainline+2024} has shown that the inclusion of UV-red (and potentially dust-obscured or quiescent) galaxies substantially modifies both the shape and normalization of the GSMF, particularly at the massive end \citep{Weibel+2024}. Similarly, \citet{Turner+2025} demonstrate that stellar masses of $z \sim 7$ galaxies may be overestimated by $\sim 0.2$--$0.3$ dex when SPS models do not fully capture the contribution of young stellar populations and binary evolution. Additionally, recent work suggests that widely used SED models may not fully capture the diversity of star formation histories, chemical enrichment patterns (e.g. $\alpha$-enhancement), or late stellar evolutionary phases required to accurately model these galaxies \citep{Hamadouche2026}. These results reveal that selection biases and SED-modeling choices must be carefully considered when comparing observed GSMFs with theoretical predictions and such effects could broaden the observational range of inferred stellar masses and potentially increase or alleviate the apparent tension with the $\Lambda$CDM expectations. 

In this work, we employ the best fitting models to the observed GSMFs, $\phi_{\ast,\text{obs}}$, as well as the true GSMFs, $\phi_{\ast,\text{true}}$, from Rodriguez-Puebla et al. (in prep., see also \citealp{Rodriguez-Puebla2024,Rodriguez-Puebla+2025}). The authors compiled and homogenize a large data set of GSMFs from the literature in the redshift range $z\sim0{-}9$, as well as for quiescent galaxies from $z\sim0{-}4.5$. As discussed by the authors, this compilation represents and update to the GSMFs presented in \cite{Rodriguez-Puebla+2017}, including recent data published by JWST. The key quantities that the authors homogenized were the IMF to that of \citet{Chabrier2003} and adopted cosmological parameters as in \cite{Planck+2015}. Nonetheless, since one of the relevant aspects of this paper is to understand the impact of the cosmological parameters, corrections will be made according to section \ref{sec:impact_cosmo_params} and the discussion around Eq. (39) from \cite{Rodriguez-Puebla+2017}. 

Finally, we notice that Rodriguez-Puebla et al. (in prep.) performed Bayesian fits the redshift evolution of the GSMFs simultaneously for all galaxies ($0\leq z \leq 9$) and of quiescent galaxies ($0\leq z \leq 4.5$). The authors assumed that quiescent galaxies are well described by a triple modified-Schechter function \citep[see also][]{Vazquez-Mata+2025}, while star-forming galaxies by a double modified-Schechter function. For details, we refer to the reader to Rodriguez-Puebla et al. (in prep.), see also Appendix \ref{sec:GSMF_fits} for a brief summary and best fitting model from that paper.

\begin{figure}
    \centering
    \includegraphics[width=\columnwidth]{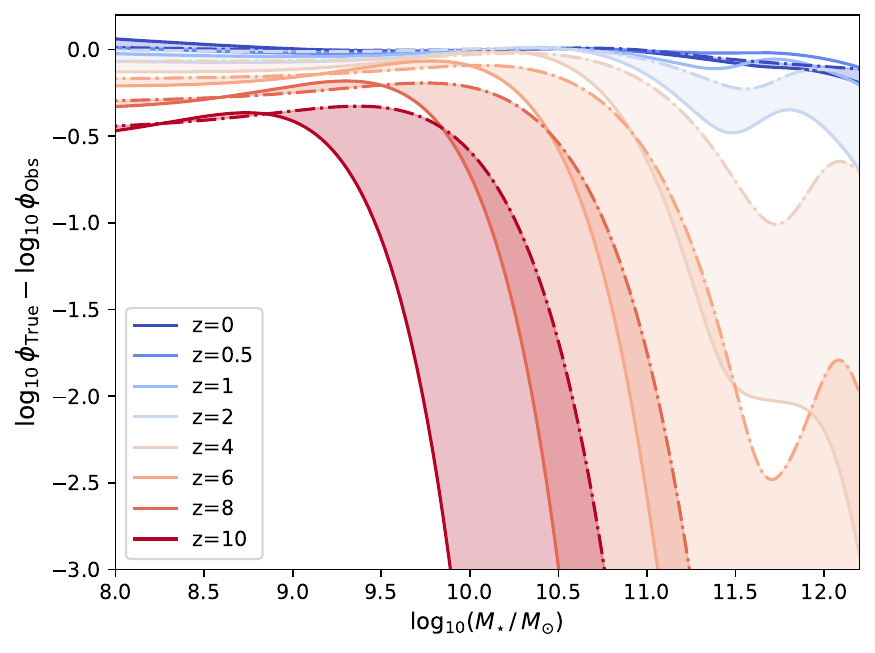}
    \caption{Logarithmic difference between the observed and true GSMF, $\phi_{\text{obs}}$ and $\phi_{\text{true}}$, accounting for halo dispersion corrections at various redshift values. The shaded regions represent the plausible range of random errors (see Figure \ref{fig:ran_ms}), with the solid line showing the resulting differences based on the model from \citet{Rodriguez-Puebla+2025}, and the dot-dashed lines showing the results using the model discussed in this section.}
    \label{fig:ratio_phis}
\end{figure}

Figure \ref{fig:ratio_phis} presents the lograithmic difference between the observed and the true GSMF, $\Delta \log\phi = \log \phi_{\ast,\text{obs}} -\log\phi_{\ast,\text{true}}$, as a function of stellar masses. This figure demonstrates two key points: i) as random errors increase with redshift, the normalization of $\phi_{\ast,\mathrm{true}}$ becomes systematically lower; ii) high-mass galaxies are more strongly affected by random errors, with differences up to $\Delta \log \phi \sim 2\,\mathrm{dex}$ at $z \sim 4\text{–}6$ for galaxies with $M_\ast \sim 3 \times 10^{11} M_\odot$.
 
From these results, we can potentially anticipate that the tension observed between stellar masses and the halo mass function, $\phi_{\text{vir}}$, reported in previous studies (sometimes referred to as the ``impossible early galaxy problem'', \citealp{Steinhardt+2016,Labbe+2023,Boylan-Kolchin_2023}) may diminish or even disappear if random errors are taken into account. We will return to this point in Section \ref{sec:connecting_massive_gals_to_halos}. 

Finally, we note that \citet{Lovell+2023} introduced a correction for the Eddington bias. As discussed in their work, this correction is an approximation that depends on the slope of the HMF and on the square of the uncertainty in stellar mass estimates, see also Section 2 of \citet{Rodriguez-Puebla+2025}. While a direct comparison between our correction and that presented by \citet{Lovell+2023} is not straightforward, we emphasize that our estimates of the intrinsic quantities are not based on an approximation but are instead obtained through the deconvolution procedure described in this section. 

\subsubsection{Halo mass function}

The main sources of random errors in the HMF estimated from cosmological $N$-body simulations arise from two primary factors: resolution limits and cosmic variance. The finite resolution of simulations affects the ability to resolve small halos, thus introducing uncertainties in their estimated abundance. In particular, low-mass halos may be artificially suppressed due to the limited number of particles used to characterize them. Cosmic variance arises from statistical fluctuations in the matter density field due to the finite sampling of large-scale structures. In simulations with limited volumes, massive halos are especially affected, as their number density depends on the presence (or absence) of the rarest peaks in the cosmic density field \citep{Despali+2016,Comparat+2017}.

In addition to these random errors, several sources of systematic uncertainties can introduce biases in the estimated HMF as we listed below:
\begin{itemize}
    \item Different studies adopt varying halo mass definitions, such as virial mass or masses defined within fixed overdensity thresholds. These choices can introduce systematic variations in the HMF: while consistent conversions between definitions typically lead to differences at the few-percent level, neglecting these effects can result in discrepancies of order $\sim 20\text{–}50\%$ in the normalization and shape of the HMF (e.g. \citealp{Tinker+2008,Despali+2016}).
   
    \item The underlying cosmological parameters (e.g., $\Omega_{\text{m}},\Omega_{\text{bar}},\sigma_8$ and $n_s$) directly affect halo abundance predictions. Small variations in these parameters can lead to significant changes in the normalization and shape of the HMF.
    \item Cosmological simulations are inherently limited by their finite volumes, which prevent them from fully capturing large-scale density fluctuations, particularly those governing the formation of the most massive halos. This can bias estimates of the high-mass end of the halo mass function, as the long-wavelength modes that drive their growth are not adequately sampled (e.g. \citet{Bagla+2005,Reed+2007,Trenti+2008}.
  
    \item Specific functional forms and best-fitting models used to describe the HMF in $N$-body simulations introduce additional sources of systematic uncertainty. Different fitting formulas may lead to variations in the predicted halo abundance.
\end{itemize}


\begin{figure*}
    \centering
    \includegraphics[width=2.0\columnwidth]{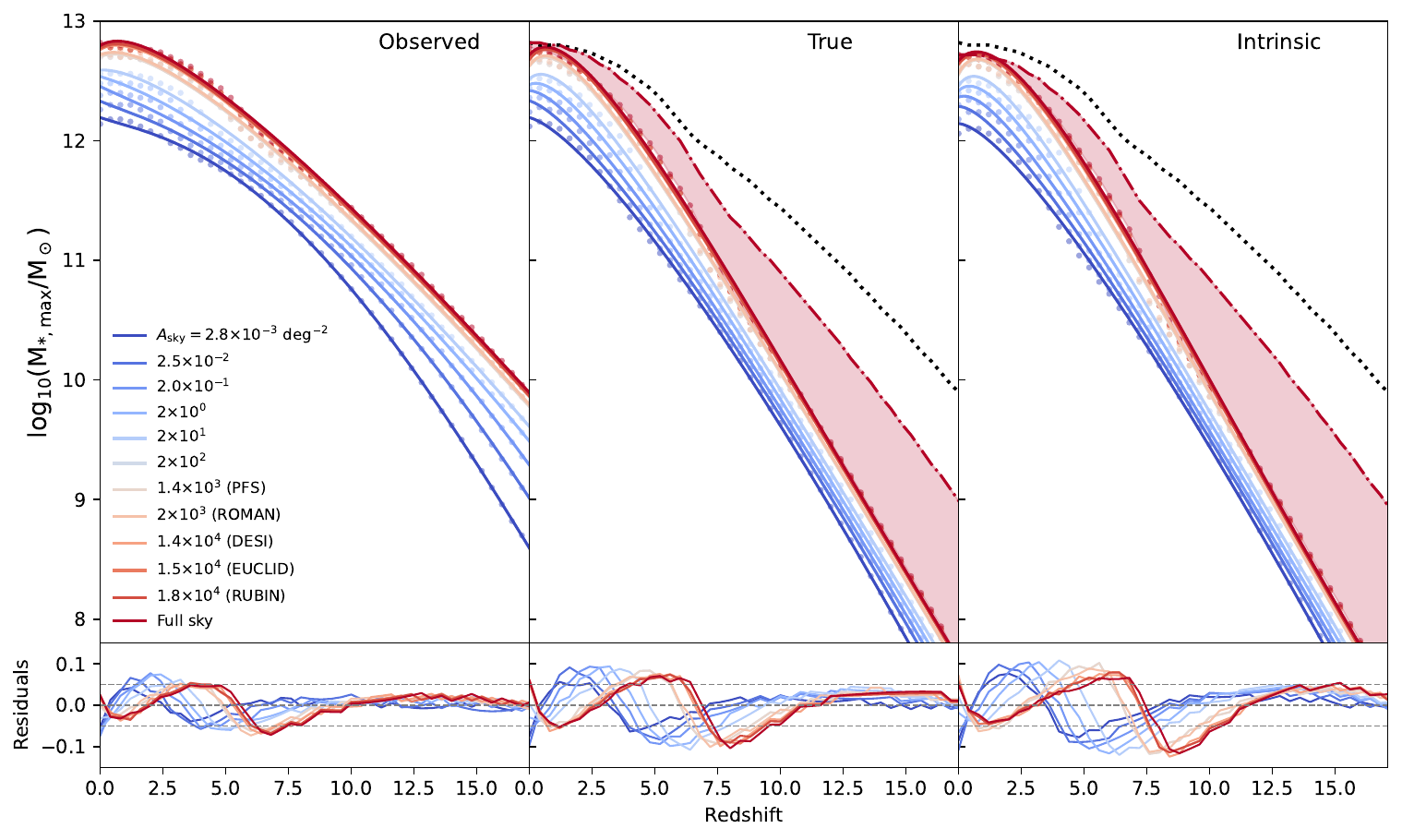}
\caption{EVS predictions for the most massive galaxies as a function of redshift for different survey sky apertures (color dotted lines) and their best fits (color solid lines, see Eq. ~\ref{eq:model_most_massive} and Appendix~\ref{sec:evs_fits}). The left, middle, and right panels present results from the EVS framework applied to the observed, true, and intrinsic GSMFs, respectively. The true GSMF accounts for the effects of Eddington bias arising from random errors in stellar mass estimates (see Figure~\ref{fig:ran_ms}), whereas the intrinsic GSMF additionally accounts for the intrinsic scatter in the SHMR. The black dotted line in the middle and right panels reproduces the predicted most massive galaxy from the observed GSMF in the full sky case (red dotted line in the left panel) for ease of comparison with the effect of Eddington bias and the intrinsic scatter of the SHMR, respectively. The shaded regions in both panels indicate the plausible range of random errors shown in Figure~\ref{fig:ran_ms}. The lower panels show the residuals with respect to the corresponding best-fit models.} 
    \label{fig:evs_5sigma_gal}
\end{figure*}


While we do not provide a thorough discussion of HMF uncertainties here, they have been analyzed in detail by \citet{Yung+2024}. In particular, these authors compared some of the most widely used HMFs from the literature, including \citet{SMT2001}, \citet{Reed+2007}, \citet{Tinker+2008}, \citet{Tinker+2010}, and \citet{Behroozi+2013}. Their findings indicate that fractional differences between these models are typically within $\sim\pm0.5$ dex at intermediate masses, but become significantly larger toward the high-mass end, where the various fitting functions diverge, and this discrepancy increases with redshift, particularly at z $\gtrsim$ 6, when compared to their new HMF fits based on the suite of \textsc{GUREFT} simulations  \citep{Yung+2024}. In this regime, different models exhibit distinct behaviors, with Sheth-Tormen-like fits tending to overpredict the abundance of the most massive halos, while Tinker-like fits generally bend toward lower abundances.
 
Therefore, in this paper, we assess the impact of different HMF models in the EVS framework by estimating the most massive halos using the mass functions of \citet{Sheth+1999,Tinker+2008,Tinker+2010,Behroozi+2013,Rodriguez-Puebla+2016,Despali+2016} and \citet{Yung+2024,Yung+2025}.

Finally, in this paper, we also use the posterior distributions of the \citet{Planck+2015} cosmological parameters to assess the impact of their uncertainties on the HMF and within the EVS framework.

\section{Empirical Predictions} \label{sec:EVS_GSMF}

Following the methodology outlined in Section \ref{sec:EVS}, we employ EVS to predict the most massive galaxies expected from the  GSMF as a function of redshift and for a given survey aperture. 

\subsection{The most massive galaxies}
\label{sec:most_massive_gal}

What is the most massive galaxy we should expect to observe at different survey apertures? As discussed earlier, EVS offer a framework for deriving the PDF of the most massive galaxy. While this is useful, it does not yield a single definitive answer. To provide a more objective definition of the most massive galaxy, and later, the most massive halo, we define it here as the galaxy corresponding to the $5\sigma$ value of the distribution predicted by the EVS framework applied to the GSMF. In the case of that the EVS of galaxies is represented by a Gaussian distribution, the probability of observing a galaxy more massive than this at given survey aperture is approximately $1:1.7 \times 10^6$. Note that, in reality, the EVS distribution of galaxies is not Gaussian; however, we use $5\sigma$ as a measure of extremeness.

Our analysis begins by calculating the expected most massive galaxy, $M_{\ast,\text{max|obs}}$, from the observed GSMF, $\phi_{\ast,\text{obs}}$. We compute these predictions in narrow redshift bins of width $\Delta z = 0.02$, such that the comoving volume $V_c$ entering the EVS formalism is evaluated over intervals $[z - \Delta z/2, z + \Delta z/2]$. 

The left panel of Figure \ref{fig:evs_5sigma_gal} shows how the stellar mass of the $5\sigma$ most massive galaxy varies with redshift across a range of survey apertures, from ultra-narrow fields to full-sky coverage (see labels), based on $\phi_{\ast,\text{obs}}$. The survey areas adopted here (see Table~\ref{tab:surveys}), particularly the largest ones, are chosen as references to current and planned surveys. These choices do not imply that such surveys will be capable of observing galaxies across the full redshift range considered in this work.

For the full-sky case, the $5\sigma$ most massive galaxy at $z \sim 0$ has a stellar mass of $M_\ast \sim 7 \times 10^{12} M_\odot$. This mass decreases slowly with redshift, reaching $M_\ast \sim 10^{12} M_\odot$ by $z \sim 7.5$, that is a slope of $d\log M_\ast/dz\sim 0.1$. At higher redshifts, the decline has a similar slope, with $M_\ast$ dropping to around $10^{10} M_\odot$ by $z \sim 16$. 

Although the data used to constrain the GSMF fits extend only up to $z \lesssim 10$, one might question the validity of extrapolating beyond this range. However, such extrapolation is reasonable in this case because the GSMF was parameterized mainly as a function of the scale factor, $a = 1 / (1 + z)$. A key advantage of this approach is that large differences in redshift correspond to relatively smaller differences in the scale factor. For example, $z \sim 10$ corresponds to $a \sim 0.09$, while $z \sim 17$ corresponds to $a \sim 0.06$. Using $a$ as the independent variable thus helps minimize uncertainties when extrapolating the GSMF at high redshifts, due to its compressed dynamic range.

In Section \ref{sec:GSMF}, we discussed the effects of random errors in the observed GSMF. There, Figure \ref{fig:ratio_phis} illustrates the logarithmic differences between the true GSMF, obtained by deconvolving the observed GSMF from random errors, and the observed GSMF itself. As we showed, the true GSMF falls off more steeply than the observed one at a given stellar mass, an effect that becomes more pronounced at higher redshifts. As we will see, these random errors significantly affect the EVS prediction of the most massive galaxy. Furthermore, since our goal is to compare these galaxies with dark matter halos, we also account for intrinsic scatter in the SHMR. This introduces a second deconvolution, as described in Section \ref{sec:GSMF} and Eq. (\ref{eq:intrinsci_gsmf}). Thus a more steeply fall off is expected for the intrinsic GSMF as we describe below. 

The middle and right panels of Figure \ref{fig:evs_5sigma_gal} show, respectively, the impact of random observational errors and the effect of accounting for the intrinsic scatter in the SHMR. In both panels, this is shown by comparing the black (representing $M_{\ast,\text{max|obs}}$) and red solid lines for $f_\text{sky} = 1$. The effect of the steep decline in the GSMF is evident in both cases, as shown by the increasing difference between the observed curves and the others. While the predictions are similar at low redshift, the discrepancy between $M_{\ast,\max|\mathrm{obs}}$ (black dotted curve) and $M_{\ast,\max|\mathrm{true}}$ (red solid curve) increases significantly towards higher redshifts, reaching $\sim 0.4\text{–}0.5$ dex at $z \sim 5$ and exceeding $\sim 1$ dex at $z \sim 10$. At even higher redshifts, the difference remains substantial, $\gtrsim2$ dex of difference.
The discrepancy is smaller when comparing $M_{\ast,\max|\mathrm{true}}$ and $M_{\ast,\max|\mathrm{int}}$ (red dashed curve), amounting to $\sim 0.1\text{–}0.2$ dex at $z \sim 5$, and increasing to $\sim 0.5\text{–}0.6$ dex at $z \sim 10$. 
Figure \ref{fig:evs_5sigma_gal} also shows how the inferred most massive galaxy evolves with survey area and redshift. As expected, the most massive galaxies identified in smaller survey areas are systematically less massive. For our smallest survey area, the $5\sigma$ most massive galaxies are typically up to a factor of $\lesssim 4$ less massive than those in the full-sky case, $f_\text{sky}=1$, at low redshifts, $z \sim 0$, but this difference grows to more than an order of magnitude at higher redshifts, $z \gtrsim 12$. 

We provide a best-fitting model for the most massive galaxy in units of solar masses using the following functional form  
for a given sky fraction: 
\begin{equation}
\begin{aligned}
M_{\ast,\max}(z,f_{\rm sky})
&=
A(f_{\rm sky})
\Bigg[
\left(\frac{a_0(f_{\rm sky})}{1+z}\right)^{\alpha_1(f_{\rm sky})\,n(f_{\rm sky})}
\\
&\quad +
\left(\frac{a_0(f_{\rm sky})}{1+z}\right)^{\alpha_2(f_{\rm sky})\,n(f_{\rm sky})}
\Bigg]^{-1/n(f_{\rm sky})}
\end{aligned}
\label{eq:model_most_massive} 
\end{equation}

where $A$ is a normalization factor, $\alpha_1$ and $\alpha_2$ control the slopes of the two power-law regimes,

$a_0$ is a characteristic redshift scale for the transition, and $n$ governs the sharpness of that transition. The corresponding best-fit parameters and auxiliary functions are listed in Table \ref{tab:evs_model_fit_gal}. The solid lines in Figure \ref{fig:evs_5sigma_gal} represent the best-fit model. As indicated in the bottom panel of the same figure, the model reproduces the EVS results with an accuracy better than $\sim0.07$ dex which is equivalent to $\sim17\%$ in the stellar mass.

\begin{figure}
    \centering
    \includegraphics[width=1.07\columnwidth]{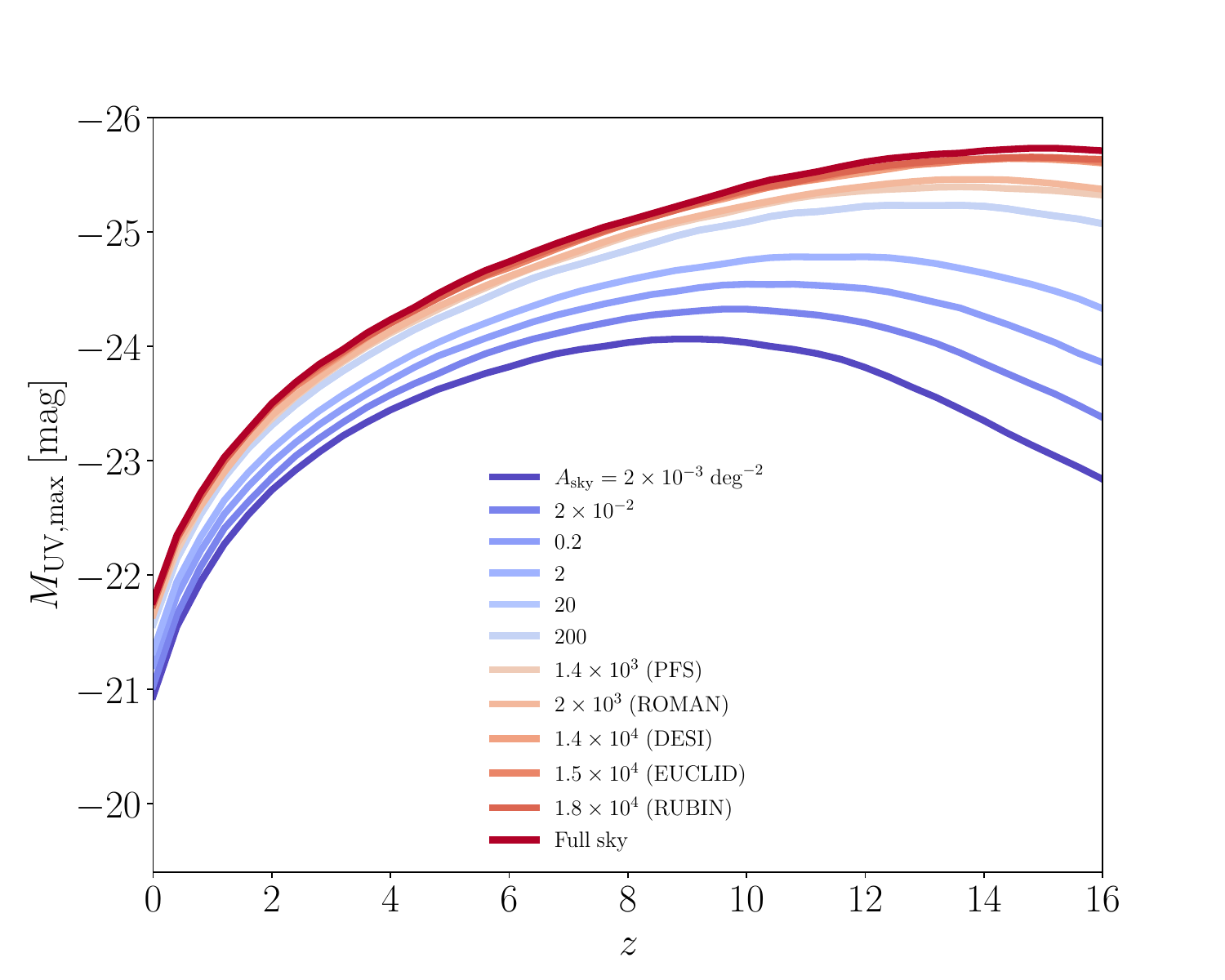}
    \caption{EVS predictions for the most luminous galaxy as a function of redshift for different survey areas, derived from the observed UV LF (see Appendix~\ref{sec:UV_LF} for details). In contrast to the most massive galaxies (Fig.~\ref{fig:evs_5sigma_gal}), whose masses increase monotonically, the brightest galaxies reach a maximum magnitude and redshift that depend on the survey area.}
    \label{fig:evs_muv}
\end{figure}


\begin{figure*}
    \centering
 
    \includegraphics[width=2.1\columnwidth]{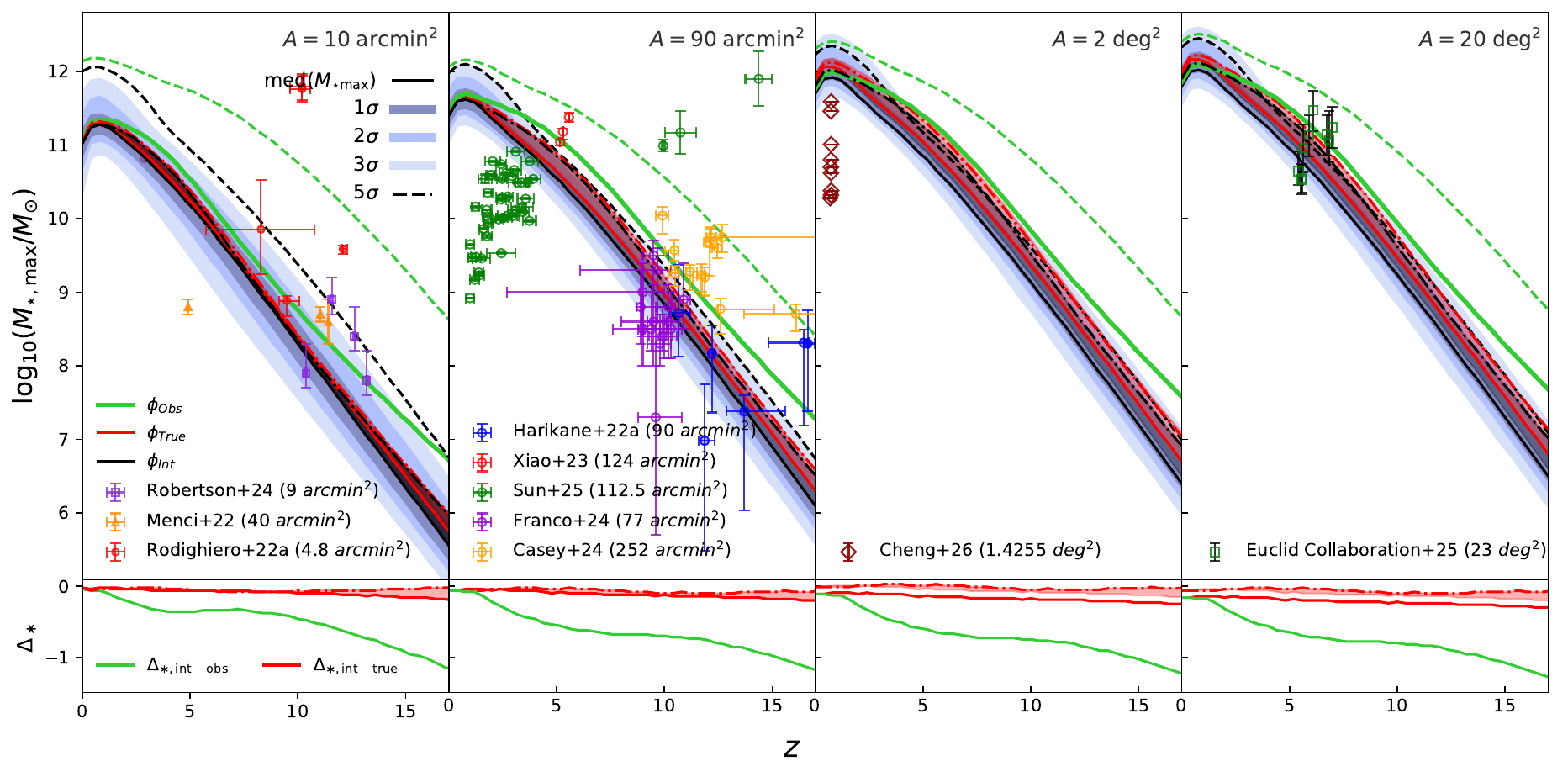}
    \caption{EVS distributions for $M_{\ast,\mathrm{max}}$ as a function of redshift for four different survey areas, derived from the observed, true, and intrinsic GSMFs. In the case of the intrinsic GSMF, the $\pm 1\sigma$, $\pm2\sigma$, and $\pm3\sigma$ confidence regions are shown as shaded areas. The dashed black line corresponds to $+5\sigma$; the same applies for the green dashed line, but for the case of the observed GSMF. The medians for the true and observed GSMFs are shown by the red and green solid lines, respectively. The black and red curves enclose the region that marks the plausible range of random errors in stellar mass estimates. All observed values have been homogenized, when necessary, to adopt the \citet{Chabrier2003} IMF. From left to right, the panels correspond to survey areas of $10 \text{ arcmin}^2$, $90 \text{ arcmin}^2$, $2 \text{ deg}^2$, and $20 \text{ deg}^2$, respectively.
The bottom panels show the logarithmic differences between the intrinsic and true stellar masses,
$\Delta_{\ast,\text{int-true}} \equiv \log M_{\ast,\text{max|int}} - \log M_{\ast,\text{max|true}}$, in red, and between the intrinsic and observed values,
$\Delta_{\ast,\text{int-obs}} \equiv \log M_{\ast,\text{max|int}} - \log M_{\ast,\text{max|obs}}$, in green.}
   
    \label{fig:phiOTI}
\end{figure*}


\subsection{The most luminous galaxies}
\label{sec:UV_brightest}

So far, we have focused on the most massive systems. However, it is also important to consider the most luminous galaxies, which are typically, but not always, the most massive due to intrinsic scatter and other effects. This comparison is presented in Fig. \ref{fig:evs_muv}. This figure shows the $5\sigma$ most luminous galaxy predicted by the UV luminosity function model described in Appendix \ref{sec:UV_LF}. Note that this model does not include corrections for dust attenuation, and therefore represents the observed UV luminosity. Briefly, we compiled a large number of UV luminosity functions (LFs), most of them from Table 1 of \citet{Rodriguez-Puebla+2020a}, along with others from more recent JWST results, see Appendix \ref{sec:UV_LF} for details. In Appendix \ref{sec:UV_LF} we provide two different fits: one that reproduces the data as observed, and another that assumes a single source of Eddington bias arising from scatter in the $L_\text{UV}$–$M_\text{vir}$ relation. This scatter is modeled as a lognormal distribution with dispersion $\sigma_\text{UV}=0.3$ dex, independent of redshift. This choice is motivated by the observed $\sim0.25$–$0.3$ dex width of the star-forming main sequence \citep{Speagle+2014}.

We find that similarly to the $5\sigma$ most massive galaxy, and as expected, the most luminous galaxies are found in larger survey areas. However, unlike the curves for the most massive galaxies, which increase monotonically towards low redshifts, the curves for the most luminous galaxies are slightly bell-shaped and decrease as a consequence of star formation being quenched in massive galaxies. For a full-sky survey, we find that the maximum UV luminosity expected for a $5\sigma$ detection is $M_\text{UV} \sim -25.7$ at $z \sim 15$. This corresponds to $\mathrm{SFR} \sim 470\,M_\odot\,\mathrm{yr}^{-1}$, assuming the standard UV--SFR conversion from \citet{Madau_Dickinson2014}, in which the UV luminosity traces the star formation rate averaged over $\sim100$–300 Myr and no dust correction is applied. For the \citet{Planck+2015} cosmology, the age of the Universe at $z\sim15$ is $\sim270$ Myr, which justifies the use of this conversion. At very high redshift ($z \gtrsim 20$), when the Universe is $\lesssim240$ Myr old, the age becomes comparable to the UV–SFR calibration timescale, and the UV continuum may lag the star formation rate, leading to underestimated SFRs. 

For our smallest area we find a maximum UV luminosity of $-24.1$ at $z \sim 9$, which would correspond to $\mathrm{SFR} \sim 108\,M_\odot\,\mathrm{yr}^{-1}$ or $\log(\mathrm{sSFR}/\mathrm{yr}^{-1}) \sim -9$. Note that both of these estimates are based on unattenuated UV luminosities and therefore should be interpreted as lower limits to the intrinsic star formation rates, particularly for such massive systems where dust attenuation may be significant.

\subsection{Comparison with JWST data}

Since the discoveries from JWST appear to challenge the $\Lambda$CDM model, in this section we compare our results with JWST data. Considering survey volumes of 10 and 90 arcmin$^2$ as representative values of some of the fields observed by the JWST, the lefts panels of Figure \ref{fig:phiOTI} show the median EVS distributions for $M_{\ast,\text{max}}$ as a function of redshift, based on the observed, true, and intrinsic GSMFs, as indicated in the labels, for these survey areas. For the intrinsic GSMF, the $\pm1\sigma$, $\pm2\sigma$, and $\pm3\sigma$ confidence regions are shown as shaded areas. The red and green solid lines indicate the medians for the true and observed GSMFs, respectively. The region enclosed by the black and red lines marks the plausible range of random errors in stellar masses.  

The leftmost panel of Figure \ref{fig:phiOTI} shows the observed values reported by \citet{Rodighiero+2023,Menci+2022,Robertson+2024} and \citet{Sun_Yan2025} compared to our EVS results assuming a survey area of $10$ arcmin$^2$, while the middle-left panel presents the observations from \citet{Harikane+2023,Xiao+2023,Franco+2024,Casey+2024}, and \citet{Sun_Yan2025} compare to a survey area of $90$ arcmin$^2$. Some of the very bright candidates from \citet{Sun_Yan2025}, particularly those assigned photometric redshifts of $z\gtrsim10$, lie well above our 5$\sigma$ EVS expectations. However, this apparent excess should be interpreted with caution, since these redshifts are based on photometric/dropout selection and may be affected by low-redshift interlopers. \citet{Sun_Yan2025} show that a Balmer 4000 \AA\ break in dusty or evolved low-redshift galaxies can mimic the Lyman-break signature expected at very high redshift. In both panels, all observational points align well with predictions from the observed GSMF, $\phi_{\ast,\text{obs}}$, which is encouraging and suggests that our analysis of the continuous distribution of $M_{\ast,\text{max|obs}}$ accurately reflects the current state of observations. However, as discussed above and in Section \ref{sec:GSMF}, $\phi_{\ast,\text{obs}}$ is significantly affected by random errors in stellar mass estimates which introduces an Eddington bias effect. 
\begin{figure}
    \centering
    \includegraphics[width=1.07\columnwidth]{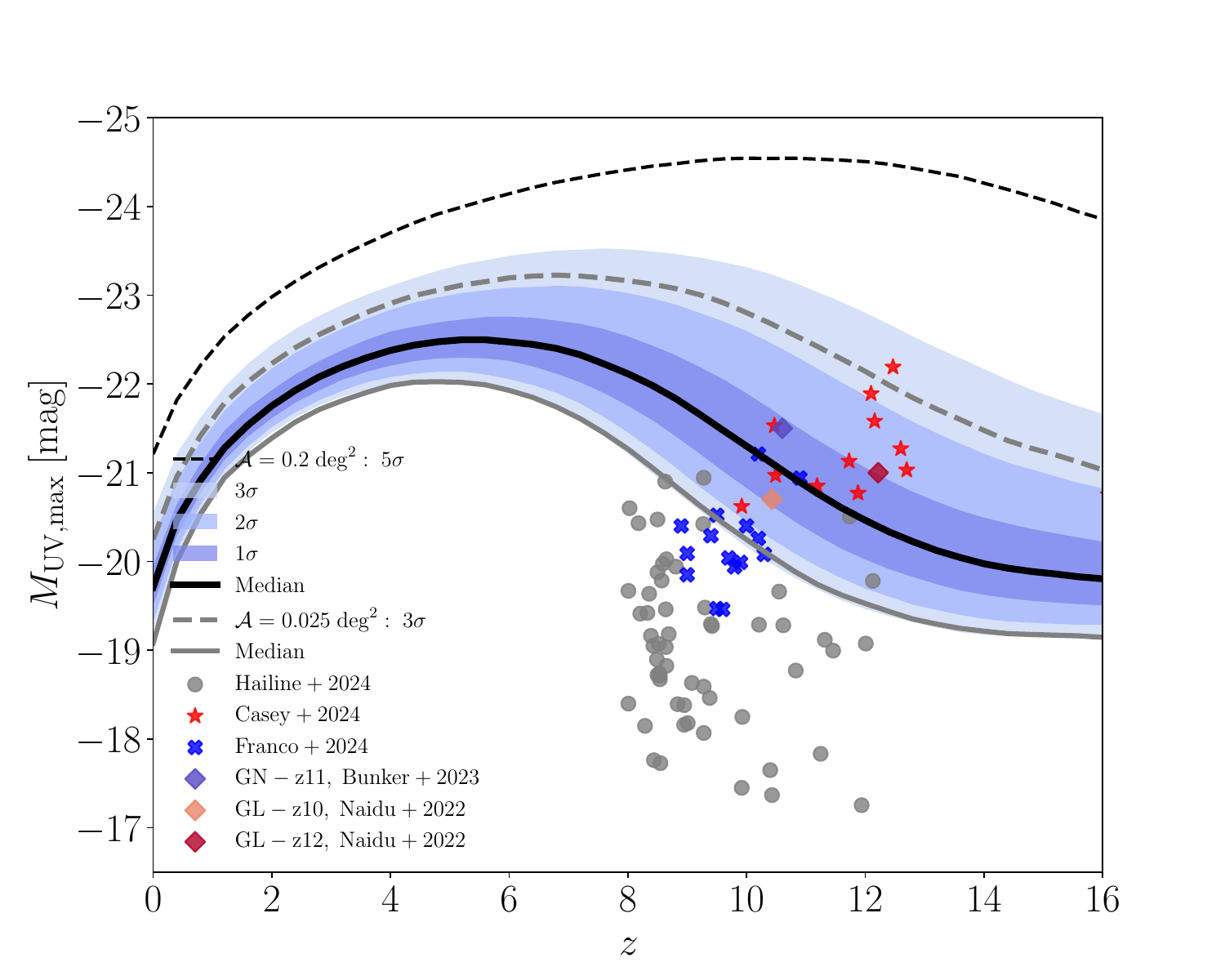}
    \caption{EVS predictions for the maximum UV magnitude, $M_{\text{UV,max}}$, as a function of redshift, derived from the observed UV LF. The solid curve shows the median for a survey area of $A = 0.2,\mathrm{deg}^2$ (COSMOS-Web), while the shaded regions indicate the $\pm1\sigma$–$\pm3\sigma$ intervals and the dotted line marks the $+5\sigma$ limit. The gray solid and dashed lines show, respectively, the median and $5\sigma$ of the EVS distribution for $A = 0.025,\mathrm{deg}^2$. Recent JWST observations of UV-bright galaxies are overplotted for comparison.}
    \label{fig:evs_muv_JWST}
\end{figure}

When accounting for random errors in stellar mass estimates, the median of the EVS distribution of $M_{\ast,\text{max|true}}$ shifts toward lower stellar masses at high redshifts. At lower redshifts, the difference between $M_{\ast,\text{max|obs}}$ and $M_{\ast,\text{max|true}}$ is minimal; with the observed median $M_{\ast,\text{max|obs}}$ being marginally higher. By $z \sim 10$, the difference reaches $\sim 0.7$ dex, approaching $\sim 1$ dex at the highest redshifts, with the observed distribution consistently yielding higher values across both survey areas. This highlights the impact of the Eddington bias on the EVS distributions of galaxies in the high-redshift universe \citep[see also][for a similar conclusion in the pre-JWST era, and more recently by \citealp{Chen+2023} in the JWST era]{Behroozi-Silk2018}. This effect is intuitive: as the Eddington bias flattens the massive end of the GSMF, the number density of massive galaxies is artificially inflated \citep[see e.g.,][]{Cattaneo+2008,Behroozi+2010,Wetzel+2010}. When comparing with the EVS distribution from the intrinsic GSMF, $M_{\ast,\text{max|int}}$, the differences are even larger relative to $M_{\ast,\text{max|obs}}$, and only slightly smaller compared to $M_{\ast,\text{max|true}}$. 

The middle-right and rightmost panels present the same information but consider larger areas, using data from \citet{Cheng2026} for a 2 deg$^2$ field (homogenized to a \citealp{Chabrier2003} IMF) and \citet{Euclid2025} for a 20 deg$^2$, respectively. Again, our empirical limits of the most massive are consistent with these survey areas. 

Figure \ref{fig:evs_muv_JWST} shows the EVS distribution for the maximum UV magnitude assuming a sky aperture of $\mathcal{A}_\text{sky} = 0.2 \ \text{deg}^{2}$. The solid line represents the median of the distribution, while the shaded regions indicate the $\pm1\sigma$, $\pm2\sigma$, and $3\pm\sigma$ ranges. The dashed black line marks the $+5\sigma$ limit. For comparison, we include data points from JWST observations. The red stars correspond to the extremely luminous candidate galaxies from \citet{Casey+2024} at $z\sim10{-}14$, based on JWST/NIRCam imaging of COSMOS-Web over a survey area of $0.28 \ \text{deg}^2$. The blue stars are from \citet{Franco+2024}, also using COSMOS-Web data. Our EVS distribution is consistent with the interpretation that the sample of galaxies from \citet{Casey+2024} represents the $\sim1{-}3\sigma$ most luminous systems at those redshifts, while for \citet{Franco+2024} these are mostly below the median and $1\sigma$. These results support the robustness in capturing the brightest galaxies within the COSMOS-Web area of our EVS analysis.

We also include data points from \citet{Hainline+2024}, based on JADES observations over a survey area of $\mathcal{A}\text{sky} = 0.125 \ \text{arcmin}^{2}$. For comparison, we show the EVS distribution assuming a sky aperture of $\mathcal{A}\text{sky} = 0.025 \ \text{deg}^{2}$ (equivalent to $0.90 \ \text{arcmin}^{2}$), which is the closest area we computed to match that of \citet{Hainline+2024}. Despite being slightly smaller, our EVS distribution remains highly consistent with their results. Results for GN-z11 \citep{Bunker+2023}, as well as GL-z10 and GL-z12 \citep{Naidu+2022}, are also shown and found to be in agreement with our EVS distributions. Note, however, that these early JWST measurements may be subject to calibration uncertainties, and their inferred luminosities could be revised as photometric calibrations improve. 

Finally, in this section, we show that our empirical constraints on the \emph{observed} GSMF and UV LF predict EVS distributions consistent with JWST observations, including those suggesting that the star formation efficiency in $\Lambda$CDM halos could be $\sim1$ or even higher. Since we cannot perform observations over comparable or larger areas, nor obtain significantly larger galaxy samples, we rely on our empirical models as robust extrapolations of what JWST or other telescopes are expected to observe at different redshifts and survey areas.

Next, we investigate the predicted most massive halos, which we will later compare to our empirical predictions from the GSMF and UV LFs.

\section{Theoretical Predictions} \label{sec:EVS_HMF}

In this paper, we adopt the reference cosmological parameters from \citet{Planck+2015}, see also Section \ref{sec:impact_cosmo_params} for further discussion, when deriving the EVS distribution from the HMF. In this paper, halo masses are defined using a redshift-dependent virial overdensity, which varies from $\Delta \sim 330$ at scale factor $a = 1/(1+z) = 1$ to $\Delta \sim 180$ as $a \rightarrow 0$ \citep[e.g.,][]{Bryan+1998}. 

\subsection{The expected most massive $\Lambda$CDM halo}
\label{sec:most_massive_halo}

\begin{figure}
    \centering
    \includegraphics[width=\columnwidth]{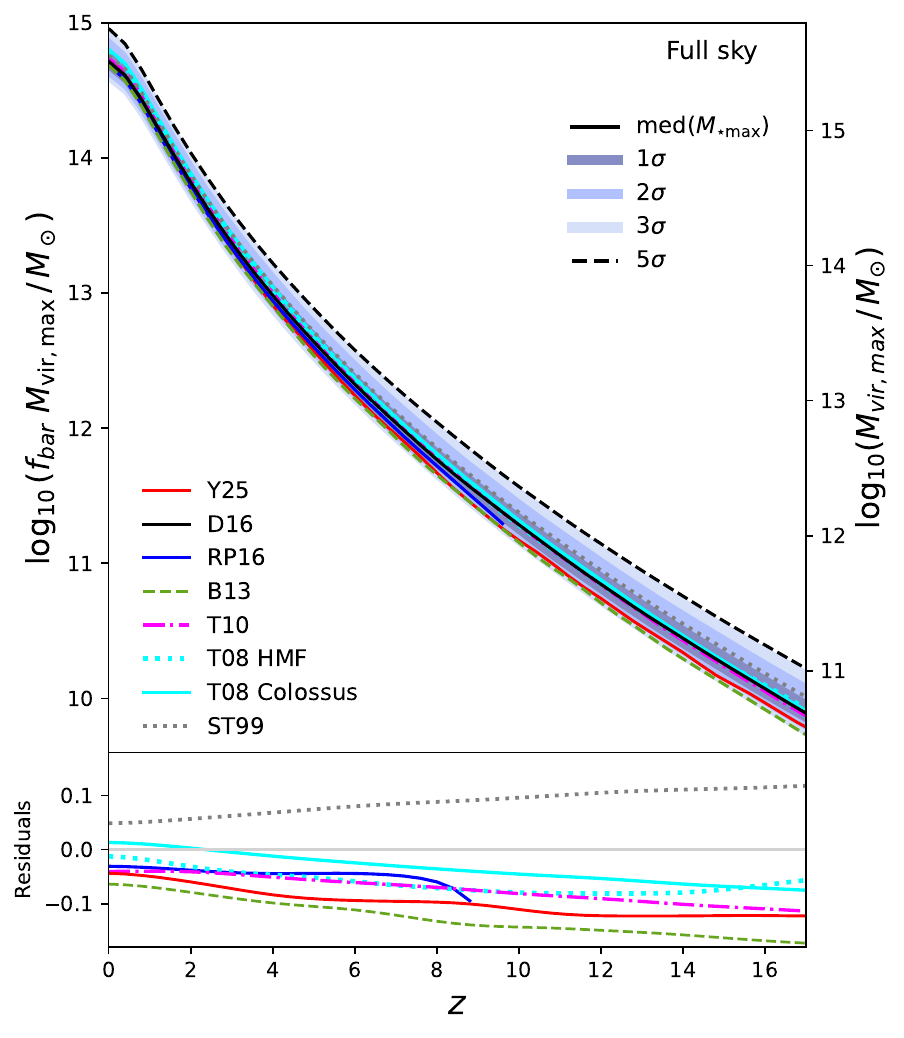}
    \caption{EVS predictions for the maximum dark matter halo mass and the corresponding baryonic mass as a function of redshift for the full sky ($f_{\mathrm{sky}}=1$), assuming $f_\text{bar}=\Omega_\text{bar}/\Omega_{m}=0.156$. The solid curve and shaded regions show the reference HMF model \citep{Despali+2016}, with the median and the $1-3\sigma$ EVS intervals, while the black dashed line marks the $5\sigma$ limit. Additional curves show results from other commonly used halo mass functions. The bottom panel displays their logarithmic differences relative to the reference model (i.e., \citealp{Despali+2016}). A dispersion of $\sim0.1$–$0.2$ dex is found among these HMFs. Note the differences in halo masses between the \citet{Tinker+2008} halo mass function as implemented by \textsc{HMF} and \textsc{COLOSSUS} codes. 
    }
    \label{fig:phimax_differences_between_authors}
\end{figure}

Figure \ref{fig:phimax_differences_between_authors} presents the median of the maximum available baryonic halo mass for the full sky ($f_\text{sky}=1$ with $M_{\text{bar,max}} \equiv f_{\text{bar}} \; M_{\text{vir,max}}$ where $f_{\text{bar}} = \Omega_\text{bar}/\Omega_{m} = 0.156$ is the universal baryon fraction, according to our adopted cosmology). The axis on the right present the corresponding halo mass, i.e., $M_{\text{vir,max}}$. The median of the EVS distribution is calculated using a set of 7 commonly adopted dark matter HMFs in the literature \citet{Sheth+1999,Tinker+2008,Tinker+2010,Behroozi+2013,Rodriguez-Puebla+2016,Despali+2016} and \citet[][see also, \citealp{Yung+2025}]{Yung+2024}. To compute the HMF, in this work we use the \textsc{COLOSSUS} Toolkit \citep{Diemer2018}. 

In Figure \ref{fig:phimax_differences_between_authors} the black solid line with shaded regions represents our reference HMF model \citep{Despali+2016}, showing, in addition, its corresponding $\pm1\sigma, \pm2\sigma$ and $\pm3\sigma$ ranges of the EVS probability distribution. The dashed line corresponds to the $+5\sigma$ of the distribution. In contrast to the results obtained from the GSMF, the EVS derived from the HMF exhibits a much narrower distribution. The remaining lines correspond to the other HMFs, as indicated by the labels. The bottom panels represent the logarithmic difference between these HMFs with respect to the \citet{Despali+2016} HMF. 

\begin{figure*}
    \centering
    \includegraphics[width=2\columnwidth]{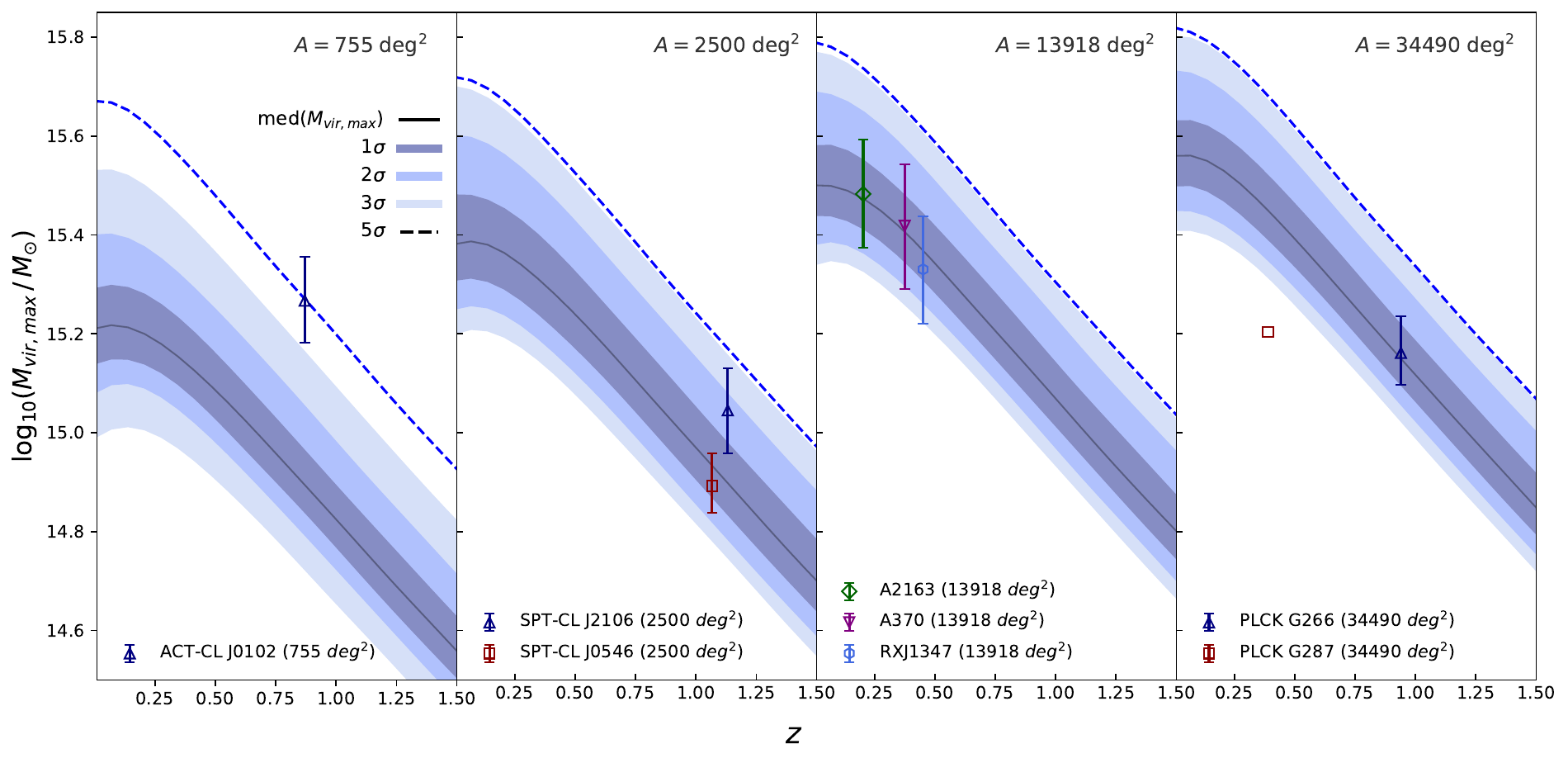}
    \caption{EVS distribution of the most massive dark matter halos over the redshift range $0 \lesssim z \lesssim 1.5$ for four survey areas (from left to right): 750, $2.5\times10^3$, $14\times10^3$, and $35\times10^3,\mathrm{deg}^2$. The solid dark line shows the median EVS prediction, while the shaded regions indicate the $\pm1\sigma$, $\pm2\sigma$, and $\pm3\sigma$ intervals; the dashed line marks the $+5\sigma$ limit. Observed extreme galaxy clusters are overplotted for comparison, including ACT-CL J0102–4915 \citep[“El Gordo”,][]{Menanteau+2012}, SPT-CL J2106 and SPT-CL J0546 \citep{Bleem+2015}, Abell 2163, Abell 370, and RXJ1347 from X-ray surveys \citep{Bohringer+2004}, and PLCKG266 and PLCKG287 from the \textit{Planck} SZ catalogue \citep{Planck+2016}, with the mass of PLCKG287 updated by \citet{Gitti+2025}. In general, our results show that observed clusters are consistent with our theoretical predictions from the HMF and that the ``impossible galaxies'' problem is unlikely to originate from the cosmological model itself, but as a result of the baryonic effiency in halos.}
    \label{fig:evs_hmf_clusters}
\end{figure*}


\begin{figure}
    \centering
    \includegraphics[width=\columnwidth]{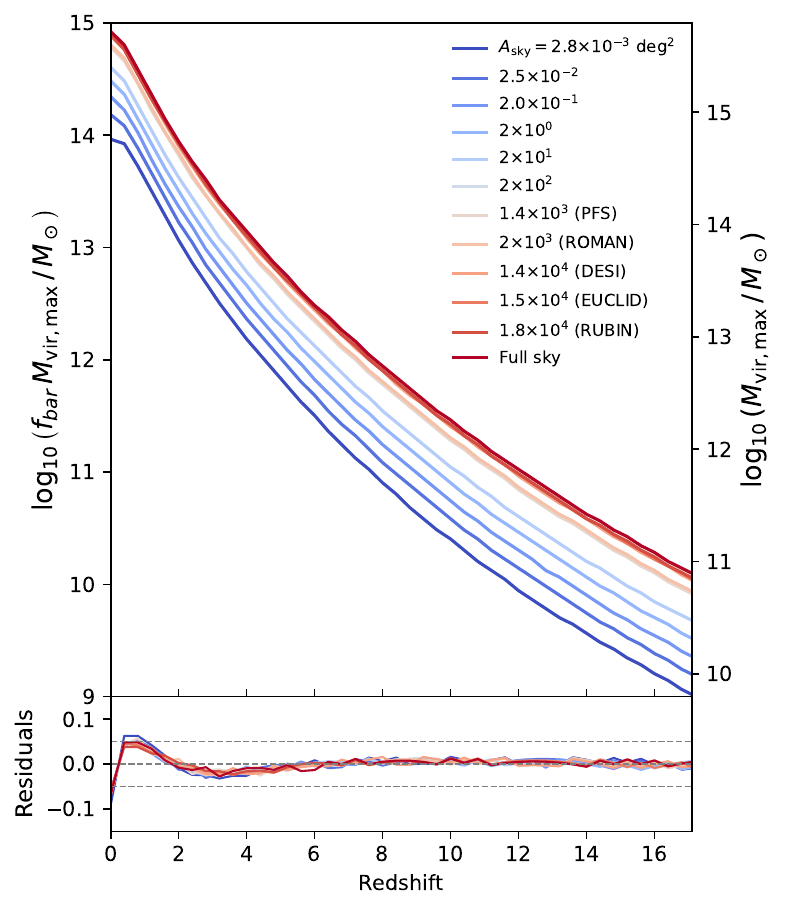}
    \caption{ EVS distribution of the $5\sigma$ halo mass for the $\Lambda$CDM HMF as a function of redshift and survey area. The survey areas are the same as in Figure~\ref{fig:evs_5sigma_gal}. The bottom panel is the residual with respect to the best-fit models to Eq. \ref{eq:model_most_massive_5s}, see also Appendix \ref{sec:evs_fits}. 
    }
    \label{fig:evs_5sigma}
\end{figure}

\begin{figure}
    \centering
    \includegraphics[width=1.07\columnwidth]{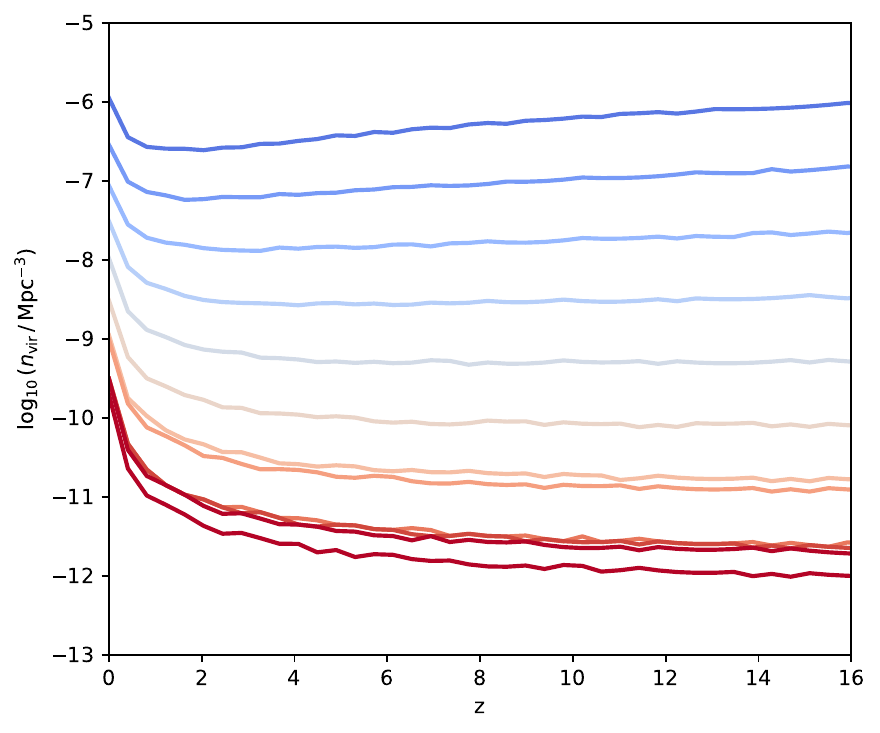}
    \caption{Redshift evolution of the number density $n_{\rm vir}$ of the most massive $5\sigma$ halos predicted by EVS for different survey areas, using the same color scheme as in Figure \ref{fig:evs_5sigma} representing the same survey areas.}
    \label{fig:number_density}
\end{figure}


The HMFs of \citet{Tinker+2008,Tinker+2010,Behroozi+2013,Rodriguez-Puebla+2016} and \citet{Despali+2016} are largely consistent with each other at all redshifts, with deviations between $\sim0.05$ and $\sim0.15$ dex at very high redshifts with respect \citet{Despali+2016}. In contrast, the HMFs by \citet{Sheth+1999} predicts the largest values for the maximum available baryonic mass distribution among all the considered HMFs. It reaches a maximum difference of $\sim0.1$ dex at the highest redshift. 

We note that \citet{Yung+2024} identified discrepancies in the \citet{Tinker+2008} halo mass function when implemented in the \textsc{hmf} \citep{Murray+2013} and \textsc{colossus} packages. We pause here and note that \citet{Lovell+2023} used the \textsc{hmf} code in their EVS analysis. As discussed above, differences between implementations may affect the high-mass end of the HMF, potentially leading to an underestimation of extreme-value statistics. A detailed investigation of the origin of these differences is beyond the scope of this work, and we refer the reader to \citet{Yung+2024} for further discussion. Here, we reproduce these discrepancies between the two implementations. We therefore caution that different numerical implementations can lead to non-negligible variations in HMF predictions.

Figure \ref{fig:phimax_differences_between_authors} shows that while the systematic differences between the HMFs are not large, they are still \emph{nonzero}. Understanding them in the context of the most massive galaxies allowed by the model is important for precision in future studies, see Section \ref{sec:expected_baryons_and_stellar_mass} for a brief discussion. In the reminder of this paper we will ignore this effects and focus more in other sources of uncertainty that may appear to be more important, such as in the observed stellar masses. As such, here we adopt the \citet{Despali+2016} HMF as our reference model, as it is a recent determination based on a large suite of cosmological $N$-body simulations used to constrain the HMF. 

Figure~\ref{fig:evs_hmf_clusters} presents the EVS distribution of dark matter halos over the redshift range $0 \lesssim z \lesssim 1.5$ for four survey areas (from left to right): 750, $2.5\times10^3$, $14\times10^3$, and $35\times10^3 \mathrm{deg}^2$. The solid dark line represents the median of the EVS distribution, while the shaded regions indicate the $1\sigma$, $2\sigma$, and $3\sigma$ intervals; the dashed line marks the $5\sigma$ limit. 

In the same panels we compare these predictions with several extreme observed galaxy clusters, including ACT-CL J0102–4915 ``El Gordo'', a Sunyaev–Zel’dovich (SZ) source that is also bright in X-rays \citep{Menanteau+2012}; SPT-CL J2106 and SPT-CL J0546, identified via their SZ signal by the South Pole Telescope \citep{Bleem+2015}; clusters A2163, A370, RXJ1347 and Abell 2163 and Abell 370 from X-ray surveys \citep{Bohringer+2004}; and PLCKG266 and PLCKG287 from the \textit{Planck} catalogue of SZ sources \citep{Planck+2016}, with the mass of PLCKG287 updated by \citet{Gitti+2025}.
 
Although not shown explicitly in Figure \ref{fig:evs_hmf_clusters}, we also examined several additional massive clusters from the literature, including XMMU J2235.3-2557 and XMMU J0044.0-2033 from the X-ray selected sample of \citet{Santos+2011}, and the infrared-selected cluster IDCS J1426.5+3508 reported by \citet{Brodwin+2016}. We find that IDCS J1426.5+3508 and XMMU J0044.0-2033 lie only 0.09 dex below the 5$\sigma$ EVS threshold, while XMMU J2235.3-2557 lies marginally above it by only 0.05 dex.

In general, the EVS distributions shown in Figure \ref{fig:evs_hmf_clusters} for the most massive $\Lambda$CDM halos are consistent with the cluster observations shown in the different panels. Nevertheless, we find that the El Gordo cluster lies $\sim0.08$ dex above the predicted $5\sigma$ value (leftmost panel). While this may suggest a mild tension with the $\Lambda$CDM model, the observational uncertainty still places the measurement within the $5\sigma$ region, and we therefore consider it consistent with $\Lambda$CDM. Overall, these results show that $\Lambda$CDM successfully reproduces the most extreme clusters reported in the literature. Consequently, some of the tensions discussed in the ``imposible galaxies'' literature are \emph{unlikely to originate from the cosmological model} itself (as our results strongly suggest, but see \citealp{Shen+2024}), but rather from the baryonic processes that regulate the conversion of baryons into stars, as we discuss later in this paper.

We compute the maximum baryonic mass available within the most massive halo. Following the definition outlined in Section \ref{sec:EVS_GSMF}, this corresponds to the $5\sigma$ value of the distribution obtained using the EVS framework applied to the $\Lambda$CDM HMF.  
Figure \ref{fig:evs_5sigma} shows the halo corresponding to the $5\sigma$ value of the $\Lambda$CDM distribution across a wide range of survey apertures, the same ones considered in Section \ref{sec:most_massive_gal}. For reference we show the areas of current and future galaxy surveys. As expected, increasing the sky coverage leads to the detection of increasingly massive halos, reaching a maximum when the entire sky is surveyed. For full-sky coverage, we find that at $z=0$, the most massive halo has a virial mass of $M_\text{vir,max} = 5.248 \times 10^{15} \ M_\odot$ or correspondingly a baryonic mass of $M_\text{bar,max} = f_\text{b} M_\text{vir,max} = 8.202\times10^{14} \ M_\odot$. The corresponding virial radius, based on the definition of \citet{Bryan+1998} and under our adopted cosmological parameters is $R_\text{vir,max} = 5.214$ Mpc. This halo would represent the largest virialized structure in the entire sphere, according to our definition.

To model the redshift evolution of the most massive halo, in units of solar masses, corresponding to the $5\sigma$ value, we adopt the functional form 
described by: 
\begin{equation}
   M_\text{vir,max}(z,f_\text{sky}) = A(f_\text{sky}) \left( 1 + \left[ \frac{a(f_\text{sky})}{1+z} \right]^{\alpha(f_\text{sky}) n(f_\text{sky})} \right)^{-1/n(f_\text{sky})},
   \label{eq:model_most_massive_5s}
\end{equation}
The functional form and best-fit parameters are summarized in Table \ref{tab:evs_model_fit}. This expression captures the evolution of the maximum halo mass expected in a given comoving volume as a function of redshift. The bottom panel of Figure \ref{fig:evs_5sigma} shows the residuals of our fit, in logarithmic units, compared to the values obtained by applying EVS to the HMF.  As shown, our model achieves an accuracy better than $\sim12\%$.  

\subsection{Number density evolution of the most massive halos}

Figure \ref{fig:number_density} presents the redshift evolution of the halo number density, $n_\text{vir}$, for the halos shown in Figure \ref{fig:evs_5sigma}, notice that we use the same color labels as in Figure \ref{fig:evs_5sigma}.  
Overall, the figure reveals two distinct evolutionary regimes for these dark matter halos, whose behavior depends on the survey area. For smaller areas, $\mathcal{A}\text{sky} \leq 20,\mathrm{deg}^2$, the number density declines slowly with decreasing redshift. For the smallest area, it decreases from $n_\text{vir}(z \approx 16) \approx 1 \times 10^{-6}$ to $n_\text{vir}(z \approx 1) \approx 3 \times 10^{-7},\mathrm{Mpc}^{-3}$. This corresponds to a decline by a factor of $\sim3.3$ over a redshift interval of $\Delta z \approx 15$, or a slope of $\Delta \log n_\text{vir} / \Delta z \approx 0.22$, implying an exponential increase in number density toward higher redshift, $n_\text{vir} \propto \exp(0.22,\Delta z)$. For larger areas, $\mathcal{A}_\text{sky} > 20,\mathrm{deg}^2$, the number density is approximately constant.

At lower redshifts ($z \lesssim 1$–4), the number density increases for all survey areas. For the full-sky case, we find $n_\text{vir} \sim 1 \times 10^{-12},\mathrm{Mpc}^{-3}$ at $z \sim 4$, rising to $n_\text{vir} \sim 3 \times 10^{-10},\mathrm{Mpc}^{-3}$ by $z = 0$, implying an evolution of $n_\text{vir} \propto \exp(-0.62,z)$.

A natural question is whether estimating the most massive halo at each redshift is equivalent to tracing the assembly history of the most massive halo at $z=0$. If that were the case, a naive expectation would be that its number density remains nearly constant across redshift. This behavior would be expected in the absence of mergers and without scatter in halo mass assembly histories \citep[][see also \citealp{Rodriguez-Puebla+2025} for a recent discussion]{Behroozi+2013f,Clauwens+2016,Wang+2023}. Figure \ref{fig:number_density} suggests that this is approximately true: the number density remains roughly constant between $z \sim 16$ and $z \sim 1-4$, for all survey areas. However, for the smallest areas, as discussed above, the number density slowly increases exponentially with redshift. This behavior is not fully understood, but it may be related to the higher matter density of the early universe and the high merger rates of massive halos during that epoch that. It is possible that some of these early massive halos subsequently reduced their accretion rates and thus slowing their mass growth at later times, and left behind from the highest-mass bin. The latter would explain the decrease in their number density. On the other hand, the increase in number density at lower redshifts $z \lesssim 1-4$ is even more puzzling.  
One plausible explanation is that galaxy surveys observe over light-cones at a fixed solid angle, $\Omega$. As redshift decreases, the effective survey volume also decreases, leading to the inference of $5\sigma$ halos that are smaller than their counterparts at neighboring redshifts. Hence their increase in their number densities.

Returning to the original question, whether tracking the most massive halo across redshift traces the assembly of the most massive halo at $z=0$, the answer appears to be approximately yes between $z \sim 16$ and $z \sim 2-4$ (depending on the area), though it remains uncertain outside this redshift range. Nonetheless, particularly in the case of $f_\text{sky}=1$, almost by definition, there are no halos larger than our $5\sigma$ halo. We therefore adopt this assumption and proceed to explore its implications. In other words, we interpret the evolution of the most massive halo with redshift as a proxy for the assembly history and progenitors of the most massive halo at $z=0$. Under this assumption, we are particularly interested in the expected baryonic content of the most massive halo, $M_\text{vir,max}$, for $f_\text{sky} = 1$.

\begin{figure*}
    \centering
    \includegraphics[width=1.02\columnwidth]{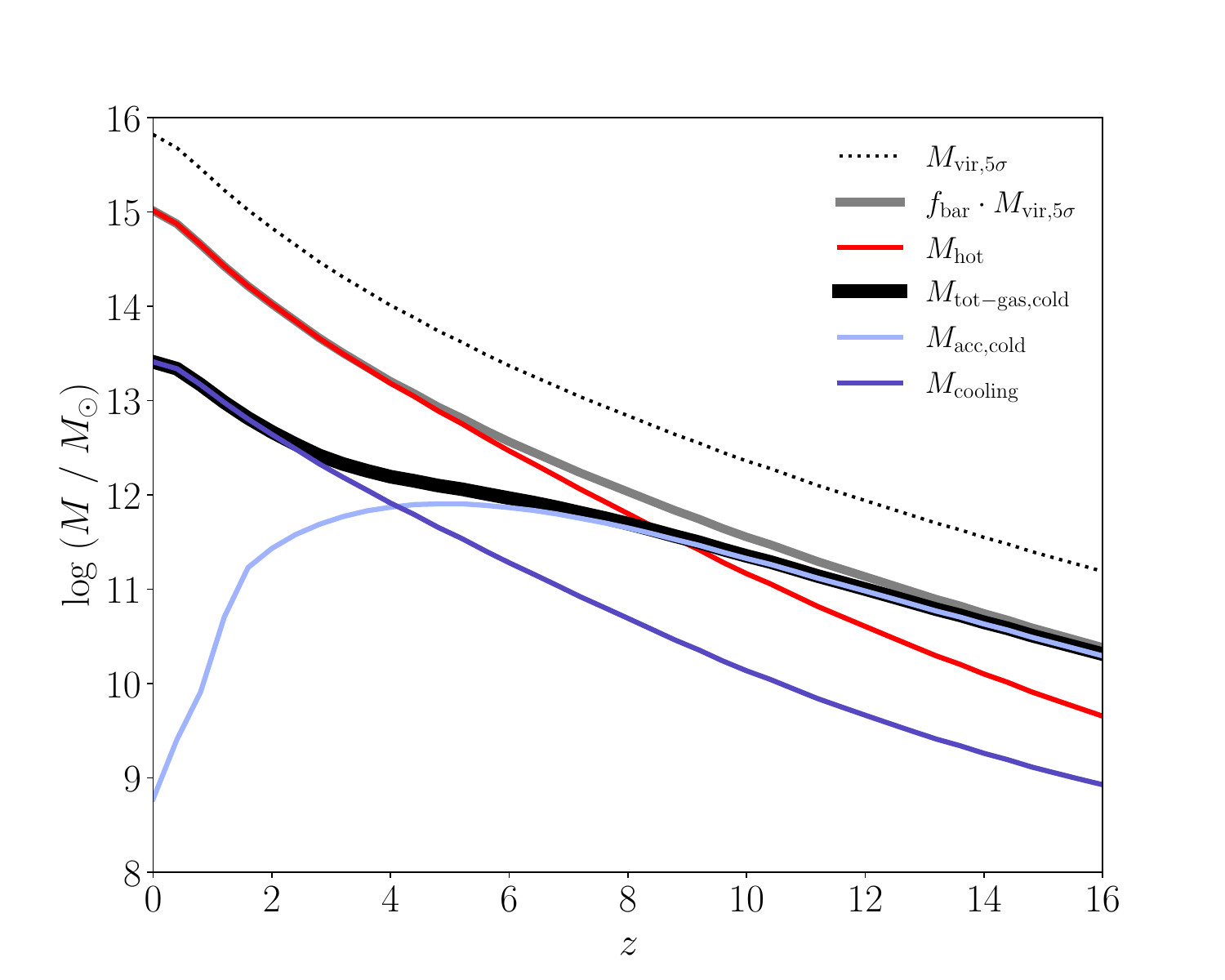}
    \includegraphics[width=1.02\columnwidth]{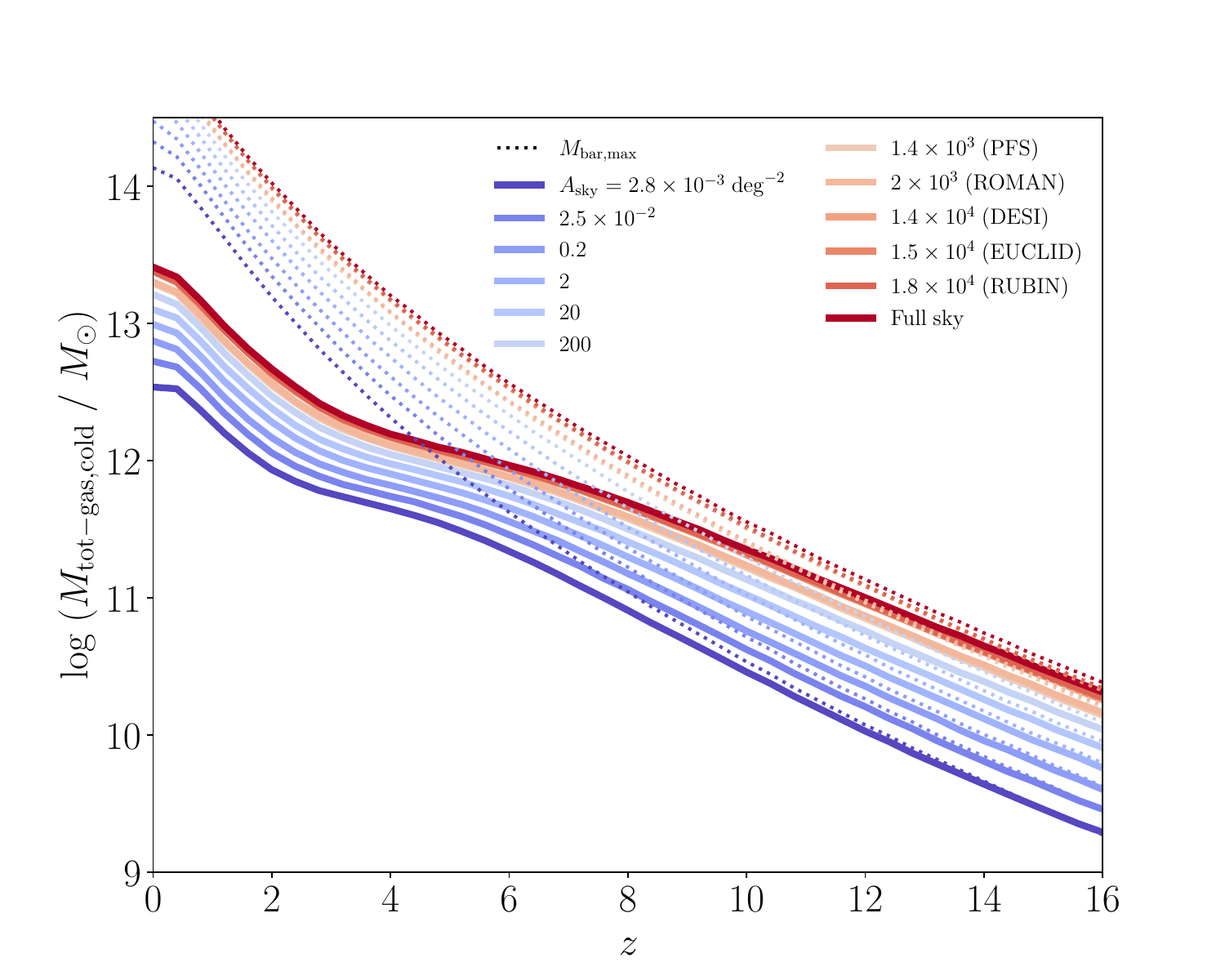}
\caption{
\textbf{Left panel:} Baryonic content of the $5\sigma$ most massive halo in a full-sky survey. The dotted line represents the mass of the most massive halo, while the gray solid line shows the maximum baryonic mass, $f_\text{bar}M_\text{vir}$. The red line indicates the gas mass in the hot phase, $M_\text{hot}$, while the thin dark- and light-violet lines show the cooling gas mass, $M_\text{cooling}$, and the accreted cold gas mass, $M_\text{cold}$, respectively. The thick black solid line represents the total cold gas mass, composed of cooling gas and cold streams. Each component is computed using the model described in the text.
\textbf{Right panel:} Total cold gas mass of the $5\sigma$ most massive halos as a function of survey area. The dotted lines show the corresponding total baryonic masses of the $5\sigma$ most massive halos.
}
    \label{fig:evs_5sigma_baryons}
\end{figure*}


\subsection{What is the expected baryonic content of the most massive halo?}
\label{sec:expected_cold_gas_mass}

The halo corresponding to the $5\sigma$ value derived from full-sky coverage, described above, is reproduced in Figure \ref{fig:evs_5sigma_baryons} and shown as a dotted line. In this subsection we examine its baryonic content and the physical conditions to which these baryons are subjected within the dark matter halo. Understanding these conditions is crucial for determining the nature of the most massive galaxies that such halos can host.

We begin by analyzing the most direct component of the baryonic content: the total baryonic mass, $M_\text{bar,max}$, shown as the gray line in Figure \ref{fig:evs_5sigma_baryons}. As discussed previously, the maximum baryonic mass available to the most massive dark matter halo at $z \sim 0$ is approximately $8\times10^{14} M_\odot$. However, this would only represent the maximum possible stellar mass of the hosted galaxy if all the baryons were in a cold phase and all the cold gas was transformed into stars, which does not happen in reality. 

In the absence of mergers, the multiphase structure of intrahalo baryons can be attributed to two reasons. Firstly, the gas accreted onto the halo can vary in temperature depending on how it was accreted. It has been well-established that most of the accreted gas is shock-heated to temperatures $T \gtrsim 10^5\,{\rm K}$ (e.g., see \citealt{Correa+2018}) once halos reach a critical mass that is nearly constant with redshift ($M_\text{shock} \sim 10^{11.8}\,{\rm M}_\odot$; \citealt{Dekel+2006}). This is termed the `hot mode' accretion. Massive halos at high redshifts can also be fed by narrow streams of dense gas that deliver cold gas directly to the central regions, known as `cold mode' accretion \citep{Keres+2005,Dekel+2009,Faucher+2011,Mandelker+2020}. The survival of these cold streams is dependent on whether the disruption via Kelvin–Helmholtz instabilities occurs on a timescale longer than the cooling time, which sets a corresponding redshift-dependent critical halo mass ($M_\text{stream}$; see \citealt{Mandelker+2020b,Daddi+2022}). The transition between these accretion regimes is found to usually occur at $z_\tau\approx1-2$ \citep{Dekel+2009,Mandelker+2020}. Additionally, radiative cooling can allow hot intrahalo gas to lose energy and flow inward toward the halo center by diminishing its thermal pressure support \citep{WhiteFrenk1991}.

We attempt to estimate the {\it maximum} cold gas that the halo can possess and ignore the heating effects from feedbacks from stellar or AGN activity. For the sake of simplicity, we also ignore the contribution from mergers. In this scenario, the fraction of accreted baryons that eventually ends up in the cold phase is the summation of the mass coming through cold streams and that from the cooling of hot gas. We model both of these processes based on the latest developments, as described below.

We begin by quantifying the contribution from cold streams through the prescription by \citet{Waterval+2025} 
inspired by two suites of state-of-the-art hydrodynamical cosmological simulations of galaxy formation: Numerical Investigation of a Hundred Astrophysical Objects (NIHAO; \citealt{Wang+2015}) and High-$z$ Evolution of Large and Luminous Objects (HELLO; \citealt{Waterval+2024}). In their analysis, the hot gas is deemed to be the gas with $T>2.5\times 10^5\,{\rm K}$ acquired via \textit{smooth accretion} by the halo, where `smooth'
implies that the gas has never been bound to any halo other than the one in question. 
The fraction of accreted gas supplied by cold streams is modeled using a double sigmoid function, enabling a smooth and continuous transition between two different accretion regimes at $z_\tau \simeq 1.23$.

\begin{equation}\label{fcold}
    f_{\text{acc, cold}} = \frac{\tau}{1+e^{-\kappa_1(\log M_\text{vir}-\log M_{\text{c},1})}}
    +\frac{1-\tau}{1+e^{-\kappa_2(\log M_\text{vir}-\log M_{\text{c},2})}},
\end{equation}
where
\begin{equation}
\begin{aligned}
    &\kappa_1 = \, -3.60+2.40\log(1+z),\\
    &\kappa_2 = \, -2.65 + 1.50\log(1+z),\\
    &\log M_{\text{c},1} = \, 11.83 - 0.18\log(1+z),\\
    &\log M_{\text{c},2} = \, 10.94 + 1.69\log(1+z),\\
    &\tau =\, \frac{1}{1+e^{10(z-1.23)}}.
\end{aligned}
\end{equation}
The first term in Eq. (\ref{fcold}) describes the $f_{\text{acc, cold}}$ at $z\lesssim z_\tau$, the second term models it for higher redshifts, and the transition between accretion regimes is modulated by $\tau$. We define $M_{1/2,\text{hot}}(z)$ as the virial halo mass at which the hot-gas fraction reaches $f_\text{cold}(z,M_{1/2,\text{hot}}) = f_\text{hot} (z,M_{1/2,\text{hot}})= 1/2$.

Next, we model the cold gas generated via radiative cooling of the hot gas. This requires
the cooling time to be shorter than the dynamical timescale \citep{White+1978,WhiteFrenk1991},
which is satisfied at radii smaller than a threshold `cooling radius' ($r_\text{cool}$) where the gas density is high enough to strike a balance between the two timescales. We denote the mass of gas that cools and flows inward -- i.e. the gas enclosed within $r_\text{cool}$ -- by $M_\text{cooling}$, and estimate it assuming an isothermal gas density profile as
\begin{equation}
M_\text{cooling} = M_\text{acc, hot} \; \frac{r_\text{cool}}{r_\text{vir}},
\end{equation}
where $M_\text{acc, hot} =(1-f_\text{acc, cold})f_\text{bar}M_\text{vir}$ is the fraction of accreted
gas in the hot phase, and $r_\text{vir}$ is the virial radius of the halo \citep[see e.g.,][for a similar treatment]{Correa+2018}. For $r_\text{cool}$, we introduce a heuristic formalism that parametrises $r_\text{cool}/ r_\text{vir}$ as a function of $M_\text{vir}$ and $z$, inspired by the latest cooling model used in the {\sc{\large galform}} semi-analytical framework \citep{Hou+2018}. Here $r_\text{cool}$ is defined
as the radius where the cooling time is equal to the time available for radiating
away the thermal energy in the shell. Unlike other cooling models in the literature, it simultaneously accounts for the contraction of the hot gas halo due to cooling in inner regions and by dark matter halo growth, and the effect of the cooling history of the hot gas on the current cooling rate. These modelling improvements have been demonstrated to be crucial for providing agreement with hydrodynamical simulations without feedback \citep{Hou+2019}. Our formalism has been calibrated to reproduce the median relations in the right column of fig. 1 in \citet{Hou+2019}, and is given below:
\begin{equation}
\frac{r_\text{cool}}{r_\text{vir}} = A(1+z)^\gamma10^{\alpha(\log M_\text{vir}-x_1)}+\frac{B(1+z)^{\delta}}{1+e^{-(\log M_\text{vir}-x_2)/s}},
\end{equation}
where
\begin{equation}
\begin{aligned}
    &A = 0.06491,\\
    &\gamma = 0.13654,\\
    &x_1 = 11.20819,\\
    &B = 0.02481,\\
    &\delta = 0.45913,\\
    &x_2 = 9.61671,\\
    &s = 0.40,
\end{aligned}
\end{equation}
and
\begin{equation}
\alpha = 1.07174 - \frac{0.48337}{1+e^{-(\log M_\text{vir}-10.59352)/0.2}}.
\end{equation}

From the above, the total amount of cold gas that reaches the galaxy and is available for star formation at any redshift is given by
\begin{equation}
    M_\text{cold} = M_\text{acc, cold} + M_\text{cooling} \approx M_\text{acc, cold} + \frac{r_\text{cool}}{r_\text{vir}} \times M_\text{acc, hot},
    \label{eq:Mcosmo-cold}
\end{equation}
where $M_\text{acc, cold}$ is the fraction of accreted gas acquired via cold streams. If galaxies were to convert this cold gas into stars with $100\%$ efficiency, and in the absence of any feedback processes, $M_\text{cold}$ would represent the maximum possible stellar mass of the central galaxy in the most massive dark matter halo. 

The left panel of Figure \ref{fig:evs_5sigma_baryons} shows the evolution of the baryonic components for the most massive halo permitted by $\Lambda$CDM across the full sky. The accreted hot gas mass, $M_\text{acc, hot}$, is plotted as a thin red solid line; the total cold gas mass, $M_\text{cold}$, as a thick black solid line; and its two components, the cooling gas mass, $M_\text{cooling}$, and the accreted cold gas mass, $M_\text{acc,cold}$, as thin dark- and light-violet lines, respectively.

At high redshifts, $z \gtrsim 5$, most of the gas accreted onto the central regions of halos arrives in cold form through narrow streams \citep{Keres+2005,Dekel+2009,Mandelker+2020}. By $z \sim 4$, however, when halos reach virial masses of $M_\text{vir} \sim M_{1/2,\text{hot}}$ most of the gas has been shock-heated and resides primarily in the hot phase. 
At $z \lesssim 3$, cooling flows begin to dominate, supplying the central galaxy with gas that becomes the main fuel for star formation at later times.

In the absence of feedback processes, the total cold gas mass, shown by the thick black line, would represent the upper limit for the stellar mass of the most massive galaxy that could form within the halo. While this prediction appears unrealistic at low redshift, since no galaxies with stellar masses as high as $M_\ast \sim 2\times 10^{13} M_\odot$ are observed, with the most massive known galaxies typically reaching $M_\ast \sim 5 \times 10^{12} M_\odot$ \citep[see e.g.,][]{Dullo_2019}. On the contrary, the situation at higher redshifts, $z \gtrsim 7-8$, may offer clues about the existence of the massive galaxies identified by JWST \citep[see e.g.,][]{Dekel+2023, Boylan-Kolchin_2023} since most of the baryon mass is available as cold gas to form stars. It is therefore essential to understand the conditions under which this gas can be converted into stars, as we will discuss in Section \ref{sec:physical_cond}.

The right panel of Figure \ref{fig:evs_5sigma_baryons} shows the total cold gas mass, assuming no feedback, for $5\sigma$ halos across a wide range of survey apertures. For comparison, the dotted lines indicate the total baryonic mass of each $5\sigma$ halo. Starting from the smallest area, $A_\text{sky}=2.8\times10^{-3}\text{ deg}^2$, up to the full sky, we find that the cold gas mass begins to diverge from the total baryon mass between $z\sim12$ and $z\sim8$, respectively. Below these redshifts, the cold gas mass changes in slope, reflecting the transition from cold-mode accretion to hot-mode cooling flows. For smaller survey areas, corresponding to less massive halos, this transition occurs at lower redshifts. In particular, for our smallest survey area, the transition takes place around $z\sim2$ while for the full sky survey this is around $z\sim4$, implying that these halos underwent the shift from cold to hot accretion more recently than their more massive counterparts.

\begin{figure*}
    \centering
    \includegraphics[width=1.6\columnwidth]{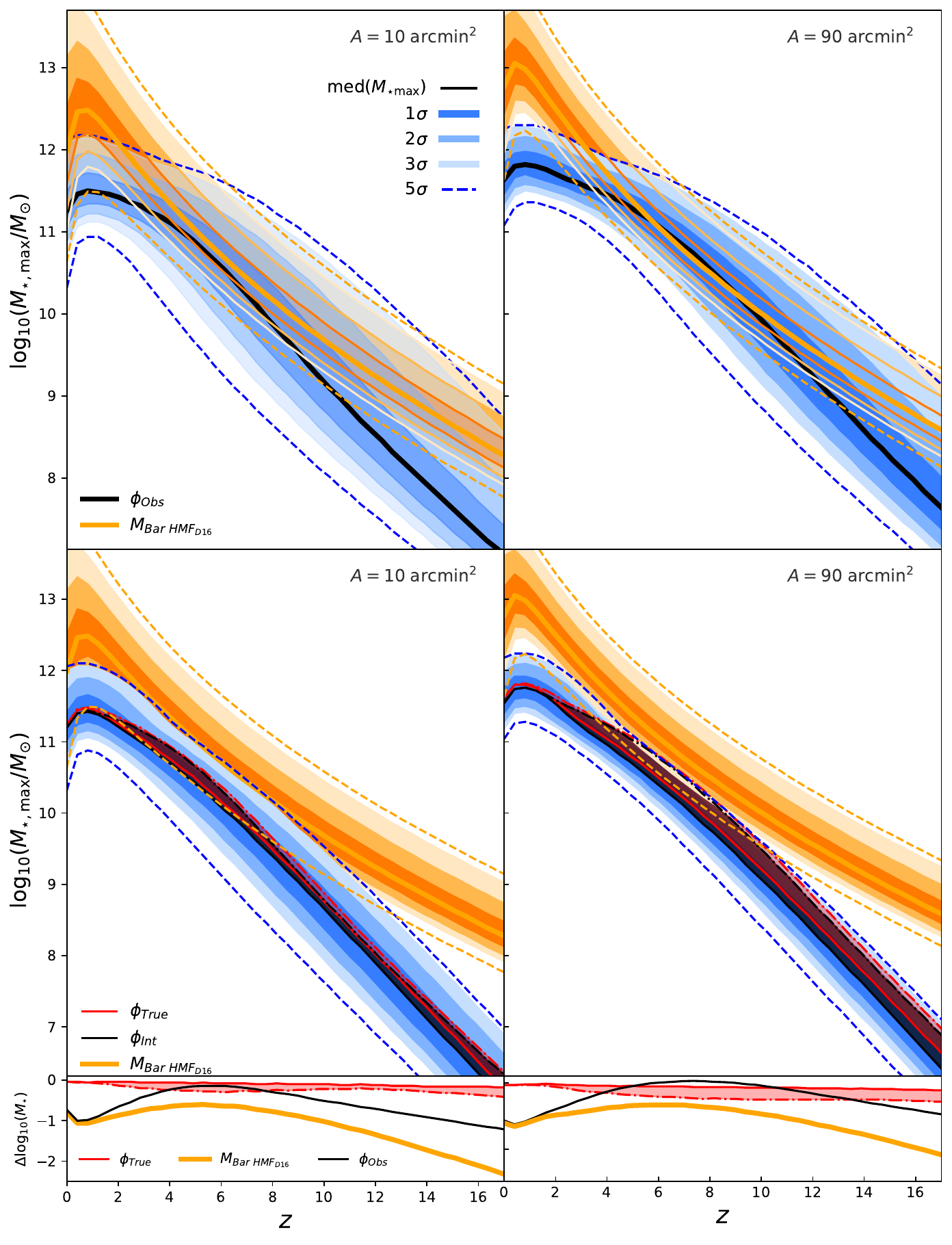}
 
\caption{
EVS distributions for $M_{\ast,\text{max}}$ and $M_\text{bar,max} = f_\text{bar} M_\text{vir,max}$. \textbf{Upper Panels:} The black solid line shows the median, while the blue shaded regions indicate the $1$–$3\sigma$ intervals and the blue dashed lines mark the $5\sigma$ limit for the EVS distribution derived from the observed GSMF in small survey areas of $10$ and $90,\mathrm{arcmin}^2$, similar to those probed by JWST. The yellow curve and shaded regions show the corresponding EVS predictions for the maximum available baryonic mass, $M_{\mathrm{bar,max}}$. In both panels, the two EVS distributions overlap over a wide range of redshifts, particularly at $3\lesssim z \lesssim 6$. This would imply that halos convert nearly $100\%$ of their baryons into stars, consistent with previous claims. \textbf{Medium Panels:} Same as above, but for the EVS distributions derived from the intrinsic GSMF, accounting for random errors in stellar mass estimates as well as the intrinsic scatter in the GSMF (black line and blue shaded regions). The red region between the red solid and dot-dashed lines shows the result obtained using the true GSMF, including different values of the random errors in stellar mass estimates. Both EVS distributions are compared with the EVS prediction for the maximum available baryonic mass, as in the upper panel. When Eddington bias is taken into account, halos generally have efficiencies below the maximum available baryonic limit, i.e., $M_{\mathrm{bar,max}} > M_{\ast,\mathrm{max}}$. Moreover, at higher redshifts, $z \gtrsim 8$, the implied efficiency of dark matter halos is significantly lower than previously suggested \citet[see, e.g.,][]{Boylan-Kolchin_2023}. The lower panels show the logarithmic difference in median values for the intrinsic case compared with the cases labelled inside the panels.
}
    \label{fig:phimax_halo_vs_gals}
\end{figure*}


\section{Examining the empirical connection between the most massive galaxies and the most massive halos}
\label{sec:connecting_massive_gals_to_halos}

In the preceding two sections, we examined separately the most massive galaxies and halos (along with their maximum baryonic and cold gas content) as a function of survey area and redshift. In this section, we combine those results to provide empirical estimates of how massive galaxies can become within the $\Lambda$CDM framework, as expected.  

\subsection{Comparison with JWST areas}

Given the attention that JWST has received over the past few years, we start our discussion by reproducing the results of Figure \ref{fig:phiOTI} in Figure \ref{fig:phimax_halo_vs_gals}, this time comparing them to the EVS distribution of the most massive halos rescaled by the universal baryon fraction, $f_\text{bar}$, as previously done in the literature \citep[e.g.,][]{Boylan-Kolchin_2023,Lovell+2023}. As mentioned earlier, our empirical results are robust extrapolations valid for any survey at any redshift. In particular, this allows us to place results from different JWST areas in the context of theoretical expectations. Since most of the curves were already presented in Figure \ref{fig:phiOTI}, we will not discuss them again here. 

The upper-left panel of Figure \ref{fig:phimax_halo_vs_gals} shows the $\pm1-3\sigma$ EVS distribution for the maximum stellar mass from the observed GSMF, $\Phi(M_{\ast,\text{max|obs}})$, while the same is presented in the middle-left panel but for the intrinsic GSMF, $\Phi(M_{\ast,\text{int|obs}})$. In this panel is plotted also the median of the maximum stellar mass and its uncertainty from the true GSMF. Both panels are for a $10 \text{ arcmin}^2$ field. In both panels, we also reproduce the theoretical EVS distribution for the maximum baryonic halo mass, $M_\text{bar,max}$, for the same field size (yellow lines and shaded areas). In both panels the color short-dashed lines encompass the $\pm5\sigma$ values of the  corresponding EVS distributions. Additionally, the bottom panel shows the differences between the medians plotted in the upper panels with respect to the median $M_{\ast,\text{max|int}}$. 

We begin our discussion by comparing $\Phi(M_{\ast,\text{max|obs}})$ with $\Phi(M_\text{bar,max})$. Overall, the EVS distributions overlap at the 3-$\sigma$ levels across most of the redshift range, particularly at intermediate redshift, $3 \lesssim z \lesssim 6$. Qualitatively, this suggests that many galaxies reach efficiencies close to $100\%$. However, as we will see below, the comparison between the two curves is not straightforward.

Focusing on their median relations, at low redshifts ($z \lesssim 2$) the observed median, $M_{\ast,\text{max|obs}}$, lies below the baryonic halo median, with a difference of about $\sim 1$ dex at $z \sim 0$. At high redshifts, $z \gtrsim 8$, the medians diverge again, reaching a difference of $\sim 1.5$ dex by $z \sim 16$. In contrast, at intermediate redshifts, $2 \lesssim z \lesssim 8$, the baryonic halo median closely approaches the observed $M_{\ast,\text{max|obs}}$.

While the above result holds, it does not account for the Eddington bias. The red lines with shaded regions in the middle left panel of Figure \ref{fig:phimax_halo_vs_gals} incorporate the observed random errors in stellar masses, as described in Section \ref{sec:random_errors}. When compared to the baryonic halo median, the effect of Eddington bias becomes evident: it reduces the number density at the high-mass end, implying less massive galaxies. At low redshifts, our results remain similar to those obtained from the observed median stellar mass. At high redshifts, however, the differences become substantial, reaching $\sim 2.5$ dex at $z \sim 16$. At intermediate redshifts, $2 \lesssim z \lesssim 8$, the baryonic halo median lies above the corrected values, yielding $ M_\text{bar,max} > M_{\ast,\text{max|true}}$.

Accounting for the scatter in the stellar-to-halo mass relation introduces a second source of Eddington bias. This additional correction, which is crucial when comparing to dark matter halos, predicts lower maximum stellar masses, $M_{\ast,\text{max|int}}$, as discussed in Section \ref{sec:EVS_GSMF}. Overall, the results remain broadly consistent with the comparison between the medians $M_\text{bar,max}$ and $M_{\ast,\text{max|true}}$. At high redshifts, however, we find slightly larger discrepancies with $M_\text{bar,max}$, even after accounting for the uncertainty in the galaxy–halo connection. This leads to a more robust conclusion: at $z \gtrsim 8$, the efficiency of dark matter halos in converting gas into stars is significantly lower than previously suggested \citep[see e.g.,][]{Boylan-Kolchin_2023}. For clarity, the bottom panels show the differences between medians relative to the median $M_{\ast,\text{max|int}}$.

The right panels of Figure \ref{fig:phimax_halo_vs_gals} present the same information as the left ones, but for an area of 90 $\text{arcmin}^2$. While most of the conclusions drawn for the 10 $\text{arcmin}^2$ field remain valid, the distributions become narrower. In the redshift range $4 \lesssim z \lesssim 8$, the baryonic halo median lies slightly closer to $M_{\ast,\text{max|int}}$ than in the 10 $\text{arcmin}^2$ case, though it never falls below it.

So far, our results suggest that much of the claimed tension between the $\Lambda$CDM model and the massive galaxies observed by JWST at high redshifts arises because random errors can artificially inflate the expected abundance of massive galaxies \citep{Chen+2023}. This effect is further amplified by the intrinsic scatter in the SHMR. Nonetheless, our results are not yet conclusive as to whether halos could convert their baryons into stars with the maximum efficiency allowed by $\Lambda$CDM, $f_\text{bar}\sim0.16$. In the following, we address this question in more detail by modeling the full EVS distributions to estimate the distribution of the maximum ratio $R = M_\ast/M_\text{vir}$.

\begin{figure*}
    \centering
    \includegraphics[width=1.\textwidth]{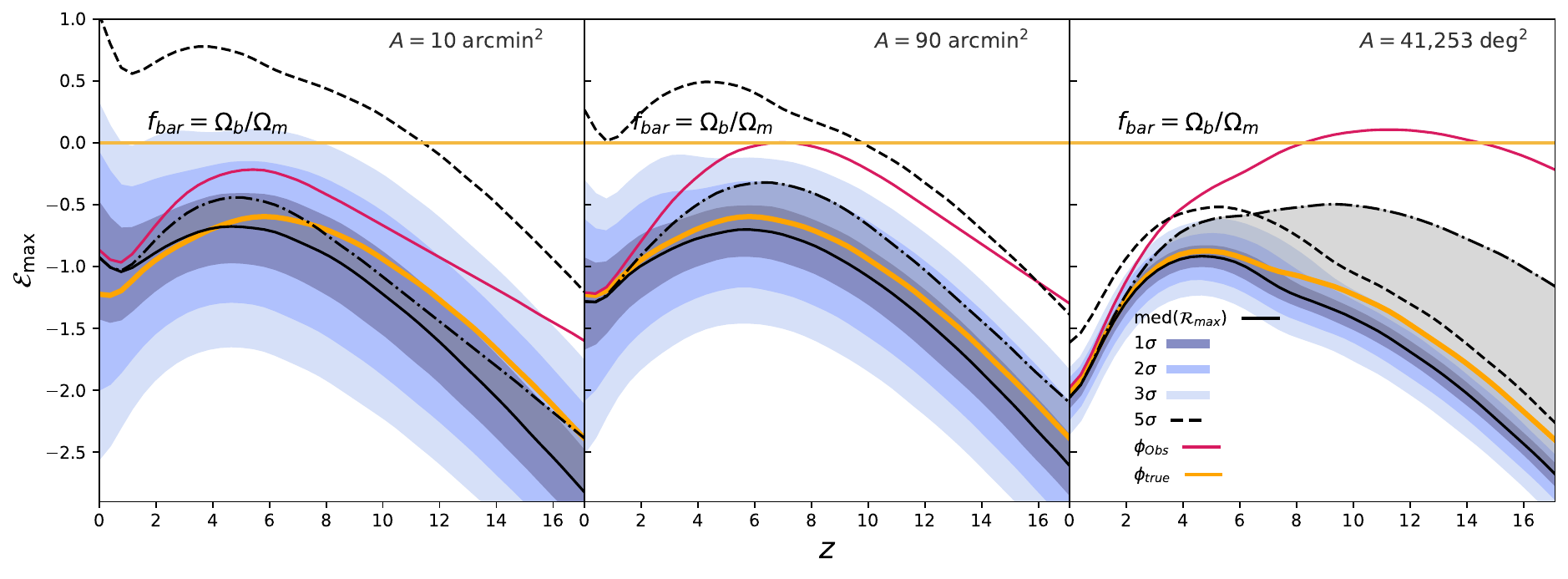}  

    \caption{ EVS distribution of the maximum cumulative star formation efficiency in $\Lambda$CDM halos in the stochastic case, computed as the convolution of the EVS distributions obtained from the GSMF and the HMF. The figure shows the redshift evolution of log$\epsilon_{\max} =$log$[M_{\ast,\max}/(f_{\mathrm{bar}} M_{\mathrm{vir},\max})]$ for different survey areas: from left to right, $10$, $90$, and $41,243$ arcmin$^2$. The red solid line shows the median of the distribution $\Phi_\text{eff}(\mathcal{E}_\text{max|obs})$, while the yellow line represents the median of the EVS distribution $\Phi_\text{eff}(\mathcal{E}_\text{max|true})$. The black line and the shaded regions around correspond to the median and $\pm1-3\sigma$ of the $\Phi_\text{eff}(\mathcal{E}_\text{max|int})$ EVS distribution, which represents the intrinsic galaxy–halo connection. The gray area encompassed between the solid and dash-dotted lines shows the uncertainty caused by the various methods used to calculate random errors in the stellar mass estimates.
}
    \label{fig:conv_R}
\end{figure*}


\begin{figure*}
    \centering
    \includegraphics[width=1.04\textwidth]{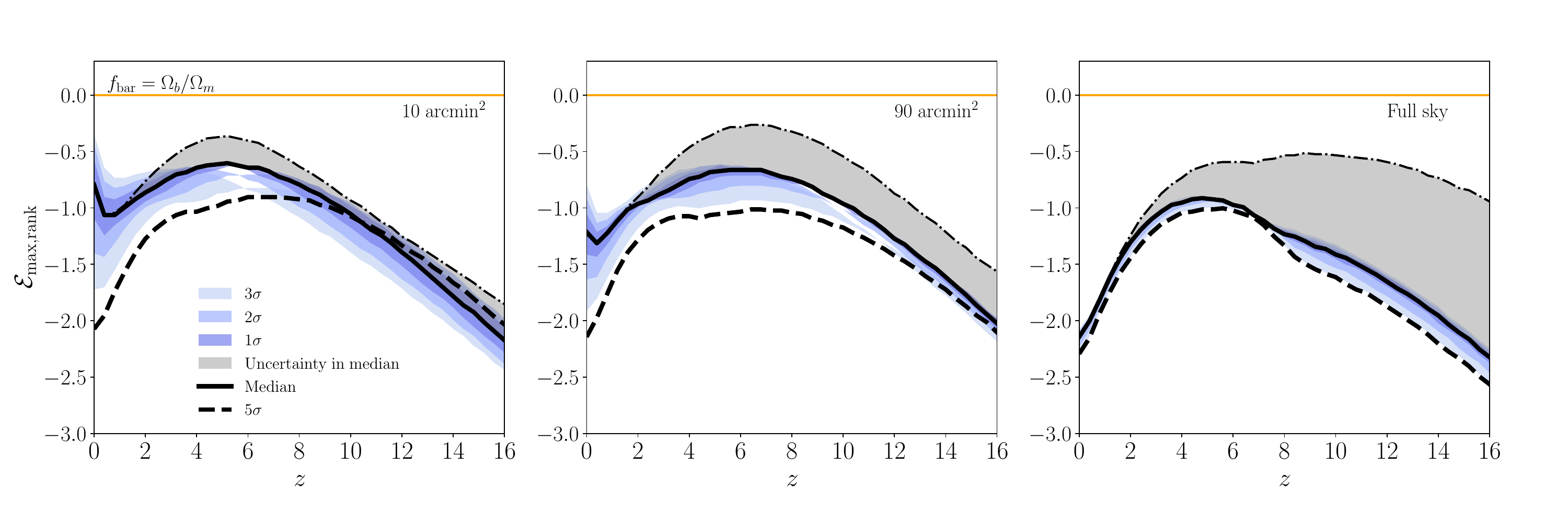}
    \includegraphics[width=1.04\textwidth]{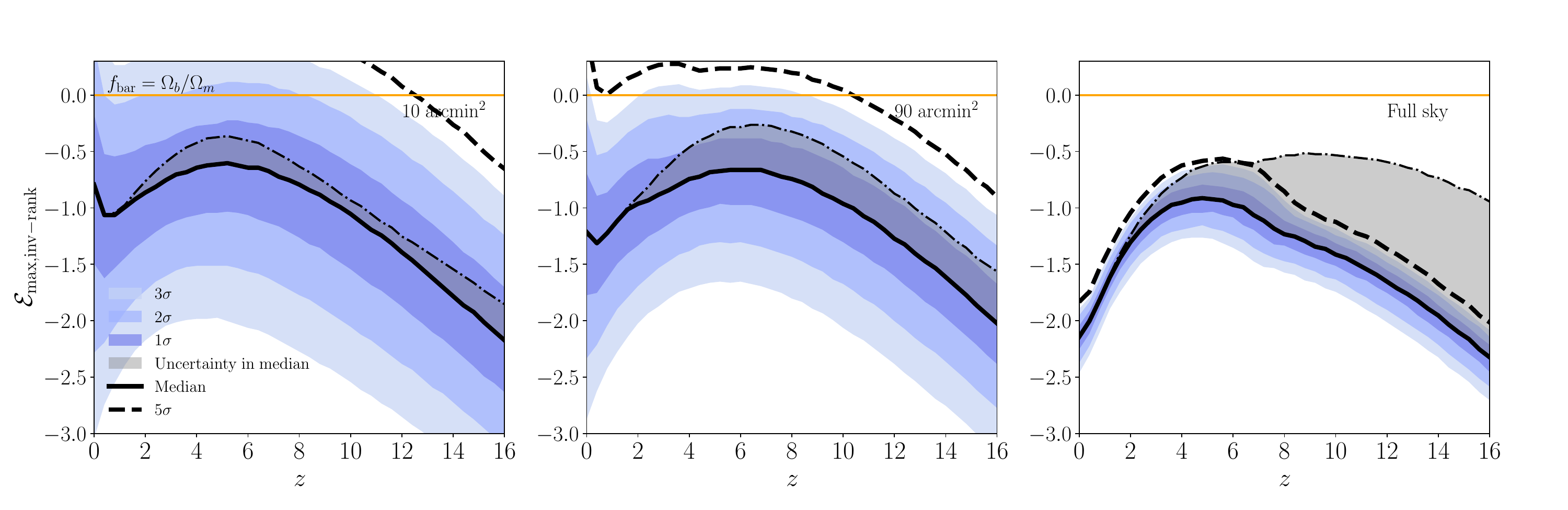}
    \caption{
Same as Figure~\ref{fig:conv_R}, but for a positively correlated rank order (upper panels) and an inverted rank order (lower panels). Only the EVS distributions that correspond to the intrinsic GSMF, $\Phi_\text{eff}(\mathcal{E}_\text{max|int})$, are plotted.
}

    \label{fig:evs_5sigma_density_ranking}
\end{figure*}


\subsection{The cumulative star formation efficiency in the most massive $\Lambda$CDM halos}

Here we explore three extreme cases on how to correlate galaxies and dark matter halos.

\subsubsection{The stochastic case}

It is important to note that computing the distribution of maximum cumulative efficiency is not as simple as taking the ratio of the curves displayed in Figure \ref{fig:phimax_halo_vs_gals} between $M_{\ast,\text{max}}$ and the baryonic mass $M_\text{bar,max} = f_\text{bar}M_\text{vir,max}$. Instead, we use Eq. (\ref{eq:SHMR_ratio_max}) from Section \ref{sec:gal_halo_conn}, which assumes that $M_{\ast,\text{max}}$ and $M_\text{vir,max}$ are independent variables. Under this assumption, the EVS distribution of the maximum stellar-to-halo mass ratio, $R_\text{max}$, or equivalently the cumulative star formation efficiency $\epsilon_\text{max} = R_\text{max} /f_\text{bar}$ (recall that $\mathcal{E} = \log \epsilon$ and $\mathcal{R} = \log R$), can be computed as a convolution of the EVS distributions of $M_{\ast,\text{max}}$ and $M_\text{vir,max}$.

Figure \ref{fig:conv_R} shows the EVS distribution of the maximum star formation efficiency in dark matter halos, $\Phi_\text{eff}(\mathcal{E}_{\text{max|j}})$, following the notation introduced in Section \ref{sec:gal_halo_conn}. We use the subscript $j$ in $\mathcal{E}_{\text{max|}j}$ to indicate the type of GSMF employed, with $j =$ `obs', `true', or `int'. That is, observed GSMF, GSMF deconvolved by random errors, and, in addition to that, by the scatter around the SHMR. 

From left to right, we show the results for survey areas of 10, 90, and 41,243 arcmin$^2$ (the whole sky), respectively. The red and yellow solid lines represent the median of the distribution from $\Phi_\text{eff}(\mathcal{E}_{\text{max| obs}})$ and $\Phi_\text{eff}(\mathcal{E}_{\text{max| true}})$, respectively. Additionally, the solid and dot-dashed black lines enclosing the gray area correspond to the median from the EVS distribution $\Phi_\ast(\mathcal{E}_{\text{max| int}})$, which provides a more accurate inference of the intrinsic galaxy-halo connection and the uncertainty around it caused by the various methods used to calculate random errors in the stellar mass estimates. The shaded areas indicate the $1$, $2$, and $3\sigma$ ranges of the distribution, while the dotted line marks the $5\sigma$ of $\Phi_\text{eff}(\mathcal{E}_{\text{max| int}})$. Finally, the thin horizontal yellow line marks the maximum star formation efficiency allowed in $\Lambda$CDM, $\epsilon = 1$.

Our results show that, as the survey area increases, the median $\mathcal{E}_{\text{max|obs}}$ approaches the maximum efficiency, $\epsilon \sim 1$, particularly at higher redshift, $z \gtrsim 7$. Since this is the median of the distribution, we expect that, at least for a full-sky survey, $\gtrsim 50\%$ of halos at high redshift will appear to have efficiencies greater than $\epsilon = 1$. In contrast, when considering the median $\mathcal{E}_{\text{max| true}}$, after correcting for Eddington bias, the (yellow) curve remains well below the maximum efficiency. Furthermore, including the scatter in the stellar-to-halo mass relation, $\mathcal{E}_{\text{max|int}}$, places most of the distribution below $\epsilon_\text{max} = 1$, with a maximum of $\epsilon_\text{max}\sim0.3$ for the $5\sigma$ halos (dot-dashed line). 

As we noted earlier, the distribution inferred using Eq. (\ref{eq:SHMR_ratio_max}) assumes that $M_{\ast,\text{max}}$ and $M_\text{vir,max}$ are independent random variables. As discussed in Section \ref{sec:gal_halo_conn}, this assumption can yield unphysical configurations in which halos exhibit cumulative star formation efficiencies greater than unity, even after accounting for random errors and the scatter around the SHMR, particularly in the case of small survey areas and in the $\sim 4-5\sigma$ tail of the distribution, as can be seen by the left and middle panels. Thus, the above is a strong indication that the galaxy-halo connection in these massive halos is unlikely to be stochastic, rather than indicating truly unphysical galaxy formation efficiencies. As we will see below, the correlated distributions are more realistic models.  

\subsubsection{The case of correlated distributions}

As described in Section \ref{sec:gal_halo_conn}, a second model for correlating the most massive halo with the most massive galaxy is to rank-order their positions at a given redshift. In this approach, the most massive halo drawn from $\Phi_\text{vir}(M_\text{vir,max})$ corresponds to the most massive galaxy drawn from $\Phi_{\ast}(M_{\ast,\text{max}})$. For brevity, hereafter we denote this as the positively correlated scatter (PCS) case. For simplicity, we will denote this as $\mathcal{E}_\text{max,rank}$. 

The upper panels in Figure \ref{fig:evs_5sigma_density_ranking} present the results for the PCS case. For simplicity, we only plot the the results of the EVS distribution when accounting for the Eddington bias and the scatter around the SHMR in the GSMF, that is, when using the intrinsic GSMF. The resulting curves in this scenario indicate that unphysical configurations of halos with $\epsilon \gtrsim 1$ are unlikely, even in the smallest survey areas. We note, however, that the dispersion increases as the area decreases, whereas in the full-sky case the dispersion around the distribution nearly vanishes. Interestingly, the $5\sigma$ halos matched to the $5\sigma$ galaxies exhibit the lowest cumulative star formation efficiency at low redshifts ($z \lesssim 8$) across all three areas. At higher redshifts, they instead correspond to the highest efficiencies, except in the case of the largest area.

Accounting for the uncertainty caused by the various methods used to calculate random errors in the stellar mass estimates (shaded region), we find that the EVS distribution for the $10 \ \text{arcmin}^2$ field predicts that massive halos have efficiencies ranging from $\epsilon_\text{max}\sim0.1$ at $z \sim 0$, peaking at $z \sim 5$ with $\epsilon_\text{max}\sim0.2$–$0.4$, and declining to $\epsilon_\text{max}\sim(0.4$–$1)\times10^{-2}$ by $z \sim 16$. Similar values are obtained for the $90 \ \text{arcmin}^2$ area, though at lower redshift the efficiency is slightly smaller than in the $10 \ \text{arcmin}^2$ case, while at intermediate and higher redshifts it is slightly larger. The main difference lies in the broader uncertainty region around the galaxy–halo connection for the $90 \ \text{arcmin}^2$ case. This is expected, since larger areas allow the EVS to sample more massive galaxies, thus probing galaxies in the exponential tail of the GSMF, where the effect of random errors is stronger, see section \ref{sec:random_errors}.

When examining the full-sky case, the efficiencies at $z \sim 0$ are even lower, with values of $\epsilon_\text{max}\sim10^{-2}$, an order of magnitude smaller than in the $10 \ \text{arcmin}^2$ area. This result is expected, since larger areas sample more massive halos that lie on the declining part of the stellar-to-halo mass ratio. At higher redshifts, the uncertainty caused by the various methods used to calculate random errors in the stellar mass estimates (gray shaded region) increases. Consequently, no clear peak in the cumulative star formation efficiency is observed at intermediate redshifts, unlike in the smaller-area cases. The uncertainties in the efficiencies can be as large as $\sim1.4$ dex, or a factor of $\sim25$, for the most massive halos at $z\sim15$. 

A third model for linking the distribution of the most massive halos with that of the most massive galaxies is to associate the most massive halo drawn from $\Phi_\text{vir}(M_\text{vir,max})$ with the least massive galaxy drawn from $\Phi_\ast(M_{\ast,\text{max}})$. This scenario is illustrated in the bottom panels of Figure \ref{fig:evs_5sigma_density_ranking} and denoted as $\mathcal{E}_\text{inv-rank}$. Although we do not examine it in detail, this model is almost destined to fail by construction. The above is evident in the smallest survey areas where it yields unphysical cumulative star formation efficiencies.  

The analysis in this section demonstrates that a careful statistical treatment of both the Eddington bias and the scatter in the SHMR is essential, as is the treatment of the correlation between the distributions of the most massive halos and the most massive galaxies. Our results robustly show that, given the current data, the link between these two variables cannot be described as if they were independent. While the PCS model provides an attractive solution, there is no conclusive evidence that it is the true distribution; the truth may instead lie somewhere in between. In this paper, we adopt the PCS model as our fiducial case, acknowledging it as a plausible option, and further explore its implications in the sections below.

\begin{figure*}
    \centering
    \includegraphics[width=1.03\columnwidth]{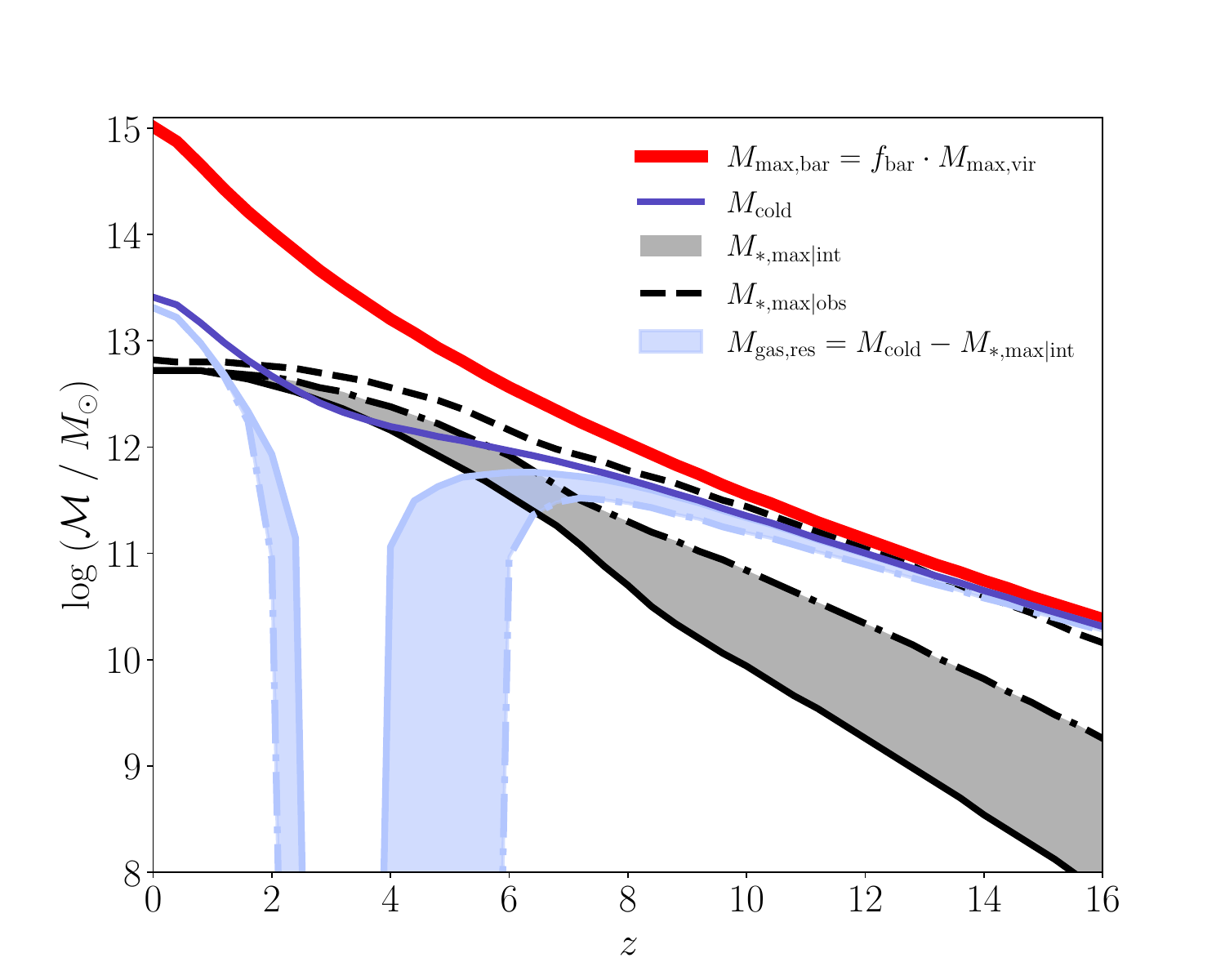}
    \includegraphics[width=1.03\columnwidth]{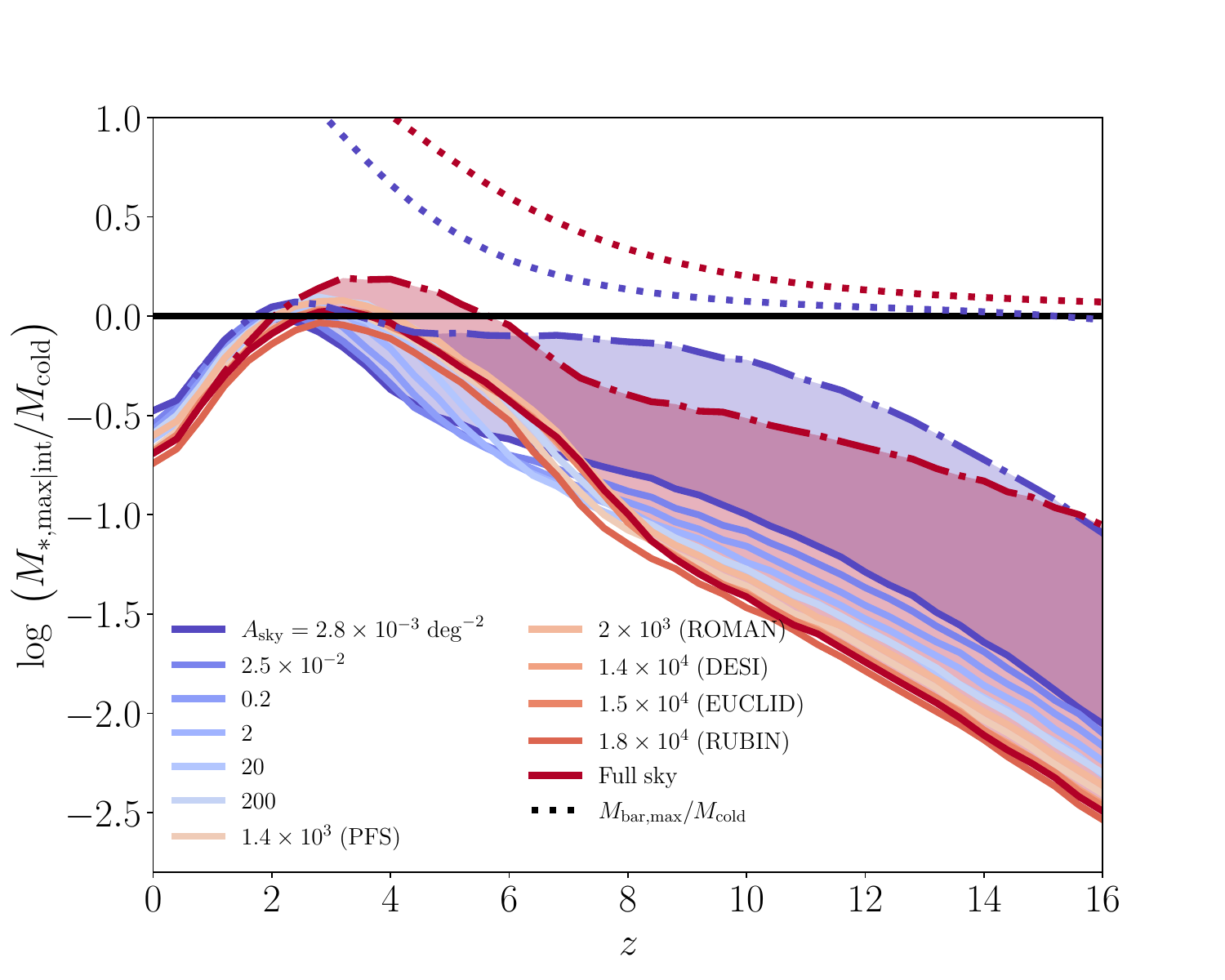}
    \caption{\textbf{Left Panel:} Baryonic content of the $5\sigma$ most massive halo in a full-sky survey. The red solid line shows the maximum baryonic mass, $f_\text{bar}M_\text{vir}$, while the violet solid line represents the total cold gas mass, composed of cooling gas and cold streams (see Section~\ref{sec:expected_cold_gas_mass}). The black dashed line corresponds to the EVS prediction based on the observed GSMF, $M_{\ast,\text{max|obs}}$. The EVS results based on the intrinsic GSMF are shown by the solid and dash–dotted lines enclosing the gray shaded region, which represents the uncertainty caused by the various methods used to calculate random errors in the stellar mass estimates. The light-violet line represents the expected residual cold gas mass, $M_\text{gas,res} = M_\text{cold} - M_{\ast,\text{max|int}}$, in the absence of feedback (see text for details). \textbf{Right Panel:} Intrinsic cumulative star formation efficiency, $M_{\ast,\text{max|int}}/M_\text{cold}$, as a function of survey area and redshift. The shaded regions between the solid and dot-dashed lines are as in the left panel. We show results only for survey areas of $0.2\times10^{-2},\mathrm{deg}^2$ and for the full sky. At intermediate redshifts, $2\lesssim z \lesssim 6$, galaxies approach maximal star formation efficiency. 
    }
    \label{fig:evs_5sigma_baryons_and_stellar_mass}
\end{figure*}

\subsection{Comparing the expected baryon content in $\Lambda$CDM halos and the empirical stellar mass of their central galaxies} \label{sec:expected_baryons_and_stellar_mass}

In Section \ref{sec:EVS_HMF}, we presented the EVS prediction for the most massive halo expected from the $\Lambda$CDM HMF. For this halo, we estimated the corresponding cold gas content, i.e., the amount of gas potentially available for star formation in the absence of feedback processes. While this represents a theoretical upper limit for the stellar mass of the central galaxy, we now compare it to our empirical estimate derived from the intrinsic GSMF, obtained after deconvolving both random errors and the intrinsic scatter in the SHMR. Following the conclusions of the preceding section, we adopt the PCS scenario. All halos and galaxies presented in this section represent the $5\sigma$ of their corresponding EVS distributions. 

The left panel of Figure \ref{fig:evs_5sigma_baryons_and_stellar_mass}, shows with the red thick line the total baryonic halo mass, $M_\text{max,bar} = f_\text{bar}M_\text{max,vir}$, the purple line shows the baryonic mass available to form stars, $M_\text{cold} = M_\text{acc,cold}+M_\text{cooling}$, and the black dashed line corresponds to the empirical estimate of the most massive galaxy based on the EVS applied to the observed GSMF, $M_{\ast,\text{max|obs}}$. Our results reproduce previous claims, based on HST \citep{Steinhardt+2016,Mason+2023} and JWST observations \citep{Boylan-Kolchin_2023,Dekel+2023,Li+2024}, that some massive galaxies require converting nearly all available baryons into stars at $z \gtrsim 8$, in tension with $\Lambda$CDM expectations, as shown before.

The $5\sigma$ most massive galaxy computed from the observacional GSMF, $M_{\ast,\text{max|obs}}$, is $\sim0.5$ dex lower than $M_\text{cold}$ at $z \sim 0$. However, it approaches to $M_\text{cold}$ and even exceeds it by $z \gtrsim 2$ and it remains above $M_\text{cold}$ until 
$z \sim 14-16$ when the two now become comparable. We note, however, that below $z\lesssim9$ the stellar mass of the most massive galaxy is $M_{\ast,\text{max|obs}}<M_\text{max,bar}$, while at higher redshifts $M_{\ast,\text{max|obs}}\sim M_\text{max,bar}$.

The new conclusion drawn from Figure \ref{fig:evs_5sigma_baryons_and_stellar_mass} is that, when comparing to $M_{\ast,\text{max|obs}}$, dark matter halos would need to be extremely efficient at converting baryons into stars from the reservoir of $M_\text{cold}$ at $z \sim 2$ and unphysically efficient at higher redshifts $z\gtrsim3$. Allowing for stellar mass growth through mergers would be an interesting avenue to explain why $M_{\ast,\text{max|obs}}\gtrsim M_\text{cold}$, in addition of other scenarios that propose higher star formation efficiencies \citep[e.g.,][]{Dekel+2023}.

In contrast, the EVS result based on the intrinsic GSMF (i.e., after deconvolving random observational errors, Eddington bias, and scatter in the SHMR), show that the maximum stellar mass at 5$\sigma$ is \textit{lower and below to} $M_\text{cold}$ across most of the redshift range (gray shaded area surrounded by the solid and dot-dashed lines in Figure \ref{fig:evs_5sigma_baryons_and_stellar_mass}). The gap between $M_{\ast,\text{max|int}}$ and $M_\text{cold}$ increases with redshift from $z \gtrsim 5$, implying lower cumulative star formation efficiencies, $M_{\ast,\text{max|int}}/M_\text{cold}$, at higher redshifts. Given the uncertainty in stellar masses, which also propagate to the galaxy-halo connection, at higher redshifts these efficiencies can be as large as $M_{\ast,\text{max|int}}/M_\text{cold}\sim 0.4- 1$ at $z\sim6$ and as small as $M_{\ast,\text{int}}/M_\text{cold}\sim(4-130)\times10^{-3}$ at $z\sim16$. At $2 \lesssim z \lesssim 6$, $M_{\ast,\text{max|int}}$ approaches, and in some cases even becomes slightly larger than, $M_\text{cold}$, coinciding with the rise of submillimeter and dusty highly star-forming galaxies \citep{Dunlop+2017,Michalowski+2017}. This result implies more star formation than cold gas available at this redshift range. 
At lower redshifts, $z\lesssim2$ the differences become larger again and by $z\sim0$ the maximum efficiency is $M_{\ast,\text{max|int}}/M_\text{cold}\sim 0.1$.

The difference $M_\text{cold}-M_{\ast,\text{max|int}}$ represents the gas that either remains in the galaxy or has been heated/expelled by feedback processes, which we denote as $M_\text{gas,res}$. In Figure \ref{fig:evs_5sigma_baryons_and_stellar_mass}, this residual gas mass is shown by the light-violet shaded region enclosed by the solid and dot-dashed lines of the same color. 
As the figure illustrates, $M_\text{gas,res}$ generally tracks $M_\text{cold}$, except in the range $2 \lesssim z \lesssim 6$, where galaxies become more efficient and $M_\text{gas,res}\sim0$. Except for that redshift range, this indicates that, overall, the cumulative efficiency of galaxy formation has remained relatively low, consistent with efficiencies observed at lower redshifts.

While the results in the redshift range $2 \lesssim z \lesssim 6$ may represent some tension with the $\Lambda$CDM model, we refer the reader to Figure \ref{fig:phimax_differences_between_authors}, which illustrates that differences among HMFs can lead to different estimates of the most massive halo. In particular, the HMF of \citet{Sheth+1999} predicts halo masses that are $\sim0.05-0.1$ dex more massive at $2 \lesssim z \lesssim 6$ 
than those inferred using the HMF of \citet{Despali+2016}. Adopting the \citet{Sheth+1999} HMF would potentially alleviate some of the tension discussed above. However, notice that \citet{Despali+2016} is already an improvement over \citet{Sheth+1999}. Therefore, we do not compare to the latter HMF.

Galaxy mergers may help alleviate the aforementioned tension at $2 \lesssim z \lesssim 6$. Because mergers contribute stellar mass formed ex-situ, directly comparing the total stellar mass with $M_\text{cold}$ may not be appropriate. In addition, mergers may increase the cold gas reservoir, particularly at high redshift when wet mergers are more common \citep{Stewart+2009b,Lagos+2018b}. For example, \citet{Lagos+2018b} showed that the frequency of wet mergers (both minor and major) increases in the \textsc{EAGLE} simulation \citep{Crain+2015} for $z\gtrsim0.5$, see their Figure 2. Likewise, \citet{Li+2023} found that the wet-merger fraction rises from $\sim34\%$ at $z\sim0$ to $\sim96\%$ by $z\sim3$, consistent with the higher cold-gas fractions of galaxies at early times predicted by semi-analytic models \citep{Guo+2013b}.

The right panel of Figure \ref{fig:evs_5sigma_baryons_and_stellar_mass} shows the \textit{intrinsic} maximum cumulative star formation efficiency, $\epsilon_{\rm max} = M_{\ast,\text{max|int}}/M_\text{cold}$ as a function of survey area and redshift, accounting for the fact that baryons within halos reside in multiple phases. As before, the efficiency is computed for the $5\sigma$ halos and galaxies under the PCS model. To assess how uncertainties caused by the various methods used to calculate random errors in the stellar mass estimates affect the inferred star formation efficiency, we include shaded regions corresponding to the survey areas of $0.2\times10^{-2} \text{ deg}^{-2}$ and the full sky cases.

Overall, the maximum star formation efficiency remains below $\epsilon_{\rm max} = 1$, with values as low as $\sim3 \times 10^{-2}$. Our results indicate a pronounced peak in star formation efficiency, with the most massive galaxies reaching maximum efficiency in the narrow redshift range $2 \lesssim z \lesssim 6$, with some regions exhibiting slightly higher values, suggesting that star formation operates near maximal efficiency within the uncertainties of the SHMR. This is, again, consistent with the epoch of peak activity in submillimeter/dusty star-forming galaxies. 

Finally, we note that the dotted lines show the ratio $M_{\mathrm{max,bar}}/M_{\mathrm{cold}}$, which is $\gtrsim 1$ at all redshifts. Moreover, this ratio is always larger than $\epsilon$, i.e., $M_{\mathrm{max,bar}}/M_{\mathrm{cold}} > \epsilon$. This provides an alternative way to show that, while galaxies can approach maximal efficiency in converting their available cold gas into stars, they are less efficient at transforming the total baryonic content available within their dark matter halos, even at the range between $2\lesssim z \lesssim 6$ when galaxies are very efficient.

\section{Discussion} \label{sec:discussion}

In this paper, we employ the EVS framework \citep{Harrison+2011,Harrison+2012,Lovell+2012} to estimate empirical probabilistic limits on the most massive galaxies in a given volume and over cosmic time by using the observed GSMF \citep{Rodriguez-Puebla2024,Rodriguez-Puebla+2025}. We compare these limits to those of the most massive dark matter halos and their corresponding maximum baryonic mass allowed by the $\Lambda$CDM model. In particular, we explore seven different HMFs commonly used in the literature, adopting the model of \citet{Despali+2016} as our reference. Overall, we find that once random errors in observed stellar masses are properly accounted for, the previously claimed tension with the $\Lambda$CDM model (namely, that inferred stellar masses approach or even exceed the baryonic mass available within dark matter halos, see e.g., \citealp{Labbe+2023,Boylan-Kolchin_2023}) is largely alleviated. This is because the \citet{Eddington1913,Eddington1940} bias artificially inflates the high-mass end of the observed GSMF. We find, however, that when dark matter halos are treated as multiphase systems, containing both hot and cold gas, and only the cold gas component is allowed to form stars in the galaxy center, the most massive galaxies at $2\lesssim z\lesssim 6$ are in reality the most efficient, reaching stellar masses of $M_{\ast,\text{max|int}}\gtrsim M_\text{cold}$.

This section focuses on discussing our results in the context of previous approaches to understanding massive galaxies in both the pre- and post-JWST eras. We also explore the physical conditions of the baryons in the most massive halos, the impact of cosmological parameters, and the predictions for future surveys.

\subsection{Comparison with other other works}

Several recent studies have investigated whether the existence of very massive galaxies and black holes at high redshifts poses a fundamental challenge to the $\Lambda$CDM paradigm.

Early work by \cite{Behroozi-Silk2018} calculated maximum allowed stellar and black hole masses by combining the cosmic baryon fraction with the cumulative halo mass function, showing that at $z>4$ galaxy number densities approach the limits predicted by $\Lambda$CDM and emphasizing the role of baryonic efficiency ceilings and observational systematics. Building on this line of work, \cite{Lovell+2023} employed the EVS formalism to model the most massive halos expected in $\Lambda$CDM and inferred corresponding stellar masses using conservative baryon-to-star conversion factors, finding that several early JWST candidates at $z>8$ lie beyond the $3\sigma$ expectation, although uncertainties in stellar mass and redshift estimates could mitigate the apparent tension. More recently, \cite{Boylan-Kolchin_2023} derived hard limits on stellar mass and stellar mass density from baryonic content arguments, reporting that some JWST-inferred stellar mass densities at $z>10$ appear to exceed the maximum allowed under $\Lambda$CDM.

In contrast to these approaches, which rely on theoretical halo mass functions and assumptions about baryon-to-star conversion efficiencies or more theoretical motivated stellar-to-halo mass mapping, our analysis adopts a fully empirical strategy by working directly with the observed GSMFs. This allows us to test the consistency of massive galaxies with $\Lambda$CDM predictions without invoking model-dependent efficiency prescriptions, providing a more observationally grounded and statistically driven assessment of the high-mass regime.

An interesting consideration is the effect of cosmic variance, which may influence predictions for the most massive galaxies in a given survey volume, particularly in small survey areas. For example, \citet[][see also \citealp{Jespersen+2025}]{Jespersen+2025A} explored the impact of combining cosmic variance with the EVS framework and found that the expected maximum stellar mass can shift by up to $\sim0.5$ dex depending on the underlying large-scale environment. While incorporating cosmic variance explicitly into our EVS framework would be an interesting extension, such an analysis is beyond the scope of the present work.

In an alternative direction, \cite{Shen+2024} proposed resolving the apparent tension through modifications to the cosmological model itself, showing that the inclusion of an Early Dark Energy (EDE) component can enhance the formation of massive halos at high redshift and improve agreement with JWST observations \citep[see also][]{Klypin+2021}. In contrast, our work remains fully within the standard $\Lambda$CDM framework, seeking to explain the observations through empirical reassessment of mass functions, uncertainties, and galaxy formation in massive halos rather than invoking new physics. This approach is further validated by \citet{Prada+2026}, who use the Uchuu-UM framework to show that $\Lambda$CDM naturally reproduces JWST observations by allowing star formation efficiency to increase with redshift (reaching $\sim3.5\%$ at $z=13$). Their findings also support our conclusion that "impossible" galaxy claims likely stem from systematic uncertainties (e.g. as AGN contamination and a lack of MIRI data) rather than a failure of the standard cosmological model. Additional support comes from \citet{Chaikin2026}, who demonstrated that the \textsc{\large colibre} simulations reproduce the observed GSMFs at $z\lesssim 12$ and the abundance of massive quiescent galaxies at $2\lesssim z\lesssim 7$ once the Eddington bias is incorporated into the simulated masses, despite the star formation efficiencies remaining within $\approx 16$ per cent.

Overall, while previous studies have variously highlighted potential tensions or proposed solutions involving cosmological modifications or specific astrophysical processes, our analysis provides a comprehensive and fully empirical evaluation of whether observed massive galaxies are consistent with $\Lambda$CDM expectations, without invoking particular scenarios or assuming non-standard physics.

\begin{figure*}
    \centering
    \includegraphics[width=\textwidth]{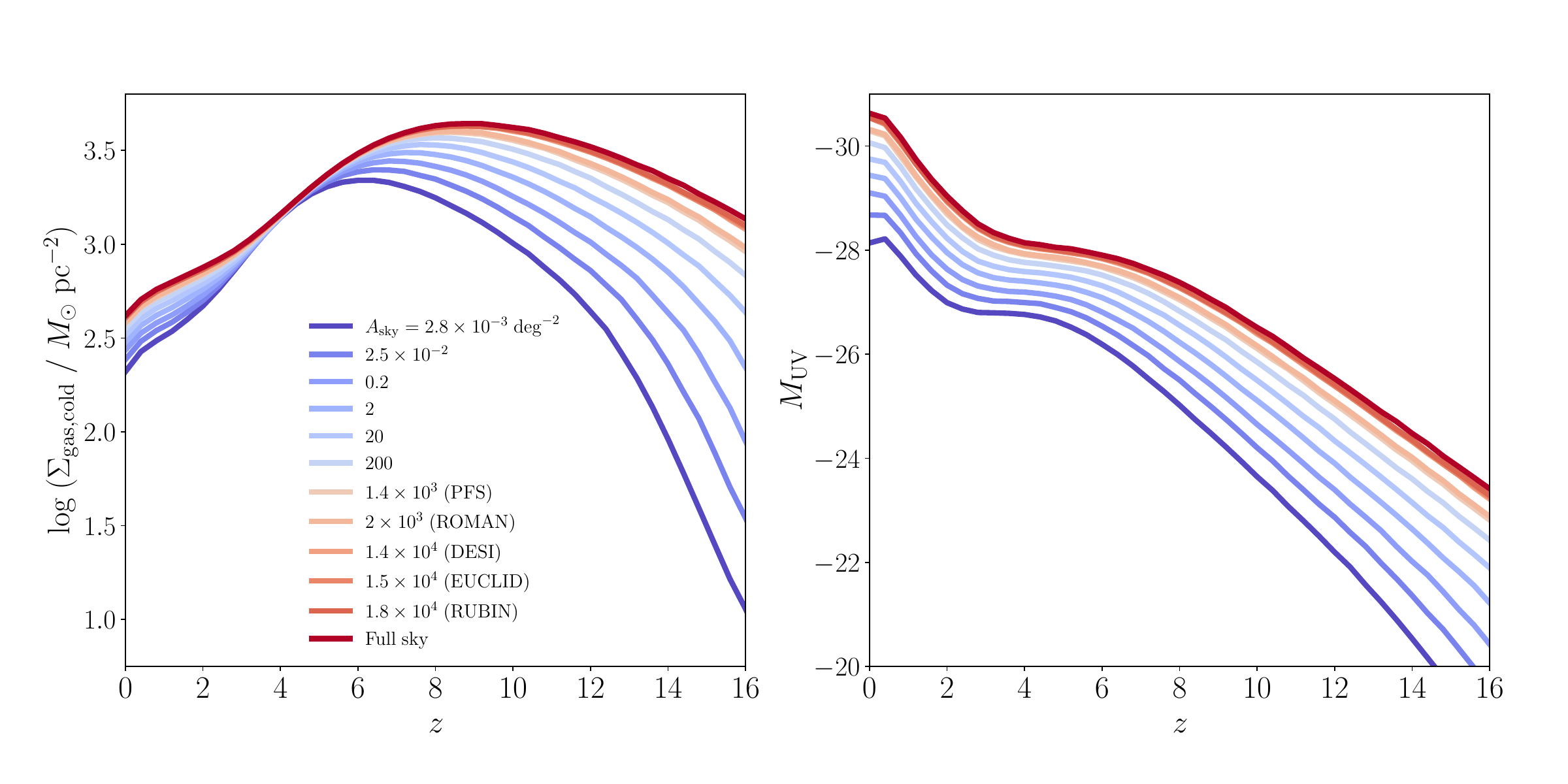}
\caption{ Physical conditions of the baryons in the most massive halo. \textbf{Left Panel:} Cold gas surface density, $\Sigma_{\mathrm{gas,cold}}$, in the most massive $\Lambda$CDM halo as a function of redshift for different survey areas.  \textbf{Right Panel:} Predicted UV luminosity of the galaxy hosted by the halo. See the text for details of the model.}
    \label{fig:evs_5sigma_density}
\end{figure*}


\begin{figure*}
    \centering
    \includegraphics[width=\textwidth]{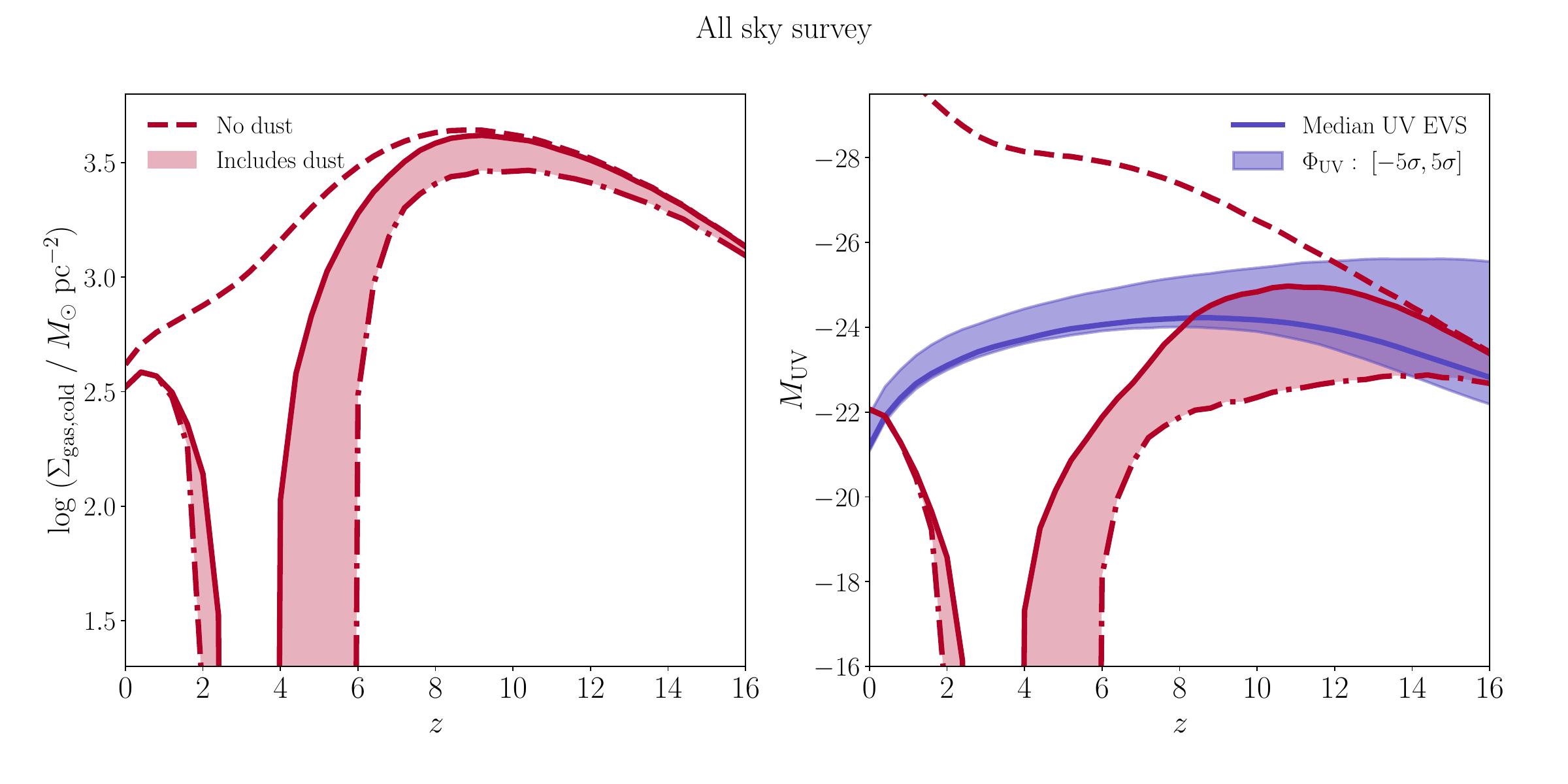}
    \includegraphics[width=\textwidth]{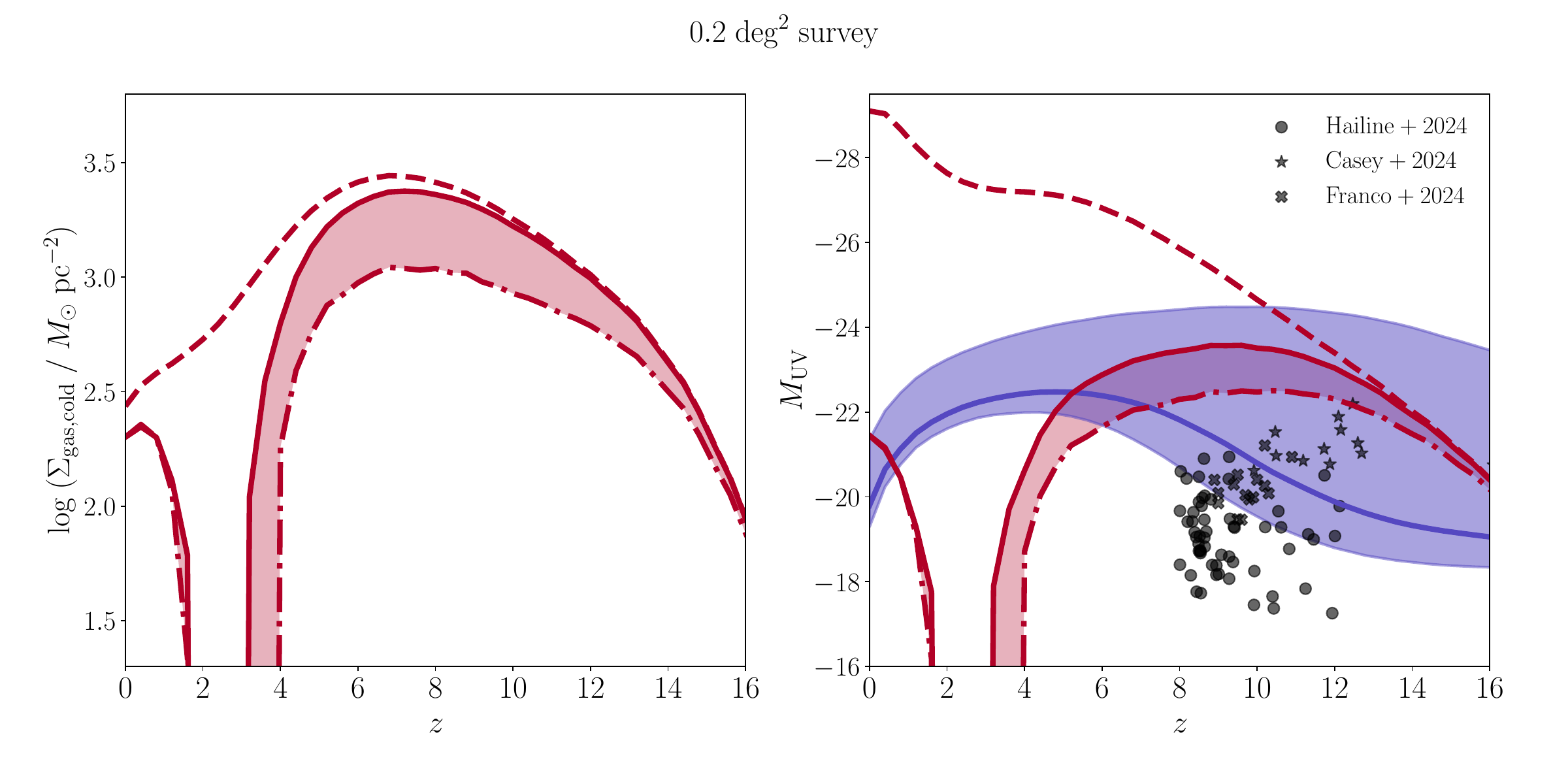}
 \caption{
Similar to Figure~\ref{fig:evs_5sigma_density}, but accounting for dust attenuation and the conversion of baryons into stars. Results are shown for a full-sky survey and for a $0.2,\mathrm{deg}^2$ field. We compare our model predictions with the intrinsic EVS distribution of UV luminosities derived from the LFs and with observations from \citet{Casey+2024,Franco+2024} and \citet{Hainline+2024}. 
}
    \label{fig:evs_5sigma_density_empirical_constrains}
\end{figure*}


\subsection{What are the physical conditions of the baryons in the most massive halos?}
\label{sec:physical_cond}

An important question that we can address in this paper, using the framework developed here, is the physical conditions of the baryons in the most massive halos.

In Section \ref{sec:expected_cold_gas_mass}, we estimated the total cold gas mass expected to be available for star formation in the absence of any form of feedback. Since this gas is already in place to potentially form stars, we now explore a complementary question: how much star formation could this gas produce?\footnote{Our estimate of the cold gas mass should be regarded as a conservative approximation. In realistic halos, additional physical processes may increase the amount of cold gas available for star formation. For instance, precipitation-driven cooling in the circumgalactic medium can allow cold gas to condense from radii larger than the classical cooling radius, thereby enhancing the gas supply reaching the central galaxy \citep{Voit2026}.} To address this, we begin by estimating the gas surface density, given by
\begin{equation}
    \Sigma_\text{gas,cold} = \frac{M_\text{cold}}{\pi R_\text{gas}^2},
    \label{eq:gas_density}
\end{equation}
where $R_\text{gas}$ is the 2D projected radius. In practice, here we use a 3D radius to estimate the area in Equation (\ref{eq:gas_density}), that is we interchange $R_\text{gas}$ by $r_\text{gas}$. We note, however, that projected sizes can be $\sim33\%$ larger than their 3D counterparts in the case of disks \citep[e.g.,][]{Behroozi+2022,Somerville+2025}. Although such a correction should be taken into account for more precise studies, our approximation is sufficient for the purposes of this analysis, as it does not affect the qualitative conclusions.
Finally, as a rough approximation, we use the cooling radius computed in Section \ref{sec:expected_cold_gas_mass} as a proxy for the radius of the cold gas $r_\text{gas} = r_\text{cool}$, with $r_\text{cool} \lesssim 0.1 \ r_\text{vir}$ \citep{Hou+2019}.
\footnote{Our definition is a factor of $\lesssim 3$–$5$ larger than that reported in \citet{Shibuya+2015} and \citet{Somerville2017} for low redshifts ($z\lesssim7$), and about a factor of $\lesssim1.6$ larger than that in \citet{Somerville+2025} for $z\sim10$. We emphasize that here $r_\text{gas}$ refers to the total gas radius, not a scale length or half-light radius.}

The left panel of Figure \ref{fig:evs_5sigma_density} shows the gas surface density of the most massive dark matter halos for different survey areas as a function of redshift. For all survey areas, these halos exhibit gas surface densities comparable to those observed in starburst galaxies \citep[$\Sigma_\text{gas}\sim 10^2 - 10^3 \text{ M}_{\odot} \text{ pc}^{-2}$,  see, e.g.,][]{Leroy+2015,Miura+2018,Kennicutt_de_los_Reyes2021}, and display a monotonic dependence of gas surface density on survey area at fixed redshift. At a fixed survey area, the gas surface density increases with redshift, reaches a maximum, and then declines. For smaller survey areas, gas surface densities decrease at high redshift ($z\gtrsim 12$) to values typical of local, normal star-forming galaxies ($\sim$ a few tenths of $\text{ M}_{\odot} \text{ pc}^{-2}$). In contrast, larger survey areas maintain high gas surface densities, comparable to those of starburst galaxies ($\sim 10^3\text{ M}_{\odot} \text{ pc}^{-2}$), and approach the critical surface density at which molecular clouds are expected to be disrupted once the momentum injected per unit stellar mass reaches $\Sigma_\text{crit}\sim2200 \text{ M}_{\odot} \ \text{ pc}^{-2}$ \citep[see e.g. the review by][see also \citealp{Thompson_Krumholz2016}]{Chevance+2023}. We additionally find that smaller survey areas peak earlier, at $z\sim6$, whereas the full-sky case peaks later, at $z\sim9$. Such extreme gas surface densities are expected at these high redshifts \citep{Dekel+2023,Somerville+2025}.

\subsubsection{The simplest model for deriving UV magnitudes}

We next infer the SFR from our model using the Kennicutt–Schmidt law combined between normal star-forming galaxies with starburst galaxies \citep{Kennicutt_de_los_Reyes2021}.
When using the best-fit relation between SFR density and total gas surface density for the combined sample of normal galaxies and starburst from \citet{Kennicutt_de_los_Reyes2021}, we can estimate the total SFR due to that gas and converted it into rest-frame UV magnitudes, using the conversion from \citet{Kennicutt+1998} and converted to a \citet{Chabrier2003} initial mass function. We notice that while the above is a popular method to related UV magnitudes to SFRs, in reality this transformation depends on the metallicity and the star formation history of the galaxy. Here we omit the above but we refer to the reader to \citet{Madau_Dickinson2014}. Additionally, for the moment we omit the fraction of UV light which can be obscured in the presence of dust \citep[see e.g.,][]{Whitaker+2017b,Rodriguez-Puebla+2020a}.

The right panel of Figure \ref{fig:evs_5sigma_density} shows the rest-frame UV magnitudes for the most massive halos as a function of the sky aperture. As a result of sustained starburst activity across all redshifts and for most of the survey areas, specially larger areas (left panel of the same Figure), the halo remains exceptionally bright in UV light. Particulary when $f_\text{sky}=1$, at high redshifts, $z \gtrsim 8$, it produces very bright galaxies with UV magnitudes between $M_\text{UV} \sim -27$ and $M_\text{UV}\sim-23$. Even for the smallest survey areas we observe bright galaxies with magnitudes  between $M_\text{UV} \sim -24$ and $M_\text{UV}\sim-20$ at the same redshift range. At intermediate redshifts, there is a plateau around at $z\sim2-7$ with $M_\text{UV}$, the length of the plateau and the UV magnitude depends on the survey area, corresponding roughly to the transition from cold accretion to hot-mode accretion. When the hot mode becomes dominant at $z \lesssim 4$ for larger areas and at $z\sim2$ for smaller areas, cooling flows become the primary source of gas in the central regions. Consequently, the UV luminosity keeps rising sharply, in some cases increasing by $\sim 3$ magnitudes at low $z$, reaching $M_\text{UV} \sim -30.5$ by $z \sim 0$ for the whole sky. Such luminosities are unrealistic compared to the EVS predictions from the UV LF, shown in Figure \ref{fig:evs_muv}. In that figure, the maximum luminosity of galaxies was $M_\text{UV} \sim -25$ at $z \sim 5.5$, well below the values found here.

\subsubsection{Deriving UV magnitudes in the presence of dust and stars}

A natural next step toward modeling more realistic luminosities is to account for the fact that a fraction of the cold gas mass, $M_\text{cold}$, has already been converted into stars as estimated from our empirical EVS, $M_{\ast,\text{max|int}}$. The residual gas mass, $M_\text{gas,res}=M_\text{cold}-M_{\ast,\text{max|int}}$, represents the gas that remains in the galaxy or has been heated/expelled by feedback processes. Another important step is to include the effects of dust attenuation.

As discussed above, the cold gas mass expected to be available for star formation in the absence of feedback should be corrected for the stars that have already formed. In other words, we update the upper limit for the available gas budget by subtracting the stellar mass of long-lived stars, $M_\ast$, yielding
\begin{equation}
\Sigma_\text{gas,cold} = \frac{M_\text{gas,res}}{\pi r_\text{gas}^2}.
\end{equation}

When converting gas density or star formation efficiency into an SFR, we account for the fact that not all star formation is traced by UV light. A fraction of the emission is dust-obscured and re-emitted in the infrared. To correct for this, we adopt the relation reported in \citet{Whitaker+2017b}, expressed as a function of stellar mass and independent of redshift, which has been shown to be a good approximation \citep[see also,][]{Rodriguez-Puebla+2020a,Nava-Moreno+2024}. 

Figure \ref{fig:evs_5sigma_density_empirical_constrains} presents the same information as Figure \ref{fig:evs_5sigma_density}, but focuses on two survey areas: the full sky and a $0.2 \text{ deg}^2$ field, the latter comparable to the COSMOS-Web survey. For this area, we compare our results with UV-selected samples from \citet{Casey+2024}, \citet{Franco+2024}, and \citet{Hainline+2024}. In both the upper and lower panels, the shaded regions enclosed by the solid and dot–dashed lines show the results obtained using $M_\text{gas,res}$ and dust. This area reflects the uncertainty in the derived stellar masses. 
For comparison, the dashed line reproduce the model based on $M_\text{cold}$, as in Figure \ref{fig:evs_5sigma_density}, which neglect the fact that some of the gas has already been converted into stars. 

We recall that the largest differences between $M_\text{gas,res}$ and $M_\text{cold}$ arise at $2 \lesssim z \lesssim 6$, when galaxies have converted most of their baryons into stars (see the left panel of Figure \ref{fig:evs_5sigma_baryons_and_stellar_mass}). In this redshift range, the reduced cold-gas supply, $\Sigma_\text{gas,cold}\sim0$ drives the instantaneous star formation efficiency to values $ \text{SFE}\sim 0$, resulting in negligible UV luminosities. Below $z\sim2$, gas cooling becomes the dominant source of cold gas, causing the gas surface density, SFE, and UV luminosity to rise again, although remaining lower than the estimates that ignore dust attenuation and are based on $M_\text{cold}$, dashed line. 

The violet shaded regions in the right panels represent the empirical full distribution of the brightest galaxy identified in the EVS analysis of the UV LF, $\Phi_\text{UV}$, as discussed in Section \ref{sec:UV_brightest}. The solid violet lines are the median of the UV EVS distributions. We use the full and the median of the distribution to avoid the discussion if there is a monotonic trend between the brightest galaxy and the most massive halo. 
When comparing our model predictions with the empirical results inferred from the observed $\Phi_\text{UV}$, we find a field-area–dependent disagreement: larger survey areas show stronger tension than smaller ones. In the full-sky case, the empirical results indicate that galaxies are brighter than predicted by the model at $z \gtrsim 8$.
The largest tension between the model and the empirical constraints occurs at $2 \lesssim z \lesssim 6$, as noted already above. For the smallest survey area, the disagreement is more modest
except again in the redshift range $2 \lesssim z \lesssim 6$. 
The results shown in this section indicate that models can predict gas densities and UV magnitudes much lower than those that seem to be required by the empirical results based on the observations. 

\subsection{Impact of cosmological paramters}
\label{sec:impact_cosmo_params}

The results from the \citet{Planck+2015} not only represent the state-of-the-art but also a highly precise determination of cosmological parameters based on observations of the CMB \citep[but see,][ for a recent determination of the cosmological paraemters]{DESI2025}, see Table \ref{tab:cosmology_planck}. These measurements combine temperature, polarization, and gravitational lensing data to constrain key parameters of the $\Lambda$CDM model.  

The Planck analysis, when combined with Baryon Acoustic Oscillations (BAO) and Type Ia Supernovae (SNe Ia) finds values for the Hubble constant $H_0$, matter density $\Omega_m$, and dark energy density $\Omega_\Lambda$  which are accurate at the level of less than $\sim1-2$ percent, rightmost column of Table \
\ref{tab:cosmology_planck}. The data also provide strong constraints on the scalar spectral index $n_s$, which characterizes primordial fluctuations, and the matter fluctuation amplitude $\sigma_8$, which is crucial for structure formation and therefore the HMF. 

The estimated values of these cosmological parameters, along with their corresponding uncertainties derived using Markov Chain Monte Carlo (MCMC) methods, are summarized in Table~\ref{tab:cosmology_planck}. These uncertainties reflect the statistical constraints from Planck CMB data combined with BAO and SNe, ensuring a robust estimation framework.

\begin{table}
    \centering
    \begin{tabular}{lcc}
        \hline
        Parameter & Estimated Value & Percentual error [$\%$] \\
        \hline
        $H_0$ (km s$^{-1}$ Mpc$^{-1}$) & $67.4 \pm 0.5$ & $0.74$ \\
        $\Omega_m$ & $0.315 \pm 0.007$ & 2.22 \\
        $\Omega_\Lambda$ & $0.685 \pm 0.007$ & 1.02 \\
        $\Omega_b$ & $0.0493 \pm 0.0003$ & 0.609 \\
        $n_s$ & $0.965 \pm 0.004$ & 0.423 \\
        $\sigma_8$ & $0.811 \pm 0.006$ & 0.740 \\
        \hline
    \end{tabular}
    \caption{Estimated cosmological parameters from BAO + SNe \citep{Planck+2015}.}
    \label{tab:cosmology_planck}
\end{table}

\begin{figure*}
    \centering
    \includegraphics[width=\textwidth]
    {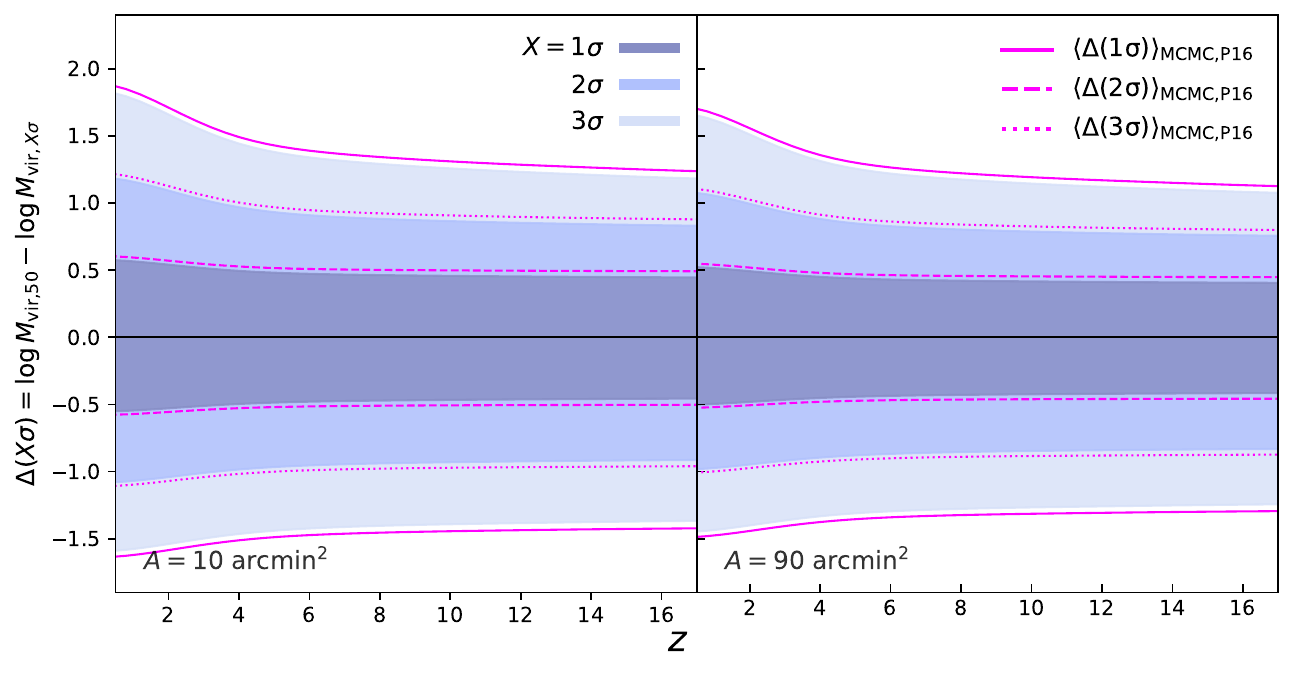}
\caption{Impact of cosmological parameter variations on the EVS predictions of the most massive halos. The shaded regions represent the differences with respect to the median prediction within the fiducial $\Lambda$CDM cosmology, showing the spread corresponding to the $1\sigma$, $2\sigma$, and $3\sigma$ levels. The magenta lines show the same quantity, but computed using the full MCMC from the \citet{Planck+2015}, i.e., computing the median for the distribution of the $1\sigma$, $2\sigma$, and $3\sigma$. Results are shown for survey areas of 10 and 90~arcmin$^2$. 
}
    
    \label{fig:phimax_mhalo_planck}
\end{figure*}

In Figure \ref{fig:phimax_mhalo_planck}, the shaded regions show the difference between the median of the EVS distribution and its $1\sigma$, $2\sigma$, and $3\sigma$ intervals, computed using the fiducial cosmological parameters from \citet{Planck+2015} and the \citet{Despali+2016} HMF. Results are shown for two of the smallest survey areas considered, 10 and 90 arcmin$^2$. We define this difference as $\Delta(X\sigma) = \log M_{\mathrm{vir},50} - \log M_{\mathrm{vir},X\sigma}$, where $X = 1, 2, 3$.

The magenta lines present the impact of cosmological parameter uncertainties on the predicted masses of the most massive halos within the EVS framework. To compute this, we evaluate the EVS distribution for each element in the publicly available \citet{Planck+2015} MCMC chains, using the corresponding cosmological parameters. We then compute $\Delta(X\sigma)$ for each sample and plot the mean of the resulting distribution using different line styles.

Overall the effect of varying cosmological parameters on the EVS predictions is negligible at all redshifts. 
This implies that variations in cosmological parameters alone are unlikely to explain potential tensions between theoretical predictions and observations of high-redshift massive galaxies.

\begin{figure*}
    \centering
\includegraphics[width=1\linewidth]{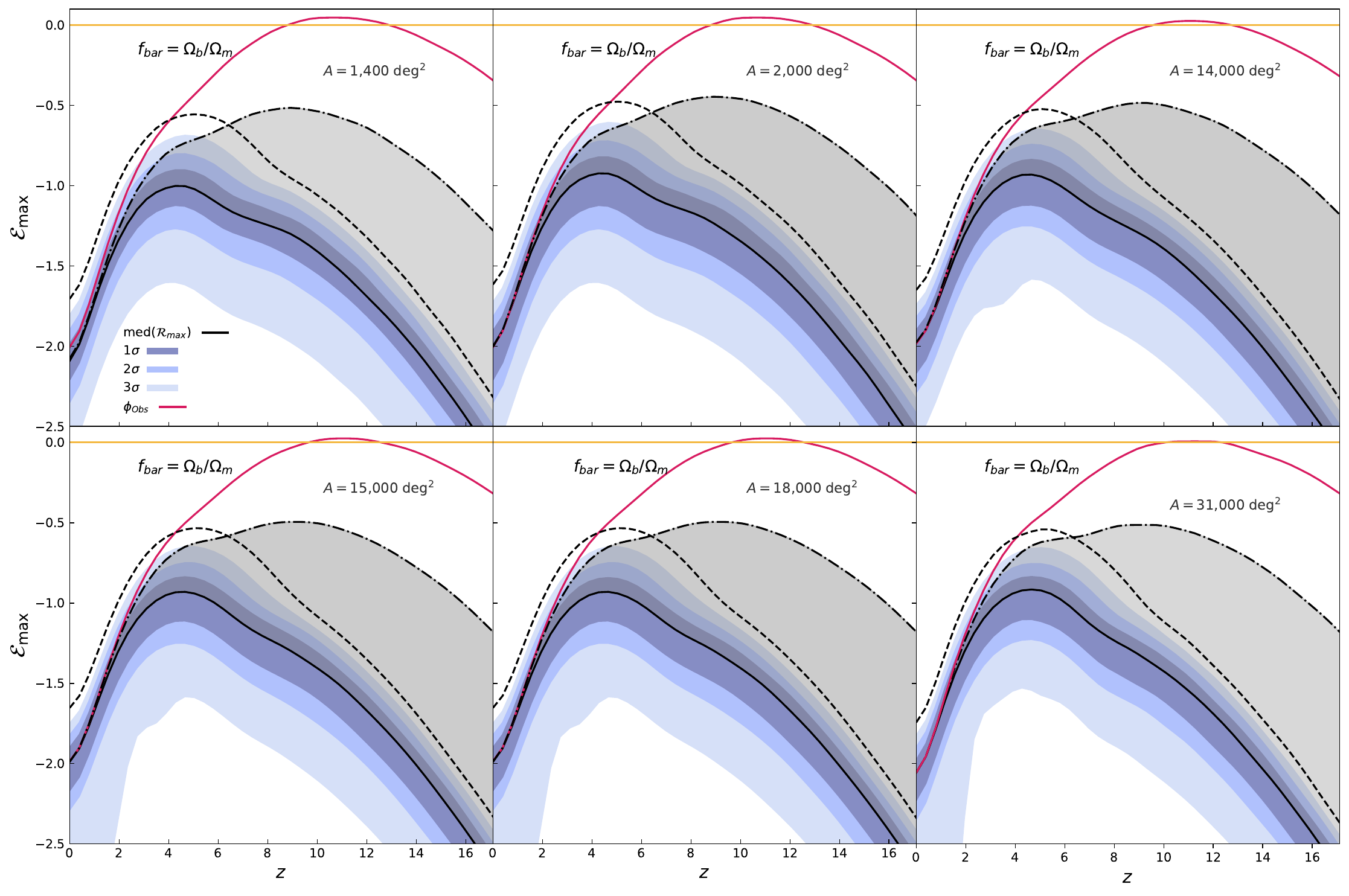}
        \caption{Comparison of the convolved distributions of halo and galaxy mass functions across the areas of different forthcoming surveys, including PFS, Roman Space Telescope, DESI, Euclid, and the Rubin Observatory, see Table \ref{tab:surveys}. For the definition of $\varepsilon_{\max}$, see Fig.~\ref{fig:conv_R}. }
    \label{fig:surveys_5}
\end{figure*}


\subsection{Predictions for future surveys}

The study of galaxy mass functions at high redshift is entering an unprecedented era, driven by upcoming large-scale surveys that will provide deeper constraints on the connection between dark matter halos and galaxy formation. These surveys will leverage the convolution of halo and galaxy mass functions using Extreme Value Statistics (EVS), allowing for a more rigorous assessment of the consistency between observed high-redshift massive galaxies and standard cosmological models.

Figure \ref{fig:surveys_5} presents the predicted EVS distributions of $\mathcal{E}_{\max} \equiv$log$[M_{\ast,\max}/(f_{\mathrm{bar}} M_{\mathrm{vir},\max})]$ across the areas of multiple upcoming surveys, including the Rubin Observatory, Roman Space Telescope, DESI, Euclid, and LSST. Each panel in the figure corresponds to a different survey, showcasing how the convolution of the  GSMF (the intrinsic one for the black solid line and blue-ish shaded regions) with the HMF varies depending on the surveyed area and depth. The figure highlights the impact of survey size on the expected detection of rare, massive galaxies. \footnote{We note that in Figure~\ref{fig:surveys_5} we present the case in which $M_{\ast,\text{max}}$ and $M_\text{vir,max}$ are treated as independent variables. This assumption is adopted to establish a conservative limit on the expected distribution in wide-area surveys.}

A key takeaway from Figure \ref{fig:surveys_5} is that larger survey areas significantly increase the probability of detecting the most extreme galaxies, as they provide a more comprehensive sampling of the tail of the mass distribution. Surveys with wider fields of view, such as LSST ($18,000\ deg²$) and Euclid ($15,000\ deg²$), will be particularly effective at detecting rare, massive galaxies that push the limits of the $\Lambda$CDM framework.
\begin{table*}
\caption{Summary of current and forthcoming Dark Energy experiments. In the Probes column, WL stands for weak lensing, LSS for large-scale structure, and SNIa for Type Ia supernovae.}
\begin{tabular}{|l|l|l|l|l|l|l}
\hline
Name & Probes & Survey Area (deg$^2$) & $\lambda_\text{range}$ & Telescope & Start Date & Duration \\
\hline
PFS    & LSS                & 1,400 & 0.38--1.26 $\mu$m  & 8.2m  & Mid-2024                & 100 nights \\
Roman  & WL \& LSS          & 2,000 & 0.48--2.3 $\mu$m   & 2.4m  & Sept. 2026 (launch)     & 5 years \\
DESI   & LSS                & 14,000 & 0.36--0.98 $\mu$m & 4m    & May 2021                & 5 years \\
Euclid & WL \& LSS          & 15,000 & 0.55--2.0 $\mu$m  & 1.2m  & Feb. 2024               & 5 years \\
Rubin  & WL \& SNIa         & 18,000 & 0.32--1.05 $\mu$m & 8.4m  & End 2025 (first light)  & 10 years \\
\hline
\end{tabular}
\label{tab:surveys}
\end{table*}


\section{Summary and Conclusions}
\label{sec:summ_and_concl}

In this paper, we explore empirical limits on the most massive and brightest galaxies as a function of survey area and cosmic time using EVS, inferred from the observed GSMF and UV luminosity function. We define the most massive galaxy (and, analogously, the most massive dark matter halo) as the corresponding $5\sigma$ value of the EVS distribution, as an objective definition of the extreme population. We account for random errors in stellar mass measurements and include the \citet{Eddington1913,Eddington1940} bias to derive EVS-based probabilistic limits for the intrinsic GSMF. We further consider an additional source of Eddington bias arising from scatter in the SHMR and derive corresponding EVS limits for galaxies expected to reside in the most massive dark matter halos. Within the framework of $\Lambda$CDM, we compare these limits with those predicted for the most massive dark matter halos using state-of-the-art HMFs, as well as with the theoretical maximum cold gas mass available for star formation. Our key results are as follows:

\begin{itemize}
    \item We have established empirical estimates of the EVS for the most massive galaxies as a function of redshift and survey area (Figure~\ref{fig:evs_5sigma_gal}; Eq.~\ref{eq:model_most_massive}). For the full sky at $z\approx0$, the expected observed $5\sigma$ galaxy, $M_{\ast,\mathrm{max|obs}}$, has a stellar mass of $M_\ast \sim 7\times10^{12} M_\odot$, while at $z\sim8$ and $z\sim16$ the expected mass decreases to $\sim 3\times10^{11} M_\odot$ and $\sim10^{10} M_\odot$, respectively. These empirical estimates are broadly consistent with recent JWST observations (Figure~\ref{fig:phiOTI}).

    \item When accounting for random errors in stellar mass estimates in the GSMF 
    (i.e., the Eddington bias) the expected most massive galaxy, denoted $M_{\ast,\mathrm{max|true}}$, is similar to the observed estimate, $M_{\ast,\mathrm{max|obs}}$, at low redshift. However, the discrepancy increases toward higher redshift, reaching a conservative difference of $\Delta_\ast = \log M_{\ast,\mathrm{max|obs}} - \log M_{\ast,\mathrm{max|true}} \sim 0.1$ dex at $z\sim5$ and nearly $\sim1$ dex at $z\sim16$ (Figure~\ref{fig:evs_5sigma_gal}). If random errors are larger, differences of up to $\sim2$ dex are possible.
    Therefore, the true masses of the most massive galaxies are smaller than they appear, especially at higher redshifts. 

    \item When accounting the intrinsic scatter around the SHMR, in addition of the random errors, the expected most massive galaxy, denoted $M_{\ast,\mathrm{max|int}}$, is systematically smaller even at low redshift, with differences of $\Delta_\ast = \log M_{\ast,\mathrm{max|obs}} - \log M_{\ast,\mathrm{max|int}} \sim 0.07$ dex at $z\sim0$ and  $\sim0.3$ dex at $z\sim5$ and nearly $\sim1.2$ dex at $z\sim16$. If random errors are larger than our conservative estimate, the discrepancy may exceed $\sim2.5$ dex at $z\sim16$. Therefore, the intrinsic maximum stellar masses permitted in the $\Lambda$CDM cosmology are significantly smaller than they appear, increasing the differences with redshift.

    \item We also provide empirical EVS predictions for the most luminous galaxies derived from the observed UV luminosity function as a function of redshift and survey area (Figure~\ref{fig:evs_muv}). These predictions are consistent with recent JWST observations (Figure~\ref{fig:evs_muv_JWST}).

\item Our results highlight that survey area is not merely an observational detail, but a fundamental parameter in determining the expected extremes of galaxy populations. The maximum stellar mass inferred from EVS scales strongly with the surveyed volume, implying that discrepancies between observations and $\Lambda$CDM predictions cannot be meaningfully assessed without explicitly accounting for survey area. In particular, small area surveys are intrinsically biased toward lower extremes, and thus cannot probe the full tail of the underlying distribution. This reinforces the need for wide-field surveys to place robust constraints on the most massive galaxies and their consistency with the $\Lambda$CDM framework.

    \item We derive theoretical EVS predictions for the most massive dark matter halos using seven state-of-the-art halo mass functions. The differences among these models are modest, reaching $\sim0.1$ dex at low redshift and up to $\sim0.2$ dex at higher redshift (Figure~\ref{fig:phimax_differences_between_authors}).

    \item We compare the EVS distribution for the most massive dark matter halos with several extreme galaxy clusters reported in the literature and find them to be consistent with the predictions of $\Lambda$CDM. We therefore conclude that if galaxies appear to be in tension with $\Lambda$CDM, this tension is unlikely to originate from the cosmological model itself, but rather from the still-uncertain processes that regulate the conversion of baryons into stars.

    \item Comparing our empirical EVS predictions for the most massive galaxies with the theoretical maximum baryonic mass available in halos, $M_{\mathrm{bar,max}} = f_{\mathrm{bar}} M_{\mathrm{vir,max}}$, we recover the previously reported tension with $\Lambda$CDM (e.g., \citep{Labbe+2023,Boylan-Kolchin_2023}), in which some observed galaxies appear more efficient than the theoretical baryonic limit. However, when accounting for random errors and scatter in the SHMR, this tension is significantly alleviated. This suggests that much of the apparent discrepancy between JWST observations and $\Lambda$CDM expectations may arise from Eddington bias artificially boosting the abundance of massive galaxies (Figure~\ref{fig:phimax_halo_vs_gals}). 

    \item Assuming a perfect rank-order correspondence between the most massive halos and the most massive galaxies, we find no galaxies with efficiencies exceeding the theoretical maximum baryonic mass, $M_{\mathrm{bar,max}} = f_{\mathrm{bar}} M_{\mathrm{vir,max}}$, when random stellar-mass errors are included, both in conservative and extreme error scenarios. If this perfect correspondence is relaxed, a residual tension with $\Lambda$CDM halos remains (Figure~\ref{fig:evs_5sigma_density_ranking}). 
\end{itemize}

Our results based on the rank-order correlation between the most massive halo and the most massive galaxy further show that:
\begin{itemize}
    \item When estimating the total cold gas mass potentially available for star formation ($M_{\mathrm{cold}}$), arising from cold and cooling flows in the absence of feedback (Section~\ref{sec:expected_cold_gas_mass}), we find that in the redshift range $2 \lesssim z \lesssim 6$ the most massive galaxies have stellar masses comparable to this reservoir ($M_{\ast,\mathrm{max|int}} \sim M_{\mathrm{cold}}$). This suggests that halos may reach near-maximal star-formation efficiency during this epoch, consistent with the emergence of submillimeter galaxies. At other redshifts, galaxies remain below this limit (Figure~\ref{fig:evs_5sigma_baryons_and_stellar_mass}). 

    \item Using the EVS distribution of the most massive halos, we infer the physical conditions of baryons within these systems. If all cooling and cold gas were available as a cold reservoir for star formation, and neglecting the effects of dust, the most massive halos would exhibit gas surface densities comparable to those observed in starburst galaxies at all redshifts and would host the brightest UV galaxies, particularly over larger survey areas (Figure~\ref{fig:evs_5sigma_density_empirical_constrains}). 

    \item When accounting for dust attenuation and for the fraction of baryons already converted into stars, based on the properties of the most massive galaxies, we infer that at high redshifts ($z \gtrsim 6$) the most massive halos host galaxies undergoing a starburst phase, characterized by high molecular-cloud star-formation efficiency and producing the brightest UV galaxies. At $2 \lesssim z \lesssim 6$, however, these halos become significantly less efficient, as a large fraction of their gas reservoir has already been converted into stars.
    
    \item We compare the modeled UV luminosities with observations of the brightest UV galaxies. We find that the model reproduces the observed luminosities at high redshift ($z \gtrsim 6$) for small survey areas. However, for larger survey areas the model shows some tension with the observed brightest galaxies, which are too bright.

\end{itemize}

\section*{Acknowledgements}
MEV and AM are supported by SECIHTI postdoctoral grants. ARP and VAR acknowledge financial support from DGAPA-PAPIIT grants IN106924 and IN106823, and CONAHCyT grant `Ciencia de Frontera' G-543. AY is supported by a Giacconi Fellowship from the Space Telescope Science Institute, which is operated by the Association of Universities for Research in Astronomy, Incorporated, under NASA contract HST NAS5-26555 and JWST NAS5-03127. ARP is grateful for inspiring conversations with Joel Primack over the last decade. 

\section*{Data Availability}
The data underlying this article will be shared on reasonable request to the corresponding author.



\bibliographystyle{mnras}
\bibliography{example} 




\appendix

\begin{figure*}
    \centering
    \includegraphics[width=\textwidth]{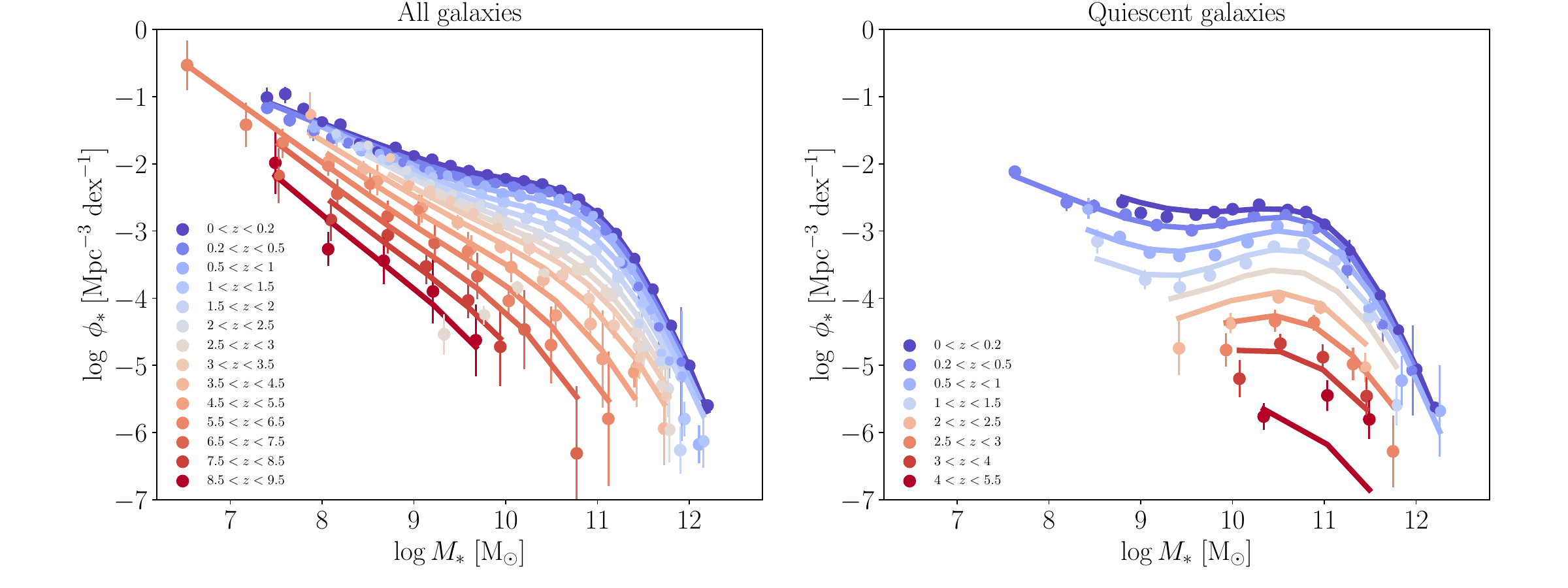}
\caption{Parametric model of the GSMF adopted in this work, based on the compilation of Rodríguez-Puebla et al. (in prep.). The intrinsic GSMF is decomposed into star-forming and quiescent components, modeled with double and triple generalized Schechter functions (left and right panels).
}
    \label{fig:gsmf_fits}
\end{figure*}


\section{The Galaxy Stellar Mass Function}
\label{sec:GSMF_fits}

In this section, we describe the GSMFs adopted in this work and their corresponding best-fit models. A detailed description of the GSMF construction will be presented in Rodríguez-Puebla et al. (in prep; see also \citealp{Rodriguez-Puebla2024,Rodriguez-Puebla+2025}). Briefly, here we followed the methodology described by \citet[][see their Section 4.1]{Rodriguez-Puebla+2017}, homogenizing the data and combining multiple observational surveys.

We assume that the intrinsic GSMF is given by the sum of the intrinsic GSMFs of star-forming and quiescent galaxies:
\begin{equation}
    \phi_{\ast,\text{intr}}(M_\ast) = \phi_{\ast,\text{intr|SF}}(M_\ast) + \phi_{\ast,\text{intr|Q}}(M_\ast),
\end{equation}
where each component is modeled as a sum of generalized Schechter functions,
\begin{equation}
    \phi_\text{sch}(M_\ast, \vec{p}) = 
    \frac{\phi^\ast}{\log e} 
    \left( \frac{M_{\ast}}{\mathcal{M}_c} \right)^{\alpha+1} 
    \exp\left[-\left( \frac{M_\ast}{{\mathcal{M}_c} }\right)^\beta\right]
\end{equation}
and $\vec{p} = (\phi^\ast,\alpha, \beta, \mathcal{M}_c)$ denotes the set of free parameters. For star-forming galaxies, we assume a double generalized Schechter function::
\begin{equation}
    \phi_{\ast,\text{intr|SF}}(M_\ast, \vec{p}_\text{SF}) = \sum_{i\leq 2}  \phi_{\text{sch},i}(M_\ast, \vec{p}_{i,\text{SF}}).
\end{equation}
whereas for quiescent galaxies we adopt a triple generalized Schechter function:
\begin{equation}
    \phi_{\ast,\text{intr|Q}}(M_\ast, \vec{p}_\text{Q}) = \sum_{i\leq 3}  \phi_{\text{sch},i}(M_\ast, \vec{p}_{i,\text{Q}}).
\end{equation}
The observed GSMF is then obtained by convolving the intrinsic GSMF with a Gaussian scatter (see Eq. \ref{eq:convo_observed_GSMF}):
\begin{equation}
    \phi_{\ast,\text{obs}}(M_\ast) = \int \mathcal{G}(x-\log M_\ast) 
     \phi_{\ast,\text{intr}}(x) 
    dx.
\end{equation}

This parametrization is motivated by the recent results of \citet{Vazquez-Mata+2025}, who showed that the local $z=0$ GSMFs of early- and late-type galaxies require multiple Schechter components. In particular, late-type galaxies are well described by a double Schechter function, with one component dominated by Scd–Irr galaxies and the other by Sc–Sa galaxies. For early-type galaxies, approximately two Schechter components are required to reproduce their GSMF.

For the first component of the star-forming galaxies is given:
\begin{equation}
    \phi_{\text{sch|SF},1} = \phi_{\text{sch-SF},1}(M_\ast,z|\alpha_1,\beta,\phi^\ast_1,\mathcal{M}_c),
\end{equation}
where $\phi^\ast_1$ is measured in units of Mpc$^{-3}$ dex$^{-1}$ and $\mathcal{M}c$ is in $M_\odot$. The parameters depend on redshift $z$ as follows:
\begin{equation}
     \begin{array}{l}
        \log  \phi^\ast_1(z) = \mathcal{Z}(\phi_{1,0},\phi_{1,1},\phi_{1,2},\phi_{1,3},z), \\
        \alpha_1 = \mathcal{Z}(\alpha_{1,0},\alpha_{1,1},0,,z), \\
        \beta_1 = \beta, \\
        \log \mathcal{M}_c = \mathcal{Z}(\mathcal{M}_{1,0}, \mathcal{M}_{1,1}, \mathcal{M}_{1,2}, \mathcal{M}_{1,3}, z),
     \end{array}
\end{equation}
where 
\begin{equation}
    \mathcal{Z}(p_0,p_1,p_2,p_3,z) = p_0 + p_1(1-a) + p_2 \log a  + p_3 z,
\end{equation}
and $a=1/(1+z)$ is the scale factor.

The second component of the star-forming GSMF isgiven by:
\begin{equation}
    \phi_{\text{sch|SF},2} = \phi_{\text{sch|SF},2}(M_\ast,z|\alpha_2,\beta,\phi^\ast_2,\mathcal{M}_c),
\end{equation}
and the dependence on redshift is given by:
\begin{equation}
	   \begin{array}{l}
		      \log  \phi^\ast_2(z) = \log \phi^\ast_1(z) + \phi_{0} \\
		      \alpha_2(z) = \alpha_1(z) + 1 ,\\
	   \end{array}
\end{equation}

For quiescent galaxies, the first component is given by:
\begin{equation}
    \phi_{\text{sch|Q},1} = \phi_{\text{sch|Q},1}(M_\ast,z|\alpha_3,\beta,\phi^\ast_3,\mathcal{M}_c),
    \label{eq:GSMF_Q_comp1}
\end{equation}
and the dependence on redshift is given by:
\begin{equation}
	   \begin{array}{l}
		      \log  \phi^\ast_3(z) = \log \phi^\ast_4(z) + \mathcal{Z}(\phi_{3,0},-2,0,0,z) \\
		      \alpha_3(z) = \alpha_1(z) ,\\
	   \end{array}
\end{equation}
Here, $\phi^\ast_3(z)$ depends on $\phi^\ast_4(z)$, which will be defined next. 

The second component for quiescent galaxies is given by:
\begin{equation}
    \phi_{\text{sch|Q},2} = \phi_{\text{sch|Q},2}(M_\ast,z|\alpha_4,\beta,\phi^\ast_4,\mathcal{M}_{c,4}),
    \label{eq:GSMF_Q_comp2}
\end{equation}
and the dependence on redshift is given by:
\begin{equation}
	   \begin{array}{l}
		      \log  \phi^\ast_4(z) = \mathcal{Z}(\phi_{4,0},\phi_{4,1},\phi_{4,2},\phi_{4,3},z) \\
		      \alpha_4(z) = \alpha_1(z) + 1 + (1 - a),\\
            \log \mathcal{M}_{c,4} = \log \mathcal{M}_{c}(z) + \mathcal{Z}(\mathcal{M}_{4,0},\mathcal{M}_{4,1},0,0,z),
	   \end{array}
\end{equation}

Finally, the third component for quiescent galaxies is given by:
\begin{equation}
    \phi_{\text{sch|Q},3} = \phi_{\text{sch|Q},3}(M_\ast,z|\alpha_4,\beta_5,\phi^\ast_5,\mathcal{M}_{c,4}),
    \label{eq:GSMF_Q_comp3}
\end{equation}
and the dependence on redshift is given by:
\begin{equation}
	   \begin{array}{l}
		      \log  \phi^\ast_5(z) = \log  \phi^\ast_4(z) +  \phi^\ast_{5,0}\\
		      \beta_5(z) = \beta + \beta_{5,0}.\\
	   \end{array}
\end{equation}

To derive our best-fitting models we compiled several GSMF from the literature. This compilation represents an update to the GSMFs presented in \cite{Rodriguez-Puebla+2017} including recent data from the James Webb Space Telescope, JWST \citep[such as][]{Navarro-Carrera+2024,Weibel+2024} and COSMOS2020 \citep{Weaver+2023}, as will be discuss in Rodríguez-Puebla et al. (in prep.).

Below we present the best-fitting models for the two prescriptions of random errors described in Section~\ref{sec:random_errors}.

Using the parametrization of $\sigma_{\ast,\text{ran}}$ from \citet[][solid line in Figure~\ref{fig:ran_ms}]{Rodriguez-Puebla+2025}, we obtain the following best-fitting parameters:
\begin{equation}
     \begin{array}{l}
        \log  \phi^\ast_1(z) = \mathcal{Z}(-3.13,1.01,2.96,0.03,z), \\
        \alpha_1 = \mathcal{Z}(-1.51,-0.10,0,-0.07,z), \\
        \beta_1 = 0.98, \\
        \log \mathcal{M}_c = \mathcal{Z}(10.43,-0.94,-3.60,-0.44, z),\\
		\log  \phi^\ast_2(z) = \log \phi^\ast_1(z) + 0.53 \\
		\log  \phi^\ast_3(z) = \log \phi^\ast_4(z) + \mathcal{Z}(-0.79,-2,0,0,z) \\
		\log  \phi^\ast_4(z) = \mathcal{Z}(-2.69,-0.02,-1.60,-0.88,z) \\
        \log \mathcal{M}_{c,4} = \log \mathcal{M}_{c}(z) + \mathcal{Z}(0.43,-0.76,0,0,z), \\
        \log  \phi^\ast_5(z) = \log  \phi^\ast_4(z) - 0.75\\
		\beta_5(z) = \beta -0.33.
     \end{array}
     \label{eq:best_fit_ran_RP25}
\end{equation}
For the second model of random errors, dot-dashed line in Figure \ref{fig:ran_ms}, we obtain: 
\begin{equation}
     \begin{array}{l}
        \log  \phi^\ast_1(z) = \mathcal{Z}(-3.06,0.63,1.89,-0.15,z), \\
        \alpha_1 = \mathcal{Z}(-1.46,-0.16,0,-0.07,z), \\
        \beta_1 = 1.01, \\
        \log \mathcal{M}_c = \mathcal{Z}(10.44,-0.87,-3.38,-0.34, z),\\
		\log  \phi^\ast_2(z) = \log \phi^\ast_1(z) + 0.42 \\
		\log  \phi^\ast_3(z) = \log \phi^\ast_4(z) + \mathcal{Z}(-0.79,-2,0,0,z) \\
		\log  \phi^\ast_4(z) = \mathcal{Z}(-2.64,-0.14,-1.98,-0.95,z) \\
        \log \mathcal{M}_{c,4} = \log \mathcal{M}_{c}(z) + \mathcal{Z}(0.38,-0.80,0,0,z), \\
        \log  \phi^\ast_5(z) = \log  \phi^\ast_4(z) - 0.89\\
		\beta_5(z) = \beta -0.35.
     \end{array}
    \label{eq:best_fit_ran_tp}
\end{equation}

Figure \ref{fig:gsmf_fits} shows the observed GSMF for the full galaxy population and for quiescent galaxies, based on the compilation presented in Rodríguez-Puebla et al. (in prep.), together with the best-fitting model from Equation (\ref{eq:best_fit_ran_RP25}). Similar figure will be obtained using the best-fit parameters in Eq. (\ref{eq:best_fit_ran_tp}) instead.

\section{The UV Luminosity Function}
\label{sec:UV_LF}

\begin{figure*}
    \centering
    \includegraphics[width=\textwidth]{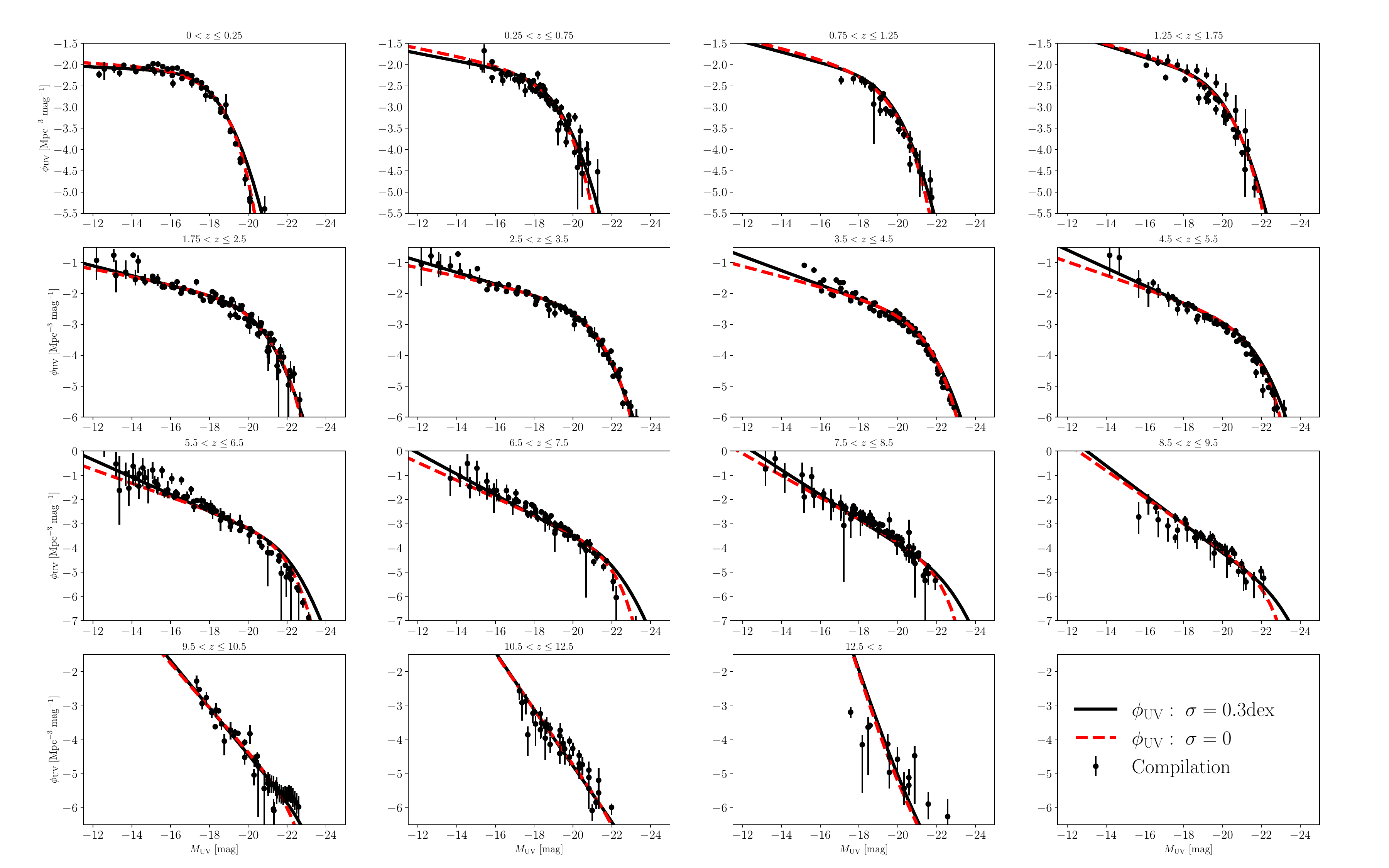}
\caption{UV luminosity functions compiled from the literature at different redshifts. Each panel corresponds to a different redshift interval. The black solid line shows the intrinsic UV luminosity function, while the red dashed line shows the observed UV luminosity function after including the scatter in the $L_{\rm UV}-M_{\rm vir}$ relation.
}
    \label{fig:UVLF_fits}
\end{figure*}


In this section, we briefly describe the UV LFs used in this work and the best-fit models adopted to estimate the EVS distribution of the brightest galaxies as a function of survey area and redshift.

As in Section~\ref{sec:gal_halo_conn}, we assume the existence of a correlation between UV luminosity and halo mass, with an associated intrinsic scatter. We model the scatter in the $L_\text{UV}$–$M_\text{vir}$ relation as lognormal, independent of redshift, with a constant dispersion of 0.3 dex. This assumption is motivated by the fact that $L_\text{UV}$ is proportional to the star formation rate (SFR), so the above relation effectively corresponds to a relation between SFR and halo mass. For the star-forming main sequence, the observed scatter is $\sim 0.25$–$0.3$ dex \citep{Speagle+2014}. We therefore assume a comparable level of scatter.

Under these assumptions, we write an expression analogous to Eq.~(\ref{eq:intrinsci_GSMF}):
\begin{equation}
    \phi_\text{UV,obs}(L_\text{UV}) = \int \mathcal{H}_\text{UV}(x-\log L_\text{UV}) \phi_\text{UV,intr}(x)dx.
\end{equation}

We assume that $\phi_\text{UV,intr}$ is well described by a modify Schechter function:
\begin{equation}
    \begin{aligned}
    \phi_\text{UV}(M_\text{UV}) = \frac{2\phi^\ast}{5 \log e} &
    10^{-0.4\left( M^{\ast} -M_\text{UV}\right)(1+\alpha)} \\
    &\exp\left( - 10^{-0.4\left( M^{\ast} -M_\text{UV}\right)\beta}\right),
    \end{aligned}
\end{equation}
where $M_\text{UV}$ is the absolute magnitude corresponding to $L_\text{UV}$, $\alpha$ is the faint-end slope of the UV LF, $M^\ast$ is the characteristic magnitude, $\phi^{\ast}$ is the normalization, and $\beta$ controls the steepness of the exponential cutoff at the bright end.

To model the redshift evolution of the UV LF, we assume that each parameter evolves with redshift according to:\begin{equation}
    \begin{aligned}
    \log \phi^{\ast} = & \mathcal{Z}(\phi_0,\phi_1,\phi_2, \phi_3,z), \\
    \alpha = &\mathcal{Z}(\alpha_0,\alpha_1,\alpha_2,\alpha_3,z), \\
    M^\ast = &\mathcal{Z}(M_0,M_1,M_2,M_3,z),    \\
    \beta = &\mathcal{Z}(\beta_0,\beta_1,\beta_2,\beta_3,z). 
    \end{aligned}
\end{equation}

To constrain the best-fitting models of the UV LFs, we include 21 studies of the rest-frame UV LFs over the redshift range $0 \leq z \leq 9$ compiled in Table 1 of \citet{Rodriguez-Puebla+2020a}, in addition to the following works: \citet{Bouwens+2022} ($1.5 \lesssim z \lesssim 9.5$); \citet{Finkelstein+2022} ($9 \lesssim z \lesssim 11$); \citet{Leung+2023} ($9 \lesssim z \lesssim 12$); \citet{Perez-Gonzalez+2023} ($8 \lesssim z \lesssim 13$); \citet{Adams+2024} ($7.5 \lesssim z \lesssim 13.5$); \citet{Casey+2024} ($9.5 \lesssim z \lesssim 12$); \citet{Donnan+2024} ($9 \lesssim z \lesssim 15$); \citet{Finkelstein+2024} ($9.5 \lesssim z \lesssim 15.5$); \citet{Robertson+2024} ($11.5 \lesssim z \lesssim 15$); \citet{Castellano+2025} ($15 \lesssim z \lesssim 20$); and \citet{Whitler+2025} ($8.5 \lesssim z \lesssim 16$). All the LFs were homogenized to the same $\Omega_\text{m}$ and $h$ values employed for this paper, see Table \ref{tab:cosmology_planck}.

Figure~\ref{fig:UVLF_fits} shows the best-fitting models obtained assuming a dispersion of $\sigma_\text{UV}=0.3$ dex (black solid line) and $\sigma_\text{UV}=0$ (i.e., no scatter, red dashed). For the case $\sigma_\text{UV}=0$, the best-fitting using the MCMC approach described in \citet{Rodriguez-Puebla+2013} parameters are:
\begin{equation}
    \begin{aligned}
    \log \phi^{\ast} = & \mathcal{Z}(-1.94,-4.84,-10.98,-1.08,z), \\
    \alpha = &\mathcal{Z}(-0.89,-3.27,-5.65,-0.47,z), \\
    M^\ast = &\mathcal{Z}(-17.33,-2.40,0.46,-0.21,z),    \\
    \beta = &\mathcal{Z}(0.89,0.54,1.98,0.21,z), 
    \end{aligned}
\end{equation}
and, for the case $\sigma_\text{UV}=0.3$ dex, the corresponding best-fitting parameters are: 
\begin{equation}
    \begin{aligned}
    \log \phi^{\ast} = & \mathcal{Z}(-1.96,-5.19,-11.17, -1.12,z), \\
    \alpha = &\mathcal{Z}(-0.89,-2.48,-4.02,-0.37,z), \\
    M^\ast = &\mathcal{Z}(-17.31,-3.64,-0.62,-0.23,z),    \\
    \beta = &\mathcal{Z}(1.67,0.63,2.18,0.43,z). 
    \end{aligned}
\end{equation}

\onecolumn

\section{Functional form for the EVS models}
\label{sec:evs_fits}

In this section we fit the EVS predictions using smooth parametric functions that capture the redshift evolution of the $5\sigma$ maximum masses. Eq \ref{eq:model_most_massive_5s} describes the evolution of the maximum halo mass predicted from the $\Lambda$CDM halo mass function, while Eq. \ref{eq:model_most_massive} provides the corresponding description for the maximum stellar mass predicted from the observed galaxy stellar mass function.

The dependence of the model parameters on survey area is summarized in Tables \ref{tab:evs_model_fit} and \ref{tab:evs_model_fit_gal}. Table \ref{tab:evs_model_fit} lists the best-fitting parameters for the analytic model describing the $5\sigma$ maximum halo mass, $M_{\mathrm{vir,max}}(z,f_{\mathrm{sky}})$, while Table \ref{tab:evs_model_fit_gal} provides the corresponding fits for the maximum stellar mass predicted from the observed GSMF.

\begin{table*}
\centering
\caption{Functional forms and best-fit parameters (with $1\sigma$ uncertainties) for the components of the stellar mass EVS model at fixed $5\sigma$ detection threshold, as a function of $f_{\mathrm{sky}}$.}
\label{tab:evs_model_fit}

\begin{tabular}{ l l l }
\hline
\textbf{Parameter} & \textbf{Functional Form} & \textbf{Best-Fit Parameters} \\
\hline \\
$A(f_{\mathrm{sky}})$ &
$\log_{10} A = A_0 - \frac{1}{n} \log_{10} \left[ \left( \frac{f_{\mathrm{sky}}}{f_t} \right)^{\alpha_1 n} + \left( \frac{f_{\mathrm{sky}}}{f_t} \right)^{\alpha_2 n} \right]$ &
$A_0 = 15.341 \pm 0.094$, $\alpha_1 = -0.102 \pm 0.032$, $\alpha_2 = -0.248 \pm 0.061$ \\
& & $\log_{10} f_t = -3.472 \pm 1.053$, $n = 10.000 \pm 175.509$ \\
\hline \\
$\alpha(f_{\mathrm{sky}})$ &
$\alpha = \alpha(f_{\mathrm{sky}})$ &
$\alpha = -8.995 \pm 0.261$ \\
\hline \\ 
$\log_{10} a(f_{\mathrm{sky}})$ &
$\log_{10} a = \alpha + \beta \cdot \log_{10}(f_{\mathrm{sky}})$ &
$\alpha = -0.804 \pm 0.009$, $\beta = -0.037 \pm 0.002$ \\
\hline \\
$n(f_{\mathrm{sky}})$ &
$n = n(f_{\mathrm{sky}})$ &
$n= 0.097 \pm 0.012$ \\
\end{tabular}
\end{table*}

\begin{table*}
\centering
\caption{Functional forms and best-fit parameters (with $1\sigma$ uncertainties) for the EVS galaxy mass model components as a function of $f_{\mathrm{sky}}$.}
\label{tab:evs_model_fit_gal}

\begin{tabular}{l l l}
\hline
\textbf{Quantity} & \textbf{Functional Form} & \textbf{Best-Fit Parameters} \\
\hline \\

$A(f_{\mathrm{sky}})$ & 
$A = 10^{A_0} \left[ \left( \dfrac{f_{\mathrm{sky}}}{f_t} \right)^{\alpha_1 n_A} + \left( \dfrac{f_{\mathrm{sky}}}{f_t} \right)^{\alpha_2 n_A} \right]^{-1/n_A}$ & 
$A_0 = 21.223 \pm 0.439$, $\alpha_1 = 4.893 \pm 0.561$ \\
& & $\alpha_2 = -1.683 \pm 0.151$, $\log_{10} f_t = 3.065 \pm 0.127$ \\
& & $n_A = 18.801 \pm 0.404$ \\
\hline \\


$n(f_{\mathrm{sky}})$ & 
$n(f_{\mathrm{sky}}) = \dfrac{n_{\infty}}{1 + \left( \dfrac{f_t}{f_{\mathrm{sky}}} \right)^m} + n_{\mathrm{floor}}$ & 
$n_{\infty} = 3.109 \times 10^{-2} \pm 2.631 \times 10^{-3}$, $m = 2.855 \pm 0.682$ \\
& & $f_t = 1.880 \times 10^{4} \pm 1.711 \times 10^{3}$ \\ \\ 
& & $n_{\mathrm{floor}} = 1.01 \times 10^{-4} \pm 2.57 \times 10^{-4}$ \\
\hline \\

$\alpha_1(f_{\mathrm{sky}})$ & 
$\alpha_1(f_{\mathrm{sky}}) = C_1 \cdot \exp\left( -\dfrac{ \left( \log_{10}(f_{\mathrm{sky}}) - \mu_1 \right)^2 }{2\sigma_1^2} \right) + C_{\mathrm{offset}}$ &
$C_1 = 1.000 \times 10^{19} \pm 0$, $\mu_1 = 160.186 \pm 36.784$ \\
& & $\sigma_1 = 18.809 \pm 4.456$, $C_{\mathrm{offset}} = -12678.394 \pm 1089.291$ \\
\hline \\ 

$\alpha_2(f_{\mathrm{sky}})$ & 
$\alpha_2(f_{\mathrm{sky}}) = C_2 \cdot \exp\left( -\dfrac{ \left( \log_{10}(f_{\mathrm{sky}}) - \mu_2 \right)^2 }{2\sigma_2^2} \right) + C_{\mathrm{offset}}$ & 
$C_2 = 1.740 \pm 0.094$, $\mu_2 = 2.846 \pm 0.077$ \\
& & $\sigma_2 = 1.949 \pm 0.151$, $C_{\mathrm{offset}} = 0.250 \pm 0.091$ \\
\hline \\

$\log_{10} a_t(f_{\mathrm{sky}})$ & 
$\begin{aligned} 
\log_{10} a_t = & C_{a1} \cdot \exp\left( -\dfrac{ \left( \log_{10}(f_{\mathrm{sky}}) - \mu_{a1} \right)^2 }{2 \sigma_{a1}^2} \right) \\
& + C_{a2} \cdot \exp\left( -\dfrac{ \left( \log_{10}(f_{\mathrm{sky}}) - \mu_{a2} \right)^2 }{2 \sigma_{a2}^2} \right) + C_{a,\mathrm{offset}}
\end{aligned}$ 
&
$\begin{aligned}
C_{a1} = 0.707 \pm 0.168, \quad \mu_{a1} = -2.939 \pm 1.176 & \\
\sigma_{a1} = 2.183 \pm 0.972, \quad C_{a2} = 2.879 \pm 0.075 \ \ \ \ & \\
\mu_{a2} = 4.492 \pm 0.027, \quad \sigma_{a2} = 0.473 \pm 0.061 \ \ \ \ & \\
C_{a,\mathrm{offset}} = -4.254 \pm 0.094 \ \ \ \ \ \ \ \ \ \ \ \ \ \ \ \ \ \ \ \  \ \ \ \ \ \ \ \ \ \  \ \ \ \ &
\end{aligned}$ \\
\hline \\

$a_t(f_{\mathrm{sky}})$ & 
$\begin{aligned} 
a_t = & D_{a1} \cdot \exp\left( -\dfrac{ \left( \log_{10}(f_{\mathrm{sky}}) - \mu_{b1} \right)^2 }{2 \sigma_{b1}^2} \right) \\
& + D_{a2} \cdot \exp\left( -\dfrac{ \left( \log_{10}(f_{\mathrm{sky}}) - \mu_{b2} \right)^2 }{2 \sigma_{b2}^2} \right)
\end{aligned}$ 
& 
$\begin{aligned}
D_{a1} &= 3.00 \times 10^{-4} \pm 1.75 \times 10^{-2} \\
\mu_{b1} &= -5.0 \pm 350.832 \\
\sigma_{b1} &= 3.709 \pm 207.327 \\
D_{a2} &= 3.45 \times 10^{-2} \pm 2.20 \times 10^{-3} \\
\mu_{b2} &= 4.608 \pm 0.098 \\
\sigma_{b2} &= 0.296 \pm 0.074 \\
\end{aligned}$ \\
\hline \\

\end{tabular}
\end{table*}




\bsp	
\label{lastpage}
\end{document}